\documentclass[11pt,a4paper,twoside, openright]{book} 

\usepackage[swedish, german, english]{babel} 
\usepackage[latin1]{inputenc} 

\usepackage{mathptmx} 
\usepackage{helvet} 
\usepackage{courier}
\usepackage[T1]{fontenc}
\usepackage{Thesis} 
\ifpdf
   \usepackage[pdftex]{graphicx}
  \else
    \usepackage[dvips]{graphicx}
\fi 

\usepackage{bbm,amsmath,amsfonts,bm}
\usepackage{boxedminipage}

\def\ket#1{\big |#1 \big \rangle}

\def\cross{\times}

\def\half{\frac{1}{2}}

\def\etal{{\em et al.}\ }

\def\cf{cf.\ }

\def\e{\mathrm{e}}

\def\ph#1{\phantom{#1}}
\def\phn{{\ph{0}}}

\def\beqn{\begin{eqnarray}}
\def\eeqn{\end{eqnarray}}

\def\abs#1{\left|#1\right|}

\renewcommand*{\d}{\mathrm{d}}
\newcommand*{\dx}{\d x}

\newcommand*{\dt}{\d t}
\newcommand*{\ds}{\d s}
\renewcommand* {\vec}[1]{\mathbf{#1}}
\def\del{\partial}
\def\e{\mathrm{e}}
\def\abs#1{\left|#1\right|}
\def\dilaton{{\mathit{\phi}}}
\def\phn{{\phantom{0}}}
\def\eqref#1{(\ref{#1})}
\renewcommand*{\i}{\mathrm{i}}

\let\br\-

\def\S{S}
\def\F{\phi}

\def\p{{(+)}}

\def\m{{(-)}}
\def\ppm{{(\pm)}}

\def\+{{+\!\!\!+}}
\def\-{=} 
\def\pp{{\mbox{\tiny${}_{\stackrel\+ =}$}}}

\def\eps{\epsilon}

\def\smallhalf{{\textstyle \frac{1}{2}}}
\def\ph#1{\phantom{#1}}
\def\phn{{\ph{0}}}

\def\d{{\rm d}}
\def\i{{\rm i}}
\def\half{{\frac{1}{2}}}

\def\+{{+\!\!\!+}}
\def\-{=} 
\def\ph#1{\phantom{#1}}
\def\phn{{\phantom{0}}}
\def\d{{\rm d}}

\ifx\bm\undefined
  \def\gen#1{{\bf #1}}
\else
  \def\gen#1{{\bf #1}} 
\fi
\def\genJ{\gen{J}}

\def\genG{\gen{G}}

\def\genU{\gen{U}}

\def\genGamma{{{\pmb \Gamma}}}

\ifx\mathbbm\undefined
  \def\genop#1{{\mathbb{#1}}}
\else
  \def\genop#1{{\mathbbm{#1}}}
\fi
\ifx\bm\undefined
  \def\genpartial{{\partial\hspace*{-1.2ex}\partial}}
  
  \def\bm\bf
\else 
  \def\genpartial{{\gen{\partial}}}
  
\fi
\def\matrix#1#2{\left(\begin{array}{#1}#2\end{array}\right)}
\def\genmatrix#1{\matrix{cc}{#1}}
\def\genMatrix#1{\matrix{ccc}{#1}}
\def\Z{\mathbbm{Z}}
\def\R{\mathbbm{R}}

\def\eq{\equiv}

\def\calB{\gen{B}}
\def\calH{\gen{H}}

\def\sign{{\rm sign}}

\def\Q{\gen{Q}}
\def\GJ{\gen{J}}
\def\GG{\gen{G}}
\def\GJt{{\tilde{\gen{J}}}}
\def\GK{{\gen{K}}}
\def\GKt{{\tilde{\GK}}}

\let\mathbb\mathbbm

\def\tsfrac#1#2{{\textstyle \frac{{#1}}{{#2}}}}
\def\chiral#1{{\mathbbm{#1}}}
\def\achiral#1{{\bar{\chiral{#1}}}}

\def\Xop#1{{\pmb #1}}
\def\Qop{\Xop{Q}}
\def\Pop{\Xop{P}}
\def\Hop{\Xop{H}}
\def\GenJ{{\Xop{J}}}
\def\Gen#1{{\Xop{#1}}}

\def\metric{G}
\def\bfield{B}
\def\metricpullback{g}
\def\wsmetric{h}
\def\H{H}
\def\Hamiltonian{\Hop}
\def\epsbos{a}

\def\B{\bfield}
\def\G{\metric}
\def\sX{u}
\def\sxi{\xi}
\def\sY{v}
\def\seta{\eta}
\def\h{\wsmetric}
\def\E{e}
\def\L{{\rm L}}
\def\chir{{\lambda}}
\def\twisted{{\chi}}
\def\Evec{{\Theta}}

\newcommand{\authorSurname}{Bredthauer} 
\newcommand{\authorFirstName}{Andreas} 
\newcommand{\authorFirstInitial}{A} 
\newcommand{\authorEmail}{Andreas.Bredthauer\@teorfys.uu.se} 
\newcommand{\dissertationTitle}{Tensionless Strings and Supersymmetric Sigma
Models} 
\newcommand{\dissertationSubtitle}{Aspects of the target space geometry}
\newcommand{\yearOfPublication}{2006} 
\newcommand{\placeOfDisputation}{Häggsalen, \AA ngstr\"{o}m Laboratory} 
\newcommand{\dateOfDisputation}{Saturday, September 30, 2006} 
\newcommand{\timeOfDisputation}{13:00} 
\newcommand{\disputationLanguage}{English} 
\newcommand{\numberOfPages}{viii+87} 
\newcommand{\placeOfPublication}{Uppsala} 
\newcommand{\ISBN}{91-554-6632-X} 
\newcommand{\series}{Uppsala Dissertations from the Faculty of Science and Technology} 
\newcommand{\serialNumber}{1214} 
\newcommand{\keywords}{ Theoretical Physics, String Theory, Tensionless Strings,
Supergravity Backgrounds, Plane Wave, Supersymmetry, Non-Linear Sigma Models,
Generalized Complex Geometry} 
\newcommand{\department}{Department of Theoretical Physics} 
\newcommand{\departmentaddress}{Ångströmlaboratorium, Lägerhyddsvägen 1, SE-752 37 Uppsala,
Sweden} 
\newcommand{\ISSN}{1651-6214} 
\newcommand{\urn}{urn:nbn:se:uu:diva-7105} 
\newcommand\abstracttext{%
In this thesis, two aspects of string theory are discussed, tensionless strings
and supersymmetric sigma models. 

The equivalent to a
massless particle in string theory is a tensionless string. Even almost 30 years
after it was first mentioned, it is still quite poorly understood. We discuss
how  tensionless strings give rise to exact solutions to supergravity and solve
closed tensionless string theory in the ten dimensional maximally supersymmetric
plane wave background, a contraction of $AdS_5\times S^5$ where tensionless
strings are of great interest due to their proposed relation to higher spin
gauge theory via the AdS/CFT correspondence.

For a sigma model, the amount of supersymmetry on its worldsheet restricts the
geometry of the target space. For $N=(2,2)$ supersymmetry, for example, the
target space has to be bi-hermitian. Recently, with generalized complex
geometry, a new mathematical framework was developed that is especially suited
to discuss the target space geometry of sigma models in a Hamiltonian
formulation. Bi-hermitian geometry is so-called generalized Kähler geometry but
the relation is involved. We discuss various amounts of supersymmetry in phase
space and show that this relation can be established by considering the
equivalence between the Hamilton and Lagrange formulation of the sigma model.
In the study of generalized supersymmetric sigma models, we find objects
that favor a geometrical interpretation beyond generalized complex geometry.
} 

\renewcommand{\listofpapers}
{\chapter*{List of Papers}
This thesis is based on the following papers, which are referred
to in the text by their Roman numerals. \vspace{13pt}
    \begin{romanlist}
 \setlength{\itemsep}{1.5ex}
 \item A.~Bredthauer, U.~Lindström and J.~Persson.
  $SL(2,\mathbb{Z})$ tensionless string backgrounds in IIB string theory.
  {\sl Classical and Quantum Gravity} {\bf 20} (2003) 3081. 
  hep-th/0303225.
 \item   A.~Bredthauer, U.~Lindström, J.~Persson and L.~Wulff.
  Type IIB tensionless superstrings in a pp-wave background.
  {\sl Journal of High Energy Physics} {\bf 0402} (2004) 051. {hep-th/0401159}.
 \item  A.~Bredthauer, U.~Lindström and J.~Persson.
  First-order supersymmetric sigma models and target space geometry.
  {\sl Journal of High Energy Physics} {\bf 0601} (2006) 144. {hep-th/0508228}.
 \item  A.~Bredthauer, U.~Lindström, J.~Persson and M.~Zabzine.
  Generalized Kähler geometry from supersymmetric sigma models.
  {\sl Letters in Mathematical Physics} (2006), {\sl in press}. {hep-th/0603130}.
 \item  A.~Bredthauer.
  Generalized hyperkähler geometry and supersymmetry.
  {\sl Manuscript} (2006). {hep-th/0608114}.
    \end{romanlist}
\vspace{13pt} \noindent {\timesTen 
}}

\newcommand{\dedication}%
{\cleardoublepage
\thispagestyle{empty}
\vspace*{\stretch{3}}
\begin{flushright}
		
		{\fontfamily{pzc}\Large\selectfont
        \emph{}}

\end{flushright}
\vspace*{\stretch{1}}} 

\begin{document}
	\pagenumbering{roman}
    
    \newcommand{\frontmattermyCS}{\titlepage\cleardoublepage
    \listofpapers}
    \frontmattermyCS
    
    \tableofcontents
    

	\cleardoublepage
	\pagenumbering{arabic}
	\setcounter{page}{1}
    \parskip 1ex
    \chapter{Introduction}

String theory is one of the most fascinating subjects modern theoretical physics
ever developed. It unifies two fundamental concepts that at first sight do not fit
together: gravity and quantum mechanics. This makes it `the' candidate
for a theory of nature. While electromagnetic, weak and strong interactions can
be described by quantum field theories to reasonable accuracy, they fail in
giving a proper description of gravity. On the other hand, we can describe
gravity at large distances by Einstein's general relativity. String theory
crosses the barrier between these two different theories with a seemingly simple
and naive idea: Why not consider one-dimensional objects, strings, as the basic
constituents of nature instead of point-like particles? 

But let us start with a short overview of the history of string theory. String
theory in the way we view it today was not invented but rather discovered. At
the end of the 1960's people were analyzing scattering amplitudes of hadronic
matter. String theory was proposed as a model for these interactions.
Scattering of relativistic strings seemed to match with the experimental data.
Unfortunately, this turned out to be wrong. String theory just was not able to
describe, for example, effects in deep inelastic hadron scattering. The correct
description was instead an ordinary quantum field theory. Quantum chromodynamics
was born after the discovery of asymptotic freedom in 1973: The strength of the
strong interaction between two quarks, the constituents of hadronic matter,
decreases as they approach each other.

At around the same time, the discovery of an excitation that had
gravitation-like interactions in the string spectrum triggered a new view of
string theory that is still valid today. Suddenly, it became a candidate for a
theory unifying the four fundamental forces. That strings have not been observed
in nature was explained by their size. The typical size of a string is at the
order of the Planck length, such that probing string theory directly would need
much higher energy than provided by experiments. String theory is finite at high
energies, in contrast to ordinary quantum field theories that all have the problem
of infinities. A lot of interesting properties were discovered after 1975.
At low energies corresponding to large distances, the gravitational interactions
resemble exactly Einstein gravity, while they obtain corrections at short
distances. This fits with the picture that general relativity breaks down below
the Planck scale where quantum fluctuations are supposed to take over. Also
supersymmetry, a symmetry that mixes bosons with fermions, was found to be
naturally included in string theory. Unfortunately, at those times, there were
too many string theories, and there did not seem to be any principle for which
one to choose.

Things changed after what is now called the first string revolution in 1985.
Since then, we know that there are only five consistent theories at quantum
level. All of them live in ten spacetime dimensions, they are called type I,
type IIA and IIB, and the two heterotic string theories with gauge groups
$SO(32)$ and $E_8\times E_8$. The problem with the extra dimensions was solved
by compactification. If the six extra dimensions are small enough, say at the
order of the Planck scale or below, we would never be able to detect them with our
experimental equipment. Supersymmetry was supposed to be unbroken at the 
compactification scale, at the size of the internal space so to speak. The
four-dimensional space should be the flat space we see and this puts very strong
constraints on the geometry of the internal six-dimensional space.

The second superstring revolution in 1995 revealed two things. First, the five
string theories are dual to each other, related by certain duality
transformations. In fact, they are perturbative expansions of one and the same
theory around different vacua. It is here, the famous M-theory enters the game.
However, despite the fact that we know it is there, not too many things are
known about it. Also, the second revolution introduced D-branes. These solitonic
objects had been known for some time but their importance to modern string
theory was first realized then. Not only is their worldvolume dynamics governed
by open strings attached to them, their existence allows for the idea that our
world might be bounded to such a brane, explaining, for example, why gravity
couples so weakly to matter.
 
Today, string theory is such a broad field of research that it is very hard to
give a complete picture of the current research. Certainly, this is not the
right place to give an introduction to string theory either. There are great
books that cover this subject \cite{Green:1987sp,Polchinski:1998rq,
Johnson:2003gi,Zwiebach:2004tj}. Also, there are some useful lecture notes
available \cite{Szabo:2002ca,Mohaupt:2002py}, just to mention some. 

The aim of this thesis is to give an introduction to the subjects that are
covered in the publications [I] to [V], tensionless strings and supersymmetric
sigma models.  This serves also as a motivation for our work. In the rest of the
thesis, we mainly focus on going through parts of our work in detail and
providing some background information for a better understanding of our results.
The list of references is not exhaustive. For a more complete list, we refer to
the papers [I-V].

In particle physics massless particles play an important role. Not only is
the photon, the carrier of the electromagnetic force, massless but particles
at very high kinetic energies can be considered as approximately massless. The
equivalent of the mass of a particle in string theory is the tension $T$ of
the string, its mass per unit length. The tensionless string first appeared when
discussing strings moving at the speed of light and is still very poorly
understood. Similar to massless particles, tensionless strings are believed to
have their place in the study of the high energy behavior of string theory. For
example, we can consider a string that rotates with higher and higher angular
momentum. As the angular momentum increases the energy gets localized around the
endpoints of the string while its core becomes tensionless. The fact that the
tension is zero turns the string basically into a collection of freely moving
particles --- it falls into pieces. However, these pieces are still connected to
each other since the string is a continuous object even in the tensionless case.
Tensionless strings have been studied for a long time, classically and
quantized, with and without supersymmetry. 

Tensionless string theory exhibits a much larger spacetime symmetry
than the tensionfull theory. The quantum theory differs drastically. In flat
space the spectrum collapses to a common zero-mass level. Especially tachyonic
states that are usually unstable and have to be banned from the physical
spectrum due to their negative mass squared, become massless and thus stable for
the tensionless string. The quantum theory has either a topological spectrum or
for the case of $D=2$ spacetime dimensions, the spacetime symmetry is retained.
There is no critical spacetime dimension for the tensionless string and the
spectrum has a huge symmetry involving higher spin gauge fields. The tensionless
string is supposed to be the unbroken phase of string theory where all states
are still considered on an equal footing and that breaks as the energy
decreasing giving rise to the different mass levels.

The tensionless string appears in various situations. The ordinary string is
approximately tensionless in a highly curved background and it appears in the
context of intersecting branes. In general quantization does not commute with
taking the tension to zero. In flat space, the common mass level has its origin
in the fact that string theory only has a single energy scale, the tension
In the tensionless limit there is no scale left. We show that tensionless
strings have a natural place in the context of supergravity. We find a
background for type IIB string theory that we are able to interpret as the
geometry sourced by a tensionless string.

The relation between higher spin gauge theory and tensionless strings can
probably be easiest understood in the context of the AdS/CFT correspondence.
If one looks at a hologram one sees a three dimensional picture that is stored in
a two dimensional area.
In string
theory this holographic principle in its most famous version states that string
theory in an Anti-de Sitter space has a dual description in
terms of a supersymmetric conformal field theory on the boundary of the space. 
This correspondence has been tested ever since it has been
conjectured back in 1997 and lead to such amazing results as that certain
sectors of the string theory are integrable models that can be treated with
solid state physics methods, but at least to my knowledge, no rigorous proof is
known. It relates the string tension to the coupling constant of the gauge
theory. Thus, the tensionless string corresponds to a vanishing gauge theory
coupling where higher spin gauge fields appear. Five dimensional anti-de Sitter
is part of a larger space where type IIB string theory is consistent,
$AdS_5\times S^5$. Unfortunately, string theory on this background is rather
difficult and not much is known about the quantum theory.
There are three known backgrounds for type IIB supergravity that are maximally
supersymmetric. That means they preserve 32 supersymmetries. These are flat
space, $AdS_5\times S^5$ and a very recently discovered so-called plane wave
background. This latter shares a lot of properties with $AdS_5\times S^5$ but
is considerably simpler. In fact, it can be derived as a certain limit of
$AdS_5\times S^5$. It turns out that closed string theory is a solvable model
on this spacetime, at least in light-cone gauge gauge, where only the physical
degrees of freedom are taken into account it has been solved and quantized.
We analyze the closed tensionless
type IIB string in this plane wave background and compare it to the tensile
case with two main results. For the first, as opposed to flat space, the
quantum theory is well-behaved and can actually be derived as a limit of the
tensile theory. This can be traced back to a scale provided by the background
itself that survives the tensionless limit. Secondly, the tension enters the
solution only in combination with this scale parameter, which is actually
related to the curvature of the space. Therefore, our result has a dual
description in terms of a tensile string in a highly curved plane wave
background.

The way string theory determines its own target space geometry is rather
intriguing. It was already mentioned in the context of compactification, that
for consistency, the internal six-dimensional manifold has to be of a certain
type. This type is determined by the fact, that we want to consider
four-dimensional space with $N=1$ supersymmetry. If the internal space is to be
Kähler, then the only choice is Calabi-Yau.  Even tough people were
aware, that there are solutions that are not Kähler, these possibilities were
not considered for a long time. For a sigma model with supersymmetry on
the worldsheet of a string, that is the area the string sweeps out in the target
manifold called spacetime, the geometry of its target space is determined by the
dimension of the worldsheet and the amount of supersymmetry. For example the
manifest $N=(1,1)$ supersymmetric sigma model admits twice this amount of
supersymmetry if the target space is bi-hermitian.  Although classified, again
the cases that are not Kähler were not considered to be of major importance.
Lately, a new mathematical concept, generalized complex geometry, was founded
that unifies complex and symplectic geometry. In fact, it smoothly interpolates
between them.  It turned out to be the right framework to discuss this
interesting relation between worldsheet supersymmetry and target space geometry
in. It was found that a subset of these new geometries called generalized Kähler
geometry is equal to the bi-hermitian geometry and moreover that it can be
completely described in terms of manifest $N=(2,2)$ supersymmetry. Generalized
Calabi-Yau is another subset and is considered in compactifications with fluxes.
Finally, generalized complex geometry might give a mathematical explanation for
mirror symmetry.  It unifies the topological A- and B-model into a single model.

Based on the fact that generalized complex geometry is related to the discussion
of supersymmetry in the sigma model phase space, we show how generalized
Kähler geometry arises very naturally in the Hamiltonian treatment of the
supersymmetric sigma model. We argue that from the physics point of view, the
relation between bi-hermitian and generalized Kähler geometry is established by
the equivalence of the Hamiltonian and the Lagrangian treatment of the sigma
model. We then go a step further and show how another subset, called
generalized Hyperkähler geometry is related to $N=(4,4)$ supersymmetry on the
worldsheet in the same way. The sigma model can be generalized by introducing
auxiliary fields. We argue how supersymmetry in such a case favors a
target geometry that is beyond generalized complex geometry. The lack of a
proper understanding of these geometries manifests itself in the absence of a
proper mathematical notion. This leaves us bound to a very simple toy-model.
However, we are able to identify the relevant geometrical objects and show how
generalized complex geometry is included in this new type of geometries.

We conclude with a summary of the publications included in this thesis.

\subsubsection*{Paper I}
In the first paper, we describe how tensionless strings give rise to background
solutions in IIB supergravity. Our starting point the geometry that is
sourced by a macroscopic string which we then accelerate to the speed of light. In
this limit, the string tension vanishes and the geometry becomes similar to a
gravitational shock wave.

\subsubsection*{Paper II}
We study the closed, tensionless IIB string in a maximally supersymmetric plane
wave background. The solution is similar to the case of non-vanishing tension.
Quantization of the tensionless string turns out to be unproblematic, as opposed
to flat space. This can be traced back to the existence of a parameter related
to the curvature of the background. We show that the tensionless string can be
derived as a certain limit of the tensile string in this background and conclude
that the limit commutes with quantization.

\subsubsection*{Paper III}
In the third paper, we discuss the condition for which a generalized $N=(1,1)$
supersymmetric sigma model admits additional supersymmetries. We find that the
involved tensors naturally group together into objects that suggest an
interpretation beyond generalized complex geometry. Since we lack a proper
understanding of this type of geometry, we are bound to a simple toy-model, such
that we only can identify the relevant geometric objects and show how
generalized complex geometry is embedded in this description.

\subsubsection*{Paper IV}
We clarify the relation between generalized Kähler geometry and bi-hermitian
geometry from a sigma model of view. We show that generalized Kähler geometry
is the condition for $N=(2,2)$ supersymmetry in a phase space formulation of
the sigma model. The relation between generalized Kähler geometry and
bi-hermitian geometry follows thus from the equivalence of the Hamiltonian and
Lagrangian formulation of the sigma model. As an application of our results, we
even discuss topological twists.

\subsubsection*{Paper V}
In this paper, we study the condition for $N=(4,4)$ supersymmetry in the
Hamilton formulation of the sigma model. We find the definition of generalized
hyperkähler geometry and define the twistor space of the generalized complex
structures.

    \chapter{String theory basics}\label{basics}
This chapter provides an elementary overview of those aspects of string
theory that are needed to understand this thesis. We also use the opportunity to
introduce our conventions and notations. For a broader introduction to string
theory, we again refer to a number of good textbooks
\cite{Green:1987sp, Polchinski:1998rq, Zwiebach:2004tj}.

The motion of a relativistic point particle with mass $m$ in spacetime is
governed by the action
\begin{align}
  S_{\rm part} = m \int \d t \sqrt{-\dot X^2}. \label{S_part}
\end{align}
Here, $X(t)$ is the position of the particle at time $t$. The action is thus
equal to the length of the particle's worldline. The action principle tells us
that classically, the particle chooses the shortest path between two points. A
string is a one dimensional object moving in spacetime. We can regard its motion
as a two dimensional worldsheet $\Sigma$ embedded in the spacetime $M$ by maps
$X: \Sigma\rightarrow M$. The worldsheet has Minkowski signature with a time
direction $\tau$ and a spacial direction $\sigma$, which we conveniently combine
into a single coordinate $\xi^a$, $a=0,1$. We use both notations on an equal
footing. Strings can be open or closed making the worldsheet either a strip or a
cylinder.  In this thesis, we mainly consider closed strings. Therefore,
$\Sigma=\mathbbm{R} \times S^1$. The compact direction is the spacial one, such
that $\sigma\simeq \sigma+\pi$. In analogy to the particle, the string moves
classically in such a way that it minimizes the area it sweeps out in spacetime.
The action is equal to the world volume of the string 
\begin{align}
  S_{\rm NG} = T\int_\Sigma \d^2\xi \sqrt{-\metricpullback},
\label{S_NG}
\end{align}
This action is called the Nambu-Goto action of the bosonic string. The factor
$T$ is the string tension and $\metricpullback$ is the determinant of
$\metricpullback_{ab} = \partial_a X^\mu \partial_b X^\nu \eta_{\mu\nu}$. This
is the pullback of the spacetime metric onto $\Sigma$.  For the moment, we
consider a string in $D$-dimensional Minkowski space. The
determinant is equal to 
\begin{align} 
\metricpullback = -\dot X^2 X^{\prime 2}
+ (\dot X\cdot X^\prime)^2.
\end{align}
We denote a derivative with respect to $\tau$ by a {\sl dot} and a
$\sigma$-derivative by a {\sl prime}.  The conjugate momenta $P_\mu =
T\sqrt{-\metricpullback}\metricpullback^{a0}\partial_a X_{\mu}$ derived from the
action are constrained:
\begin{align} 
  P_\mu X^{\prime \mu} = 0, && P_\mu P^\mu + T^2
    \metricpullback\metricpullback^{00}
   = 0. \label{Virasoro-NG}
\end{align}
These are the Virasoro constraints. Here, $\metricpullback^{ab}$ is the inverse
of $\metricpullback_{ab}$.  There is an equivalent way to write the string
action that avoids taking the square root of the fields and incorporates the
Virasoro constraints. It makes use of the worldsheet metric $h_{ab}$ and is
given by
\begin{align}
  S_{\rm Poly} = -\frac{T}{2} \int \d^2\xi
\sqrt{-h}h^{ab}\partial_a X^\mu \partial_b X^\nu \eta_{\mu\nu}. \label{S_Poly}
\end{align}
This action was found by Brink, Deser, di Vecchia, Howe and Zumino
\cite{Brink:1976sc,Deser:1976rb} but is usually known as the Polyakov action
\cite{Polyakov:1981rd,Polyakov:1981re}. This action is a special case of a sigma
model that maps one space into another, in this case the worldsheet $\Sigma$
into spacetime. The way the worldsheet is embedded in spacetime does not depend
on how we choose to parametrize it, the action is invariant under
reparametrizations of the worldsheet 
\begin{align}
  \delta(a)X^\mu = a^a\partial_a X^\mu, && \delta(a)\wsmetric^{ab} = a^c\partial_c
\wsmetric^{\,ab}
  - \partial_c a^{(a}\wsmetric^{\,c|b)}. \label{repsym}
\end{align}
Here, $A^{(ab)}= A^{ab}+A^{ba}$ denotes symmetrization in the indices $a$ and
$b$. We define symmetrization and antisymmetrization ($A^{[ab]}=A^{ab}-A^{ba}$)
without a factor. Local Weyl transformations generate an additional symmetry
of the worldsheet. They are parametrized by scalar functions on the
worldsheet $\Lambda(\sigma,\tau)$ and multiply
the worldsheet metric by a factor while leaving $X^\mu$ invariant
\begin{gather} 
  \delta(a)\h^{\,ab} = \Lambda(\sigma,\tau)\wsmetric^{\,ab}.
\end{gather}
The field equation for $\wsmetric^{ab}$ requires the two-dimensional energy
momentum tensor to vanish
\begin{align}
  T_{ab} = (\del_a X^\mu \del_b X^\nu  - \half
\wsmetric_{ab}\wsmetric^{cd}\del_c X^\mu \del_d X^\nu)\eta_{\mu\nu}
\stackrel{!}{=} 0. \label{Virasoro}
\end{align}
This is a consequence of the reparametrization invariance and it can be used to
integrate out the worldsheet metric and obtain back the Nambu-Goto action, since
it tells us that the determinant of $\metricpullback_{ab}$ is given by 
\begin{align}
  \metricpullback = \frac{1}{4}\wsmetric(\wsmetric^{ab}\metricpullback_{ab})^2.
\end{align}

We can use reparametrization invariance and Weyl symmetry do choose a
conformally flat worldsheet metric, $\wsmetric_{ab}=\eta_{ab}$. This choice is
called the conformal gauge.  Worldsheet light-cone coordinates $\xi^\pp = \tau
\pm \sigma$ correspond to left and right moving modes on the string. We denote
the worldsheet indices by $\+$ and $\-$ in order to distinguish them from
fermionic worldsheet indices $+$ and $-$ which we introduce in the discussion of
supersymmetry. In these coordinates,
the string action becomes
\begin{align}
  S=\frac{T}{2}\int \d^2\xi \partial_\+ X^\mu \partial_\- X^\nu \eta_{\mu\nu}.
\label{LCG-action}
\end{align}
This must be supplemented by requiring the energy momentum tensor $T_{ab}$ to
vanish. This is now a constraint. $T_{ab}$ is traceless and in coordinates
$\xi^{\pp}$, the constraints are given by $T_{\+\+}=T_{\-\-}=0$.  Since the
conjugate momenta are $P_\mu = T\eta_{\mu\nu}\dot X^\nu$, we recover exactly the
Virasoro constraints \eqref{Virasoro-NG}.

After choosing a conformally flat worldsheet metric there is still some gauge
freedom left.  We may choose light-cone coordinates $X^\pm =
\frac{1}{\sqrt{2}}(X^0 \pm X^{D-1})$, $X^I$, $I=1\ldots D-2$ on the target
space. The equation of motion for $X^\mu$ are the wave equations
\begin{align}
  \partial_\+\partial_\- X^\mu = (\partial_\sigma^2 - \partial_\tau^2) X^\mu = 0.
\end{align}
The remaining symmetry is given by reparametrizations of the worldsheet of the form
\begin{gather}
  \tau \rightarrow f^\p(\tau+\sigma) + f^\m(\tau-\sigma), \cr
  \sigma \rightarrow f^\p(\tau+\sigma) - f^\m(\tau-\sigma), \cr
  \wsmetric^{ab} \rightarrow (\partial_\+ f^\p\partial_\-f^\m)^{-1}\wsmetric^{ab}.
\end{gather}
Herein, $f^\p$ and $f^\m$ are arbitrary functions that leave the form of the metric
$\wsmetric^{ab}=\eta^{ab}$ invariant. After such a transformation, the new time
coordinate satisfies the one-dimensional wave equation
$(\partial_\sigma^2-\partial_\tau^2)\tau_{\rm new} = 0$. Since $\tau$ and $X^+$
both satisfy the wave equation, we can use the remaining gauge freedom to relate
them to each other by fixing 
\begin{align} 
  X^+(\sigma,\tau) = \frac{p^+}{T}\tau.
\end{align} 
The constant $p^+$ is the conjugate momentum for $X^+$. This gauge is called the
light-cone gauge and we see that $X^+$ and $X^-$ completely decouple from the
action. $X^-$ can be determined by the Virasoro constraints which in light-cone
gauge read
\begin{align}
  p^+X^{-\prime} + T\dot X^I X^{\prime}_I = 0, &&
  2p^+\dot X^- + T(\dot X^I \dot X_I + X^{I\prime}X^\prime_I) = 0.
\label{Virasoro-constraints}
\end{align}
One concludes that there are only $D-2$ physical bosonic degrees of freedom of
the string given by the transverse components $X^I$.

\section{Non-linear sigma model}

String theory is a special case of a non-linear sigma model. In general, such a
model embeds one space into another. It consists of a base manifold $\Sigma$ and a
target manifold $M$ and a map 
\begin{align}
  X^\mu: \Sigma \rightarrow M
\end{align}
that stands for the embedding. The case where $\Sigma$ is a two dimensional
worldsheet is very special, since it allows for conformal invariance of the
worldsheet. Of course, there is no need for the target manifold to be flat. It can be a
curved spacetime with metric $\metric_{\mu\nu}(X)$, but it can also be supported
by a two-form $\B_{\mu\nu}(X)$ and a scalar field $\dilaton$ called the dilaton.
Putting everything together, we obtain the most general action for a bosonic
string
\begin{align}
  S = -\half \int \d^2\xi
\Big(T(\sqrt{-\wsmetric}\wsmetric^{ab}\metric_{\mu\nu} +
\eps^{ab}\bfield_{\mu\nu}) \partial_a X^\mu \partial_b X^\nu + 8\pi
\sqrt{-\wsmetric}R\phi\Big), \label{S_NLSM} 
\end{align}
where $R$ is the two-dimensional Ricci scalar for $\wsmetric$. We see
that we can obtain \eqref{S_Poly} as a special case of it with a worldsheet
periodic in the spacial direction $\Sigma=S^1\times \R$. The part involving the
dilaton arises as a one loop effect, while the first two terms form the
celebrated non-linear sigma model. In conformal gauge when the worldsheet metric
is chosen to be conformally flat, the non-linear sigma model action reads 
\begin{align}
  S_{NLSM} = \half \int \d^2\xi \partial_\+ X^\mu \partial_\- X^\nu
   \big(\metric_{\mu\nu}(X) + \bfield_{\mu\nu}(X)\big). \label{S_NLSM_LCG}
\end{align}
Metric and $\bfield$-field can be conveniently
combined into a single tensor $\E_{\mu\nu} = \G_{\mu\nu} + \B_{\mu\nu}$. The
field strength for $\B$, $\H=\d\B$ is explicitly given by 
\begin{align}
  \H_{\mu\nu\rho} = \half\big(\B_{\mu\nu,\rho} + \B_{\nu\rho,\mu} +
    \B_{\rho\mu,\nu}\big).
\end{align}
Indices separated by a comma denote partial spacetime derivatives
$\B_{\mu\nu,\rho} =\partial_\rho \B_{\mu\nu}$. It is important to stress that
the action \eqref{S_NLSM_LCG} does not depend on $\B$ but on its field strength
$\H$ only. This is seen easiest by invoking Stokes theorem. If we assume that
$\Sigma$ is the boundary of some three-dimensional worldsheet $\Sigma^3$,
$\Sigma = \partial \Sigma^3$ and denote the pullback of $\bfield$ onto the
worldsheet $\Sigma$ by $\varphi^*(\B)$, we find
\begin{align}
  \int_\Sigma \varphi^*(\bfield) = \int_{\Sigma^3} \varphi^*(\H). \label{WZterm}
\end{align}
The term involving $\B$ respective $\H$ is called a Wess-Zumino term. It is
indeed possible to consider the more general case when $H$ is closed but not
exact.

The study of sigma models in general differs somewhat from the discussion of
string theory. We regard \eqref{S_NLSM_LCG} as a field theory for $X^\mu$.
If we want to discuss string theory, we have to make use of the Virasoro
constraint \eqref{Virasoro} as well.  From the field theory point of view, the
Lagrangian formulation and the action principle is just one way to study the
sigma model. Equivalently, we can change to a phase space formulation and
describe the worldsheet dynamics in terms of a Hamiltonian.

In the phase space formulation, the base manifold has one less dimension as
compared to the Lagrangian formulation. The phase space of a worldsheet of the
two dimensional sigma model with spacial periodic boundary conditions on the
worldsheet can be identified with the cotangent bundle $T^*LM$ of the loop space
$LM = \{X : S^1 \rightarrow M\}$ \cite{Alekseev:2004np}. The loop space
consists of vector fields $X^\mu(\sigma)$ embedding the spacial direction of the
worldsheet into the manifold. $X^\mu$ is periodic in $\sigma$:
$X^\mu(\sigma+\pi)=X^\mu(\sigma)$. With this, points in $T^*LM$ are given by 
pairs $(X^\mu, \pi_\mu)$ where $\pi_\mu$ is a section of the cotangent bundle at
$X$. When considering a string moving in spacetime, we can
parametrize its current position and conjugate momentum by a
such a pair $(X^\mu(\sigma),P_{\mu}(\sigma))$.

Momentum and fields are conjugated by means of a two form, the canonical symplectic
structure
\begin{align}
  \omega = \int_{S^1} \d \sigma \delta X^\mu\wedge \delta P_\mu.
\label{symplectic-form}
\end{align}
It yields the Poisson bracket
\begin{align}
  \{F,G\} = \int_{S^1} \d\sigma F\left( 
  \frac{\overleftarrow\delta}{\delta P_\mu}
  \frac{\overrightarrow\delta}{\delta X^\mu} -
  \frac{\overleftarrow\delta}{\delta X^\mu}
  \frac{\overrightarrow\delta}{\delta P_\mu}\right) G.
\end{align}
In phase space, we can consider generators for the symmetries of the worldsheet.
The generator of $\sigma$-translations is given by
\begin{align}
  \Xop{P}(a) = -\int \d\sigma P_\mu \del X^\mu, \label{P-bos}
\end{align}
where $\del \eq \del_\sigma$. It acts on the field via the Poisson bracket
\begin{align}
  \delta(a)X^\mu = \{X^\mu, \Xop{P}(a)\} = a\del X^\mu, &&
  \delta(a)P_\mu = \{P_\mu, \Xop{P}(a)\} = a\del P_\mu.
\end{align}
In the presence of a closed three form $H\in \Omega^3(M)_{cl}$, the symplectic 
structure is twisted in the following way:
\begin{align}
  \omega_H = \int_{S^1} \d \sigma \big(\delta X^\mu\wedge \delta P_\mu +
H_{\mu\nu\rho}\partial X^\mu \delta X^\nu \wedge \delta X^\rho\big).
\label{omega-H}
\end{align}
This is the case when the Wess-Zumino term \eqref{WZterm} is present in the
action of the sigma model. It yields a twisted version of the Poisson bracket,
denoted by $\{F,G\}_\H$.  Also, $\Xop{P}(a)$ gets twisted appropriately. The
details are part of the appendix of [IV]. If not otherwise stated, we always
assume that $\H$ is the field strength for $\B$. The symplectic structure is
invariant under transformations of the kind
\begin{align}
  X^\mu \rightarrow X^\mu,&& P_\mu \rightarrow P_\mu + \bfield_{\mu\nu}\partial X^\nu.
\label{bosonic-b-transformation}
\end{align}
This is a symmetry of the symplectic structure if $\bfield$ is a closed two-form, 
$\bfield\in \Omega^2(M)_{cl}$.  If $\bfield$ is not closed, such a
transformation twists the symplectic structure by $\d \bfield$. This will be an
important fact in the discussion of supersymmetric sigma models and generalized
complex geometry in chapter \ref{susysm}. To describe dynamics, the phase space
is accompanied by a (canonical) Hamiltonian. It is the generator of time
evolution. The Hamiltonian corresponding to \eqref{S_NLSM_LCG} with $\B=0$ is derived 
in by a Legendre transformation with respect to the worldsheet coordinate
$\tau=\xi^0$. With $P_\mu = \G_{\mu\nu}\dot X^\nu$ we can rewrite the action
\eqref{S_NLSM_LCG} in phase space
\begin{align}
  {S}_g = \int \d t \d\sigma \Big( P_\mu \dot X^\mu - \frac{1}{2}\big(
P_\mu P_\nu G^{\mu\nu}+ \partial X^\mu \partial X^\nu \G_{\mu\nu}\big)\Big).
\end{align}
The first part yields a presymplectic form, the so-called Liouville form
\begin{align}
  \Theta = \int \d\sigma P_\mu \delta X^\mu, \label{Liouville}
\end{align}
whose differential is the symplectic form $\omega = \delta\Theta$
\eqref{symplectic-form}. The second part is the Hamiltonian
\begin{align}
  \Hamiltonian(P, X) = \frac{1}{2}\int \d\sigma \Big(P_{\mu} P_\nu G^{\mu\nu} + \partial X^\mu
\partial X^\nu G_{\mu\nu}\Big).
\end{align}
The $\B$-field can be included using the $\B$-transformation
\eqref{bosonic-b-transformation}. The second term in \eqref{S_NLSM_LCG}
can be obtained in two different ways. One can perform the transformation on the
presymplectic form \eqref{Liouville}, such that 
\begin{align}
  \Theta_\bfield = \int \d\sigma (P_\mu + \bfield_{\mu\nu}\partial X^\nu)\delta
X^\mu.
\end{align}
This results in a twisting of the symplectic structure with
$\omega_{\H} = \delta \Theta_\bfield$. Acting with the inverse transformation on the Hamiltonian
generates the same term
\begin{align}
  \Hamiltonian_\bfield = \frac{1}{2}\int \d\sigma \Big( (P_\mu - \bfield_{\mu\rho} \partial
X^\rho)\G^{\mu\nu}(P_\nu - \bfield_{\nu\sigma}\partial X^\sigma) - \partial X_\mu
\partial X^\mu \Big).
\end{align}
The difference is that in the first way, $P_\mu$ denotes the
physical momentum, while for the second, it is the canonical momentum for $X^\mu$.
The physics described by the Hamiltonian is the same as for the action
\eqref{S_NLSM_LCG}. Consequently, also here, only $\H$ is important and not
$\bfield$.  Assigning the contribution from the $\bfield$-field to the
symplectic structure is thus the preferred choice. This makes it possible
to also discuss twists with closed but not exact three forms. We will see later, that
this is a crucial point in the $N=(1,1)$ supersymmetric version of the sigma
model.  There, the twisted Hamiltonian contains an additional, purely fermionic
piece proportional to the flux $\H=\d \bfield$ that cannot be removed by a
$\bfield$-transformation of the form \eqref{bosonic-b-transformation}.

Let us consider a vector field $\sX^\mu(X)$ and a one-form field $\sxi_\mu(X)$ on the
target manifold $M$. We can associate the following current to it:
\begin{align}
  J_{\sX+\sxi}(\sigma) = \sX^\mu(X(\sigma)) P_\mu(\sigma) +
    \sxi_\mu(X(\sigma)) \partial X^\mu(\sigma).  \label{Current}
\end{align}
These types of currents play an important role in the discussion of symmetries for a wide
class of two dimensional sigma models and have been studied in
\cite{Alekseev:2004np}.  We already saw that the current
\begin{align}
  J_P(\sigma) = P_\mu\partial X^\mu
\end{align}
yields the generator of $\sigma$-translations \eqref{P-bos}. The Poisson
bracket of two currents of the form \eqref{Current} has two parts 
\begin{multline}
  \{ J_{\sX+\sxi}(\sigma), J_{\sY+\seta}(\sigma') \} = \cr J_{[\sX+\sxi,
\sY+\seta]_c}(\sigma)\delta(\sigma-\sigma') + \half (\sX^\mu\seta_\mu +
\sY^\mu\sxi_\mu) \delta'(\sigma-\sigma').
\end{multline}
The first part is this kind of current associated to 
the Courant bracket of $\sX+\sxi$ and $\sY+\seta$ 
\begin{align}
  {[}\sX+\sxi, \sY+\seta{]}_c = {[}\sX,\sY{]} + L_\sX\seta - L_\sY \sxi -
  \frac{1}{2}\d(i_\sX\seta - i_\sY\sxi). \label{Courant-from-current}
\end{align}
Here, $L_\sX\,\cdot = \d (i_\sX\, \cdot) + i_\sX \d\,\cdot$ is the Lie
derivative and $i_\sX \sxi = \sX^\mu \sxi_\mu$ is the contraction of a vector
field and a one-form. The Courant bracket reduces to the ordinary Lie bracket
when restricted to vector fields $\sX$ on $TM$. 
 
\section{Worldsheet supersymmetry}\label{ws-susy}
If we quantize string theory with the action \eqref{S_Poly}, or even in the more
general background with \eqref{S_NLSM}, the physical spectrum only contains
bosons. Since nature contains also fermions and string theory is supposed 
to eventually describe fundamental physics, we must include
fermions. A way for doing that is to consider supersymmetry. Supersymmetry is
the only possible non-trivial extension of the Poincaré algebra. If $P_\mu$ is
the generator for spacetime translations and $M_{\mu\nu}$ generates Lorentz
rotations then the spacetime symmetries consistent with a relativistic quantum
field theory are generated by
\begin{gather}
  [P_\mu,P_\nu] = 0, \hspace*{1cm} [M_{\mu\nu},P_\rho] =
\frac{1}{2}\eta_{\rho[\mu}P_{\nu]}, \cr
  [M_{\mu\nu},M_{\rho\sigma}] = \frac{1}{2}\eta_{\rho[\mu}M_{\nu]\sigma} -
   (\rho\leftrightarrow\sigma).
\end{gather}
For example, the commutator of a translation and a rotation is a translation.
To consider supersymmetry, we introduce a generator $Q_\alpha$ that satisfies
\begin{align}
  \{Q_\alpha, Q_\beta\} = \Gamma^\mu_{\alpha\beta}P_{\mu}, \label{QQ=Pintro}
\end{align}
where $\{,\}$ is the anticommutator and $\Gamma^\mu$ are matrices satisfying the
Clifford algebra
\begin{align}
  \Gamma^\mu\Gamma^\nu + \Gamma^\nu \Gamma^\mu = -2\eta^{\mu\nu}\Xop{1}.
\end{align}
Supersymmetry can be introduced in various ways into string theory. We can think
of supersymmetry on the worldsheet, on the target manifold, or both and we can
vary the amount of supersymmetry. To make things clear, we consider a sigma
model in flat Minkowski space and worldsheet supersymmetry. Supersymmetry is a symmetry 
that relates bosons and fermions. In \eqref{QQ=Pintro} we see that the
anticommutator of two objects with half integer statistics gives a bosonic
object which has integer spin. For worldsheet supersymmetry, we introduce fields
$\psi^\mu_\alpha = (\psi^\mu_+, \psi^\mu_-)$ that behave as real,
anticommuting two-dimensional spinors on the worldsheet and transform as a vector under the
Lorentz group of the target manifold: 
\begin{align}
  \psi^\mu_+ \psi^\nu_- = -\psi^\nu_- \psi^\mu_+.
\end{align}
In our notation, worldsheet spinor indices are denoted by
$\alpha,\beta,\ldots=+,-$.  We introduce two-dimensional Dirac matrices that
satisfy the Clifford algebra
$\{\gamma^a,\gamma^b\} =
-2\eta^{ab}\gen{1}$. With these preliminaries, we can write down the action
\begin{align}
  S = -\frac{1}{2}\int \d^2\sigma \Big( \partial_a X^\mu \partial^a X^\nu
 - \half\i \bar\psi^\mu \gamma^a\partial_a \psi^\nu\Big)\eta_{\mu\nu},
\label{susy-action}
\end{align}
where $\bar\psi = \psi^t\gamma^0$. This action is a supersymmetric extension of
\eqref{LCG-action}. The supersymmetry transformations are parametrized by a
constant anticommuting spinor $\eps$ \begin{align}
  \delta(\eps) X^\mu = \bar\eps\psi^\mu, &&
  \delta(\eps) \psi^\mu = -\half\i \gamma^a\partial_a X^\mu \eps,
\end{align}
where the contraction of spinor indices is implicit. The expression
$\bar\eps\psi^\mu$ is a shorthand notation for
$\eps_{\alpha}(\gamma^0)^{\alpha\beta}\psi^\mu_\beta$. Indeed, this
transformation relates the bosonic field $X^\mu$ to the spinor $\psi^\mu$.
The equations of motion for the spinors $\gamma^a\partial_a \psi^\mu = 0$ show
that $\psi_\pm^\mu$ are left and right moving components 
\begin{align}
  \partial_\+ \psi_-^\mu = 0, && \partial_\- \psi_+^\mu = 0. \label{spinor-eom}
\end{align}
For our purposes, it is useful to go to a Dirac matrix free
notation.  To this end, we define contraction of spinor indices according to
the `up-left-down-right' rule and raise and lower them with the antisymmetric
tensor
\begin{align}
  C_{+-}=-C^{+-}=\i,&&\psi^\mu_\alpha = (\psi^\mu){}^\beta
  C_{\beta\alpha},&&(\psi^\mu){}^\alpha =
  C^{\alpha\beta}\psi_\beta^\mu.
\end{align}
With the Dirac matrices explicitly given by
\begin{align}
  \gamma^0 = \genmatrix{0&-\i \\\i&0},&& \gamma^1 = \genmatrix{0&\i\\ \i&0},
\end{align}
we write out the second term in the supersymmetric action to find
\begin{align}
  S = \half\int \d^2\xi \Big( \partial_\+ X^\mu \partial_\- X^\nu + \i(
  \psi_-^\mu\partial_\+ \psi_-^\nu + \psi_+^\mu\partial_\- \psi_+^\nu)
  \Big)\eta_{\mu\nu}. \label{S_SUSY}
\end{align}
The supersymmetry transformations leaving this action invariant are
\begin{gather}
  \delta(\eps) X^\mu = (\eps^-\psi_-^\mu + \eps^+\psi_+^\mu), \cr
  \delta(\eps) \psi_-^\mu = -\i \eps^-\partial_\- X^\mu, \cr
  \delta(\eps) \psi_+^\mu = -\i \eps^+\partial_\+ X^\mu.
\end{gather}
Infinitesimal translations of the worldsheet 
$\xi^a \rightarrow \xi^a+ a^a$ act on the fields as $\delta X^\mu =
a^b\partial_b X^\mu$. According to \eqref{QQ=Pintro} the commutator of two
supersymmetry transformations gives a translation.
\begin{align}
  [\delta(\eps_1),\delta(\eps_2)]X^\mu = 2(\eps_1^+\eps_2^+\partial_\+ +
   \eps_1^-\eps_2^-\partial_\-)X^\mu.
\end{align}
Concerning the spinor fields, the corresponding relation is only satisfied
on-shell, i.e.~by imposing the equations of motions \eqref{spinor-eom}. This can
be amended by introducing an auxiliary field.  A particularly useful way to
implement supersymmetry is via superspace \cite{Gates:1983nr}. It incorporates
the auxiliary field and makes supersymmetry manifest. To this end, one
introduces additional directions on the worldsheet. The number of these
directions depends on the amount of supersymmetry. In the present case, the
worldsheet is extended by two such directions $\theta^\alpha$, $\alpha=+,-$.
They are anticommuting
\begin{align}
\{\theta^\alpha,\theta^\beta\} = 0
\end{align}
and usually called Grassmann coordinates. A superfield $\Phi^\mu$ is a map from
this extended (super-)worldsheet $\hat\Sigma$ into the target manifold, 
\begin{align}
  \Phi(\sigma,\tau,\theta^+,\theta^-):\hat\Sigma \rightarrow M.
\end{align}
For each Grassmann direction, there is a generator of supersymmetry. These are
odd differential operators
\begin{align}
   Q_\pm = \i\frac{\partial}{\partial\theta^\pm} +\theta^\pm\partial_\pp. 
\end{align}
$Q_\pm$ generates a supersymmetry transformation since $Q_\pm^2 =
-\partial_\pp$.  There are two more independent odd differential operators that
one can define:
\begin{align}
 D_\pm = \frac{\partial}{\partial\theta^\pm} +\i\theta^\pm\partial_\pp. 
\end{align}
They act like covariant derivatives for $\theta^\pm$ and satisfy the following 
algebraic relations together with $Q_\pm$:
\begin{align}
  Q_\pm^2 = -\i\partial_\pp && D_\pm^2 = \i \partial_\pp && \{D_\pm, Q_\pm\} =
0. \label{susy-alg}
\end{align}
Geometrically, this means that ``flat'' superspace has torsion.

This formulation makes supersymmetry manifest, since the whole supermultiplet is
described by a single superfield $\Phi^\mu$ and supersymmetry transformations
are generated by $Q_\pm$ acting simply on $\Phi^\mu$. The worldsheet coordinates
transform as 
\begin{gather}
  \delta(\eps)\xi^\+ = -\i(\eps^+Q_+ + \eps^-Q_-)\xi^\+ = -\i\eps^+\theta^+, \cr
  \delta(\eps)\xi^\- = -\i\eps^-\theta^-, \hspace*{1cm}
  \delta(\eps)\theta^\pm = \eps^\pm. \label{worldsheet-susy-trans}
\end{gather}
The transformation of the superfield $\Phi^\mu$ is given by
\begin{align}
  \delta(\eps)\Phi^\mu = -\i (\eps^+Q_+ + \eps^-Q_-) \Phi^\mu
\end{align}
To write down an action which incorporates the manifest supersymmetry, we
notice that the transformation of any function of the form $L(\Phi,D_+\Phi,D_-\Phi)$
under \eqref{worldsheet-susy-trans} is a total derivative. Therefore, the action
\begin{align}
  S= \half\int \d^2\xi \d^2\theta D_+\Phi^\mu D_-\Phi^\nu \eta_{\mu\nu}
\label{S-manifest-N11}
\end{align}
is manifestly supersymmetric. The variation of $S$ under
\eqref{worldsheet-susy-trans} is a total derivative and vanishes for
a topologically trivial worldsheet. The action is a
straightforward generalization of \eqref{LCG-action}. The $\d\theta$
integrals are Berezin integrals and can be evaluated as
\begin{multline}
  S=\half\int \d^2\xi \d^2\theta D_+\Phi^\mu D_-\Phi^\nu \eta_{\mu\nu} \cr =
\half\int
\d^2\xi \Big(D_+D_- (D_+\Phi^\mu D_-\Phi^\nu \eta_{\mu\nu})\Big)|_{\theta^\pm
= 0}.
\end{multline}
We define the components of $\Phi^\mu$ with the help of the covariant
derivatives $D_\pm$:
\begin{align}
  &X^\mu = \Phi^\mu|,&&\psi_\pm^\mu =
  (D_\pm\Phi^\mu)|,&&F^\mu = (D_+D_-\Phi^\mu)|.
\end{align}
The {\sl bar} denotes, that we set $\theta^+=\theta^-=0$ in the expression. $X^\mu$
and $F^\mu$ are bosonic, while $\psi_\pm^\mu$ are a worldsheet spinor.
Integrating out the Grassmann directions in the action yields its component form
\begin{align}
  S = \half \int \d^2\xi \Big(\partial_\+ X^\mu \partial_\- X^\nu +
\i\psi_+^\mu\partial_\-\psi_+^\nu + \i\psi_-^\mu\partial_\+\psi_-^\nu -
F^\mu F^\nu\Big)\eta_{\mu\nu}.
\end{align}
$F^\mu$ is an auxiliary field. It has algebraic equations of motion,
$F^\mu=0$, and substituting them in the action recovers
\eqref{susy-action}.

If one solves the equations of motion and tries to write down a consistent
quantized theory, then one finds that the spectrum has to be truncated in a
certain way. Interestingly enough, this truncation yields spacetime
supersymmetry and therefore even spacetime fermions. However, we do not persue
in this direction. Instead, we turn directly to a discussion of supersymmetry in
spacetime.

\section{Spacetime supersymmetry} \label{superspace}
We introduce spacetime supersymmetry in the same way as worldsheet supersymmetry
by extending the target space to superspace. To this end, we introduce a number
of Grassmann coordinates $\theta^{A\alpha}$ where $A=1\ldots N$ counts the
number of supersymmetries and $\alpha$is the (spacetime) spinor index.  We are
only interested in the case where the target manifold is ten dimensional
Minkowski space. A spinor of the ten dimensional Lorenz group $SO(1,9)$ has 32
complex components. The $32\times
32$ dimensional Dirac matrices $\Gamma^\mu$ satisfy the ten-dimensional Clifford
algebra
\begin{align}
  \{\Gamma^\mu, \Gamma^\nu\} = 2\eta^{\mu\nu}\gen{1}.
\end{align}
Under supersymmetry, the coordinates $x^\mu$ and $\theta^A$ are transformed into
each other similar to the case of worldsheet supersymmetry
\eqref{worldsheet-susy-trans}
\begin{gather}
  \delta(\eps) x^\mu = \i \bar\eps^A \Gamma^\mu \theta^A, \cr
  \delta(\eps) \theta^A = \eps^A, \hspace*{1cm}
  \delta(\eps) \bar\theta^A = \bar\eps^A,
\end{gather}
where $\eps^{A\alpha}$ is a constant spinor. One may check that these transformations
satisfy a supersymmetry algebra of the form \eqref{QQ=Pintro}.  The
simplest supersymmetric extension of the action \eqref{S_Poly} is given by 
\begin{multline}
  S=-\frac{1}{2}\int\d^2\xi \Big(\sqrt{-h}h^{ab}T\hat\Pi_a^\mu \hat\Pi_b^\nu
\eta_{\mu\nu}
\cr
  +2\i\eps^{ab}\eta_{\mu\nu}\partial_a
X^\mu(\bar\theta^1\Gamma^\nu\partial_b\theta^1 -
\bar\theta^2\Gamma^\nu\partial_b\theta^2) \cr
  -2\eps^{ab}\eta_{\mu\nu}\bar\theta^1\Gamma^\mu\partial_a\theta^1
   \bar\theta^2\Gamma^\nu \partial_b \theta^2\Big). \label{S_TSSUSY}
\end{multline}
Here, $\hat\Pi_a^\mu = \partial_a X^\mu - \i\bar\theta^A\Gamma^\mu \partial_a
\theta^A$. As in the discussion of worldsheet supersymmetry, the contraction of
spinor indices is implicit. Besides being supersymmetric, the action has a local fermionic
symmetry called $\kappa$-symmetry
\begin{gather}
  \delta\theta^A = 2\i\Gamma^\mu\hat\Pi_a^\nu\eta_{\mu\nu} \kappa^{Aa}, \hspace*{1cm}
  \delta X^\mu = \i\bar\theta^A \Gamma^\mu \delta \theta^A, \label{kappa-sym}
\end{gather}
where $\kappa$ satisfies
\begin{align}
  \kappa^{1 a} = P^{ab}_- \kappa^1_b, &&
  \kappa^{2 a} = P^{ab}_+ \kappa^2_b, &&
  P^{ab}_\pm = \half(h^{ab}\pm \eps^{ab}/\sqrt{h}).
\end{align}
In addition to \eqref{kappa-sym}, the metric transforms 
\begin{align}
  \delta (\sqrt{-h}h^{ab}) = -16\sqrt{-h}(P^{ac}\bar\kappa^{1b}\partial_c
\theta^ 1 + P^{ac}_+ \bar\kappa^{2b}\partial_c \theta^2).
\end{align}
The $\kappa$-symmetry allows us to make the following gauge choice for the fermions
\begin{align}
  \Gamma^+ \theta^A = 0,
\end{align}
where $\Gamma^{\pm} = \frac{1}{\sqrt{2}}(\Gamma^0\pm \Gamma^9)$. This is
sometimes also called fermionic light-cone gauge. We are only interested in type
IIB string theory which has two real spacetime supersymmetries. We implement
this by choosing Majorana-Weyl spinors. The Majorana condition reduces the 32
complex components to 32 real ones. The Weyl condition for the spinors is given
with the help of $\Gamma^{11}=\Gamma^0\cdot\ldots\cdot \Gamma^9$:
\begin{align}
  \Gamma^{11}\theta^A = \pm\theta^A,
\end{align}
For type IIB theories, both spinors have the same chirality,
i.e.~$\Gamma^{11}\theta^A = \theta^A$.  The Dirac matrices decompose into chiral
and anti-chiral representations
$\gamma^\mu$ and $\bar\gamma^\mu$
\begin{align}
  \Gamma^\mu = \genmatrix{0&\gamma^\mu \\ \bar\gamma^\mu & 0}.
\end{align}
The components are given by
\begin{align}
  \gamma^\mu = (1,\gamma^I, \gamma^9), && \bar\gamma^\mu = (-1, \gamma^I, \gamma^9)
\end{align}
with $\gamma^\mu = (\gamma^\mu)^{\alpha\beta}$ and $\bar\gamma^\mu =
(\gamma^\mu)_{\alpha\beta}$. We assume that
\begin{align}
  \Gamma^{11} = \genmatrix{1&0 \\ 0& -1}
\end{align}
and that $\gamma^\mu$ and $\bar\gamma^\mu$ are real and symmetric. The positive
chirality condition reduces the number of components of $\theta^A$ to 16, given
by
\begin{align}
  \theta^A = \matrix{c}{\theta^{A\alpha} \\ 0}, && A=1,2,&& \alpha=1\ldots 16.
\end{align}
In this notation, the conditions for the fermionic light-cone gauge become 
\begin{align}
  \bar\gamma^+\theta^A = 0.
\end{align}
Imposing fermionic lightcone gauge leaves us with 16 components in total. The
connection to worldsheet supersymmetry can be seen in the following way: After
going to lightcone gauge and fixing $\kappa$-symmetry, the equations
of motion for the remaining degrees of freedom are given by
\begin{align}
  \partial_\+\partial_\- X^I = 0, && \partial_\+ \theta^1 = 0, && \partial_\-
  \theta^2 = 0.
\end{align}
These are exactly the same as those  for $X^I$, $\psi^I_\pm$
from the action \eqref{S_SUSY} in the previous section. However, we should
mention that the exact relation between the two different pictures is not just
established by relabeling $\theta^{A\alpha}$ into $\psi^I_\pm$. It is a bit more
involved since the $\theta^A$ transform as spacetime spinors while $\psi^I$ is a
spacetime vector and a worldsheet spinor.
 
\section{Low energy effective theory}
\label{effective-theory}\label{sec:I:macro_sustrings}

When choosing a conformally flat worldsheet metric, we made use of the Weyl
symmetry of the worldsheet and had to impose the Virasoro constraints by hand.
In left and right moving worldsheet coordinates, $T_{\+\-}=T_{\-\+}$ vanishes
due to the tracelessness of the energy momentum tensor. For a curved
spacetime, this is only true in $D=26$. If we go beyond the classical level and
consider a quantum theory then the two-dimensional energy momentum tensor
acquires an anomaly except for the case when the so-called $\beta$-functions 
of the background fields $\G_{\mu\nu}$, $\B_{\mu\nu}$ and $\dilaton$ vanish.
In $D=26$ dimensions and to lowest order in the string scale $\alpha^\prime =
4\pi T^{-1}$. The conditions for this are are given by
\begin{align}
  \beta^\metric_{\mu\nu} &= \alpha^\prime \big( R_{\mu\nu} + 2\nabla_\mu \nabla_\nu \dilaton
    -  \H_{\mu\rho\sigma} \H_\nu^{\ph{\nu}\rho\sigma} \big) = 0,
    \nonumber \\
  \beta^\bfield_{\mu\nu} &= \alpha^\prime \big( -\nabla^\rho \H_{\rho\mu\nu}
    + 2\nabla^\rho \dilaton \H_{\rho\mu\nu} \big) = 0,  \cr
  \beta^\dilaton &= \alpha^\prime \big( 
    - \smallhalf \nabla^2\dilaton + (\nabla\dilaton)^2 - {\textstyle
\frac{1}{6} \H^2}
    \big) = 0. \label{eqn:beta-functions}
\end{align}
All solutions to these equations yield consistent string backgrounds.
The most remarkable feature of this set of equations is that they can be
derived as the equations of motion for the background fields $\metric, \bfield$
and $\dilaton$ from the spacetime action (in $D=26$ dimensions)
\begin{align}
  S = \frac{1}{2\kappa^2}\int \d^D x \sqrt{-\metric}\e^{-2\dilaton} \Big[
    R + 4\nabla_\mu \dilaton \nabla^\mu \dilaton - \frac{1}{3}\H^2 
    \Big]. \label{eqn:space-time-action}
\end{align}
This action describes the interaction of massless modes of the bosonic closed string in
the long-wavelength limit, hence it is the corresponding low-energy effective
theory. Here, $\kappa$ is the $D$-dimensional gravitational Newton's constant. 
For supersymmetric theories, this result gets modified, the analysis however
goes through in the same way.
%
%
All supergravity theories share \eqref{eqn:space-time-action} as part of the bosonic
part of the action. Supersymmetric string theory, however, requires a $D=10$
dimensional target space. Finding consistent supergravity backgrounds was a
major activity in the 1990s that lead for example to the discovery of D-branes.
In 1990, Dabholkar \etal \cite{Dabholkar:1990yf} found a solution that was
identified as the geometry of a heterotic superstring
\begin{gather}
  \d s^2 = A^{-3/4}[-\dt^2+(\dx^1)^2]+A^{1/4}(\d x^I)^2, \cr
  \bfield_{01}=e^{2\phi}=A^{-1},\hspace{0.5cm} 
  A=1+\frac{Q}{3r^6}, \label{eqn:Dabholkar-ds2}
\end{gather}
where $Q$ is the $\bfield$-charge carried by the string and $x^I=(x^2,\ldots,x^9)$ are the
directions transverse to the string with $r^2=x^Ix_I$. The solution becomes
singular at $r=0$ and does not satisfy the equations of motion at these points.
It is precisely this singularity that was interpreted as a macroscopic heterotic
string. Later, after the discovery of S-duality, this solution was also
identified as the geometry of a type I string \cite{Dabholkar:1995ep,
Hull:1995nu}. S-duality relates the weakly coupled sector of one string theory
to the strongly coupled sector of another, in this particular case, it relates
the heterotic string to the type I string. In chapter \ref{ch:T=0_Strings} we
will see that the fundamental string and the D1-brane of IIB theory, which is
also known as the D-string, yield similar solutions.


\chapter[Tensionless strings in plane waves]
{Tensionless String Theory in a Plane Wave Background} \label{chapter:pp-wave}

In this chapter we study the tensionless closed string on the maximally
supersymmetric plane wave. This background to type IIB supergravity was found
by Blau et.al.~\cite{Blau:2001ne} as the ten dimensional equivalent to a family
of 11d supergravity solutions called Kowalski-Glikman spaces
\cite{Kowalski-Glikman:1984wv}. It is supported by a constant selfdual five form
that is directly related to the curvature of the spacetime It has parallel and
planar wave fronts. Therefore, this background is sometimes also called a
pp-wave. It is one of the three known maximally supersymmetric background for
type IIB supergravity and is related to the other two. It is a Penrose limit of
$AdS_5\times S^5$ on one side \cite{Blau:2002dy,Blau:2002mw} and becomes flat
space in the limit when the flux vanishes. 

The AdS/CFT correspondence originally conjectured by Maldacena
\cite{Maldacena:1997re} and later clarified in
\cite{Witten:1998qj,Gubser:1998bc} underlies the desire to understand string
theory in $AdS_5\times S^5$. The plane wave is a step in this direction. Metsaev
and Tseytlin showed that closed string theory in light\-cone gauge in this
background is an integrable model and provided its solution classically and at
the quantum level \cite{Metsaev:2001bj,Metsaev:2002re}. The AdS/CFT
correspondence reduces to the BMN correspondence which relates certain parts of
the string spectrum to planar diagrams on the gauge theory side
\cite{Berenstein:2002jq,Plefka:2003nb}. This correspondence is not as strict as
the AdS/CFT correspondence but it holds at least to first order in the expansion
of $AdS_5\times S^5$ over the plane wave \cite{Parnachev:2002kk,Callan:2003xr}.
In $AdS_5\times S^5$ the tensionless string is supposed to be related to higher
spin gauge theory
\cite{Vasiliev:1999ba,Haggi-Mani:2000ru,Sundborg:2000wp,Sezgin:2002rt,
Bonelli:2003zu,Lindstrom:2003mg,Savvidy:2003fx,Engquist:2005yt}. Part of this
relation should survive the limit to the plane wave. In [II] we study the
tensionless closed string in light\-cone gauge on the plane wave background and
find that it can be obtained as a well-behaved limit of the results of
\cite{Metsaev:2002re}.  This behavior is traced back to the existence of a
background scale which is related to the flux and allows for a reinterpretation
of our results as the ordinary, tensile string moving in an infinitely curved
plane wave background in accordance to
\cite{deVega:1994hu}.

This chapter proceeds in the following way. It starts out with a short
introduction to the tensionless string and issues in flat space. We then present
how the plane wave is obtained from $AdS_5\times S^5$ and review the solution of
closed string theory in this background before turning to the tensionless limit
of this theory. We conclude this chapter with some remarks on the more general
situation of a homogenous plane wave background.

\section{Tensionless strings in flat space} \label{section-classical-null-action}

The classical tensionless string was first mentioned in \cite{Schild:1976vq}
when strings that move with the speed of light turned out to have zero tension.
This makes it a candidate for the description of the high-energy behavior of
string theory \cite{Gross:1987kz}. Here, we follow the lines of
\cite{Karlhede:1986wb,Isberg:1993av} where the classical and quantized bosonic
tensionless string in flat space were discussed.  The tensionless superstring
has been studied in \cite{Barcelos-Neto:1989ms,Lindstrom:1990qb}. 

The action for a point particle is
given by \eqref{S_part}.  By introducing an auxiliary field $e$, an einbein, the
action can be brought into the form
\begin{eqnarray}
  S_{\rm part, P} = \int \dt \Big( e \dot X^2 + e^{-1} m^2 \Big). \label{S-partP}
\end{eqnarray}
As long as $m \neq 0$, it is possible to gauge away $e$ using its (algebraic)
field equations and rewrite the action in the first form. On
the other hand, the massless particle action is obtained by taking $m\rightarrow
0$. The equivalent of \eqref{S-partP} in string theory is the Polyakov action
\eqref{S_Poly}. To understand how to take the limit $T\rightarrow 0$, we have to
understand how the Nambu-Goto and the Polyakov action are related to each other.
The Nambu-Goto action was given in \eqref{S_NG}:
\begin{align}
  S = T \int \d^2\xi \sqrt{-\metricpullback}, \label{S_NG2}
\end{align}
where $\metricpullback$ was the determinant of $\metricpullback_{ab}=\del_a X^\mu \del_b X^\nu
\eta_{\mu\nu}$. The conjugate momenta to $X^\mu$ are $P_\mu = \frac{\del L}{\del
\dot X^\mu} = T\sqrt{-\metricpullback}\metricpullback^{00}\dot X_\mu$ where
$\metricpullback^{ab}$ is the
inverse of $\metricpullback_{ab}$. The momenta are constrained by the Virasoro
constraints \eqref{Virasoro-NG}
\begin{align}
  P^2 + T^2\metricpullback \metricpullback^{00} = P\cdot X' = 0. \label{Virasoro-P}
\end{align}
The Hamiltonian is given by these constraints, since the
canonical Hamiltonian vanishes due to the diffeomorphism invariance of the
worldsheet. If we introduce $\lambda$ and $\rho$ as Lagrange multiplies for the
constraints then we can write down the phase space action corresponding to the
Nambu-Goto action
\begin{eqnarray}
  S_{\rm PS} = \frac{1}{2}\int \d^2 \xi \Big( P_\mu \dot X^\mu - \lambda (P_\mu P^\mu + T^2
 \metricpullback\metricpullback^{00}) -
    \rho P_\mu X^{\prime \mu} \Big).
\end{eqnarray}
The momenta can be integrated out using their (algebraic) field equations. This
yields the configuration space action
\begin{align}
  S_{\rm CS} = \frac{1}{4\lambda}\int \d^2\xi \Big( \big(\dot X^\mu \dot X^\nu - 2\rho
\dot X^\mu X^{\prime \nu} + \rho^2 X^{\prime \mu} X^{\prime
\nu}\big)\eta_{\mu\nu} - 4\lambda^2 T^2\metricpullback\metricpullback^{00}\Big). \label{S_CS}
\end{align}
\vspace*{-0.2cm}
This is the Polyakov action \eqref{S_Poly} with 
$h^{ab}={\small \left(\begin{array}{cc}-1&\rho\\ \rho &
4\lambda^2T^2-\rho^2\end{array}\right)}$
\vspace*{-0.2cm}
\begin{align}
  S_{\rm Poly} = -\frac{T}{2}\int\d^2\xi \sqrt{-h}h^{ab}\del_a X^\mu \del_b
  X^\nu \eta_{\mu\nu}.
\end{align}
The constraints \eqref{Virasoro-P} are, of course, the Virasoro constraints. 
On the other hand, we can take the limit $T\rightarrow 0$ in the configuration
space action. This limit is not covered by the Polyakov action since
$h^{ab}$ becomed degenerate. Instead we can introduce a contravariant vector density
$V^a =\frac{1}{\sqrt{2}\lambda}(1, \rho)$ and obtain the  action
for the tensionless string:
\begin{eqnarray}
  S_{T=0} = -\frac{1}{2}\int \d^2 \xi V^a V^b \del_a X^\mu \del_{b} X^\mu \eta_{\mu\nu}.
    \label{eqn:S_T=0}
\end{eqnarray}
This action has a reparametrization symmetry 
\begin{gather}
  \delta(\epsbos) X^\mu = \epsbos^a\partial_a X^\mu, \cr
  \delta(\epsbos) V^a = - V^b\del_b \epsbos^a + \epsbos^b
  \del_b V^a + \smallhalf \del_b \epsbos^b V^a
\end{gather}
for a small parameter $\epsbos$. It allows to gauge away one of the components
of $V^a$. A particularly useful gauge is the transverse gauge $V^a = (v, 0)$ in
which the action takes the form
\begin{eqnarray}
  S_{T=0, {\rm tg}} = - \frac{v^2}{2} \int \d^2 \xi \dot X^\mu \dot X^\nu \eta_{\mu\nu}. 
\label{eqn:S_T=0:tg}
\end{eqnarray}
Apart from the $\d\sigma$ integral, this action looks like the action of a
massless particle.  As in the tensile case, the action \eqref{eqn:S_T=0:tg} is
still not completely gauge fixed. The residual symmetry that is left is 
\begin{align}
  \delta \tau = f^\prime(\sigma)\tau +g(\sigma),&&\delta\sigma = f(\sigma).
\end{align}
Here, $f$ and $g$ are arbitrary functions of $\sigma$. Again, this allows us to go to
light\-cone coordinates $X^\pm = \frac{1}{\sqrt{2}}(X^0\pm
X^{D-1}),~X^I,~I=1\ldots D-2$ and fix light\-cone gauge
by choosing $X^+ = \frac{p^+}{v^2}\tau$. The light\-cone action of the
tensionless string in flat space is given by
\begin{eqnarray}
  S_{\rm LC}=\frac{v^2}{2}\int \d^2\xi \dot X^I \dot X^I. \label{eqn:S_LC:flat}
\end{eqnarray}
We may compare this action to \eqref{LCG-action}. Taking the tensionless limit
amounts to replacing $T$ by $v^2$ and putting all $\sigma$-derivatives to zero.
This rule of thumb can be stated more exactly. In order to take the limit
$T\rightarrow 0$, we split the tension according to $T=\lambda v^2$, where
$\lambda$ is a dimensionless parameter to be taken to zero and $v$ has the
dimension of energy. Introducing a new worldsheet time $t = \tau/\lambda$, the
action \eqref{LCG-action} becomes
\begin{align}
  S_{\rm LC}=\frac{v^2}{2}\int\d t \d\sigma \big(\dot X^I \dot X^I - \lambda^2
X^{I\prime}X^{I\prime}\big). \label{LCG-action-T=lambda}
\end{align} 
Clearly, $\lambda\rightarrow 0$ amounts in \eqref{LCG-action-T=lambda} becoming
\eqref{eqn:S_LC:flat}. The original worldsheet parametrized by $\sigma$ and
$\tau$ is now a null surface.  The classical equations of motion
obtained from the gauge fixed action \eqref{eqn:S_LC:flat} are
\begin{eqnarray}
  \ddot X^I = 0. \label{eqn:free-part}
\end{eqnarray}
By fixing the transverse gauge, the equations of motion for $V^a$ become the
constraint equations
\begin{align}
  &\dot X^I \dot X^I - 2\frac{p^+}{v^2} \dot X^- = 0, &&
  &\dot X^I X^{\prime I} - \frac{p^+}{v^2} X^{\prime -} = 0.
\end{align}
These are the equivalent of the Virasoro constraints
\eqref{Virasoro-constraints}.  Also for the tensionless string, the physical
degrees of freedom are the transverse components $X^I$. At each value of
$\sigma$, $X^I$ is a solution to \eqref{eqn:free-part}. The string literally
splits into infinitely many massless particles whose motion is restricted to be
transverse to the string.

The action \eqref{eqn:S_T=0} has a global conformal spacetime symmetry.
Dilatations are given by the scale transformation
\begin{align}
  \delta(\lambda)X^\mu = \lambda X^\mu, && \delta(\lambda)V^a = -\lambda V^a,
\end{align}
and the conformal boost, or special conformal transformation, has the form
\begin{align}
  \delta(b)X^\mu = (b_\nu X^\nu) X^\mu - \half X^2 b^\mu, &&
  \delta(b)V^a = - (b_\nu X^\nu)V^a.
\end{align}
There is no critical dimension for a consistent quantum theory in flat space
\cite{Lizzi:1986nv}. However, the conformal symmetry survives quantization only
in $D=2$ spacetime dimensions. In any other dimension,
the conformal algebra acquires an anomalous term which provides a selection rule
for the physical states: The spectrum is hugely restricted and becomes
topological \cite{Isberg:1992ia,Gustafsson:1994kr,Saltsidis:1995qr}. This
strengthens the view of the tensionless string as the unbroken, topological
phase of string theory. 

The vacuum state of the tensionless theory differs from the tensile case. It has
more the form of a particle vacuum than a string vacuum. To obtain the quantum
theory, we can proceed and introduce canonical commutation relations
\begin{align}
  {[}X^I(\sigma_1), P^J(\sigma_2)] = \i \delta^{IJ}\delta(\sigma_1-\sigma_2),&& 
  {[}X^-, p^+] = -\i. \label{com-rel}
\end{align}
We saw that $X^I(\sigma)$ is a collection of infinitely many degrees of freedom
parametrized by $\sigma$. Therefore, the quantum theory has to be modified
\cite{Isberg:1993av} by regularizing the $\delta$-function. 
As long as there is little tension left, we would introduce left and right
movers
\begin{align}
  \alpha_n^I = \frac{p_n^I}{\sqrt{T}} - \i n \sqrt{T}x_n^I, &&
  \tilde \alpha_n^I = \frac{p_n^I}{\sqrt{T}} +\i n \sqrt{T}x_n^I, && n \neq 0,
\end{align}
where $x_n^I$ and $p_n^I$ are the Fourier modes of $X^I$ and their conjugate
momenta $P^I$. We would then define the vacuum state by the requirement that is
annihilated by the positive frequency modes 
\begin{align}
  \alpha_n^{1I}\ket{0}_0 = \alpha_n^{2I}\ket{0}_0=0, && n = 1,2,\ldots.
\label{T-vac}
\end{align}
In the limit $T\rightarrow 0$, this implies
\begin{align}
  p_n^I\ket{0}_0 = p_{-n}^I\ket{0}_0= 0.
\end{align}
From \eqref{T-vac} we read off that also the $x_n^I$ annihilate the vacuum state 
$\ket{0}_0$ for all values $n\neq 0$. This is inconsistent with the commutation 
relations \eqref{com-rel}. The most natural possibility is therefore to choose a
translation invariant vacuum state for tensionless string 
\begin{eqnarray}
  P^I\ket{0}_0 = 0,
\end{eqnarray}
while keeping $X^I\ket{0}_0$ unspecified.

\section{Plane wave geometry from $AdS_5\cross S^5$}
Here, we show how the plane wave geometry arises as a Penrose limit of
$AdS_5\times S^5$. In any neighborhood of a null geodesic it is possible to
choose coordinates in which the line element takes the special form
\begin{align}
  \d s^2 = \d x^+ \d x^- + a (\d x^+)^2 + k_I \d x^+ \dx^I + f_{IJ} \dx^I \dx^J,
\end{align}
This observation goes back to Penrose \cite{Penrose:1972} and is true as long as the
neighborhood does not contain intersections of neighboring geodesics.
The coordinates $x^+$ while $x^-$ parametrize a particle traveling along the
geodesic while $x^I$ are coordinates transverse to it. Recently, this limit was
extended to include the supergravity fields in ten and 11 dimensions
\cite{Gueven:2000ru}. For the type IIB supergravity background $AdS_5\times
S^5$, this is the (constant) dilaton $\dilaton$ and the self-dual five form field
strength $F_5$. The line element of $AdS_5\times S^5$ is a combination of the
part coming from $AdS$ and from the five sphere $\ds^2 = \ds_{AdS}^2 +
\ds_{S^5}^2$. The radii of both subspaces are equal. Anti-de Sitter space is
embedded in $\mathbb{R}^{2,4}$ as
the hypersurface
\begin{align}
  x_0^2 - x_1^2 - x_2^2 - x_3^2 - x_4^2 + x_5^2 = R^2.
\end{align}
There are a number of appropriate coordinates to parametrize $AdS$ space. 
We use so-called global coordinates
\begin{gather}
  x_0 = R \mbox{cosh}(\rho) \sin(t), \hspace*{1cm}
  x_5 = R \mbox{cosh}(\rho) \cos(t), \cr
  x_I = R \mbox{sinh}(\rho) \omega_I,\hspace*{1cm} I=1,2,3,4. \label{AdS-coords}
\end{gather}
The coordinates $\omega_I$ parametrize the unit three sphere $\omega_I^2=1$. 
In these coordinates, the line element of $AdS$ space is given by
\begin{align}
  \ds^2_{AdS} = R^2 \Big[ -\dt^2\mbox{cosh}^2(\rho) + \d\rho^2 +
\mbox{sinh}^2(\rho)\d\Omega_3^2 \Big].
\end{align}
It is obtained by substituting \eqref{AdS-coords} into the line element of
$\mathbb{R}^{2,4}$
\begin{align}
  \ds^2_{2,4} = -\dx_0^2 -\dx_5^2 + \dx_I \dx_I.
\end{align}
Analogously, we embed the five-sphere into flat six-dimensional space
$\mathbb{R}^6$ by
\begin{align}
  x_0^2 + x_1^2 + x_2^2 + x_3^2 + x_4^2 + x_5^2 =R^2
\end{align}
and choose coordinates
\begin{gather}
  x_0 = R \cos(\theta)\sin(\psi), \hspace*{1cm}
  x_5 = R \cos(\theta)\cos(\psi), \cr
  x_I = R \sin(\theta)\omega_I^{\prime 2}, \hspace*{1cm} I=1,2,3,4.
\end{gather}
Again, $\omega'_I$ parametrize the remaining unit three sphere. The metric for
$S^5$ is 
\begin{align}
  \ds^2_{S^5} = R^2 \Big[ \d\psi^2\cos^2(\theta) + \d\theta^2 +
\sin^2(\theta)\d\Omega_3^{\prime 2} \Big].
\end{align}
The five form field strength $F_5$ is given by
\begin{align}
  F_5 = \frac{2}{R}(\d \mbox{Vol}(AdS_5) + \d \mbox{Vol}(S^5)).
\end{align}
The plane wave geometry is obtained by considering a particle that moves along
the $\psi$ direction of $S^5$ and is located at the origin in the $\theta$ and
$\rho$ directions $\rho=\theta=0$.  The Penrose limit zooms into the region near
the particle's trajectory \cite{Blau:2002mw}. To this end, we introduce new
coordinates 
\begin{align}
  x^+ = \smallhalf (t+\psi),&& x^- = - R^2 (t-\psi),&&
   \rho = \frac{r}{R},&&  \theta = \frac{y}{R},
\end{align}
and blow up the radius of the $S^5$, $R\rightarrow \infty$. In this limit,
$\omega_I$ together with $r$ parametrize points $\vec{r}$ in $\mathbb{R}^4$.
The same is true for $\vec{y}=(y,\omega_I^\prime)$. With the identification 
$\vec{x} \equiv (\vec{r}, \vec{y})$ the metric becomes
\begin{eqnarray}
  \ds^2 = 2\dx^+ \dx^- - x^2 \dx^{+2} + \dx^I \dx^I.
\end{eqnarray}
The index $I$ runs over the transverse coordinates $1...8$ and the five form
becomes proportional to a constant 
\begin{align}
  F_{5;+1234} = F_{5;+5678} = \frac{f}{2}.
\end{align}
All other components vanish.
The rescaling $x^- \rightarrow x^-/f$ and $x^+ \rightarrow f x^+$ brings the
plane wave metric to the form 
\begin{align}
  \ds^2 = 2\dx^+ \dx^- - f^2 x^2 \dx^{+2} + \dx^I \dx^I.
  \label{eqn:ds2:plane-wave}
\end{align}
This particular combination of the metric and $F_5$ is a maximally
supersymmetric type IIB background \cite{Blau:2001ne}.

\section{String theory in plane wave geometry} \label{sec:string-theory:plane-wave}
The action for type IIB string theory in the plane wave background was found in
\cite{Metsaev:2001bj}. Metsaev and Tseytlin studied and quantized the closed
string solution in the Green-Schwarz formulation \cite{Metsaev:2002re}. The
action for the superstring in fermionic light\-cone gauge
$\bar\gamma^+\theta^A=0$ is given by
\begin{multline}
  S = -\frac{1}{2}\int \d^2\xi \Big(T\sqrt{-h}h^{ab}\Big(2\partial_a X^+
\partial_b X^- -    f^2 X^I X_I \partial_a X^+ \partial_b X^+ + \partial_a X^I
\partial_b X^I \cr +2\i
\partial_aX^+\big(\theta^1\bar\gamma^-\partial_b\theta^1
+\theta^2\bar\gamma^-\partial_b\theta^2 -2f \partial_b
X^+\theta^1\bar\gamma^-\Pi\theta^2 \Big) \cr -2\i\eps^{ab}\partial_a
X^+\Big(\theta^1\bar\gamma^-\partial_b \theta^1
-\theta^2\bar\gamma^-\partial_b \theta^2\Big)\Big) \label{S-ppwave-original}
\end{multline}
$\theta^A$ are Majorana-Weyl spinors as in section \ref{superspace}. The action is the
equivalent of \eqref{S_TSSUSY} for the plane wave background.  The term
with $\Pi$ is a reminiscent of terms that involve $F_5$. $\Pi$ and $\Pi'$ 
satisfy $\Pi^2 = \Pi'^2 = 1$ and are given by
\begin{align}
  \Pi^\alpha{}_\beta =
(\gamma^1\bar\gamma^2\gamma^3\bar\gamma^4)^\alpha{}_\beta, &&
  \Pi'^\alpha{}_\beta =
(\gamma^5\bar\gamma^6\gamma^7\bar\gamma^8)^\alpha{}_\beta.
\end{align}
It is useful to choose a conformally flat worldsheet metric and use the remaining
reparametrization invariance on the worldsheet to fix light\-cone gauge in
spacetime with $X^+=p^+\tau/T$.  In this gauge, the action reduces drastically
\begin{multline}
  S_{LCG} = \frac{T}{2} \int \d^2\xi \Big(
    \partial_\+ X^I \partial_\- X^I - m^2 X^I X^I \cr
    + 2\i\frac{p^+}{T}\big(\theta^1\bar\gamma^- \partial_\+ \theta^1
    -\theta^2\bar\gamma^-\partial_\-\theta^2\big)
    -4\i m \frac{p^+}{T}\theta^1\bar\gamma^-\Pi\theta^2\Big).
    \label{S-ppwave-LCG}
\end{multline}
For convenience, we introduce the dimensionless parameter $m=p^+f/T$. After choosing the
conformally flat worldsheet metric, the Virasoro constraints have to be imposed
by hand. In light\-cone gauge, they read
\begin{align}
  p^+ X^{-\prime} + \i p^+(\theta^1\bar\gamma^-\partial_\sigma \theta^1
    + \theta^2\bar\gamma^-\partial_\sigma \theta^2) + T\dot X^I X^{I\prime} = 0,& \cr
  2 p^+ \dot X^- + 2i p^+  (\theta^1\bar\gamma^-\dot\theta^1
    + \theta^2\bar\gamma^-\dot\theta^2) - 4\i m p^+ \theta^1
      \bar\gamma^-\Pi\theta^2 & \cr
      -m^2TX^IX^I + T\dot X^I \dot X^I + T X^{I\prime}X^{I\prime}&=0. \label{Virasoro-constraints-LCG}
\end{align}
They can be used to derive $X^-$, which does not enter the action any more. The 
equations of motion for the transverse coordinates are
\begin{gather}
   \del_\+\del_\- X^I + m^2 X^I = 0, \cr
   \del_\+\theta^1 - m\Pi\theta^2 = 0, \hspace*{1cm}
   \del_\-\theta^2 + m\Pi\theta^1 = 0. \label{eqn:eom:plane-wave}
\end{gather}
The solution with closed string boundary conditions $X^I(\sigma+\pi) = X^I(\sigma)$
is
\begin{multline}
   X^I(\sigma,\tau)=
    \cos( m\tau) x^I_0+\frac{1}{mT}
    \sin( m\tau) p^I_0\cr 
  +\i\sum_{n\neq 0}\frac{1}{ \omega_n}\big\{
    \alpha_n^{1I}\e^{-\i(\omega_n\tau-2n\sigma)}
    +\alpha_n^{2I}\e^{-\i(\omega_n\tau+2n\sigma)}\big\}, \nonumber
\end{multline} \vspace*{-0.7cm}
\begin{multline}
  \theta^{1}(\sigma,\tau)= \cos( m\tau)\theta^{1}_0+
    \sin( m\tau)\Pi\theta^{2}_0\cr 
  +\sum_{n\neq 0}  c_n\big\{
    \theta^{1}_n\e^{-\i( \omega_n\tau-2n\sigma)}
    +\i\Pi\theta^{2}_n{\textstyle \frac{ \omega_n-2 n}{ m}}
    \e^{-\i( \omega_n\tau+2n\sigma)}\big\}, \label{eqn:sol:tension:plane-wave}
\end{multline}
and similar for $\theta^2$. The frequencies $\omega_n$ and the coefficients $c_n$ 
are given by
\begin{eqnarray}
 \omega_n=\mbox{sign}(n)\sqrt{ m^2+4n^2}\mbox{~~~and~~~}
  c_n=
    {\textstyle\frac{1}{\sqrt{1+\left(\frac{\omega_n-2 n}{ m}\right)^2}}}.
    \label{omegan-cn}
\end{eqnarray}
The canonical momenta for $X^I$ and $\theta^{A}$ are
\begin{align}
  P^I = T\dot X^I, && \pi^A_\alpha = -\i p^+ (\theta^A
  \bar\gamma^-)_{\alpha}.
\end{align}
The equal time Poisson brackets for $X^I,~P^I,~\theta^A$ and $\pi^A_\alpha$
yield the brackets for the oscillators 
\begin{gather}
  \{p^I_0, x^J_0\} = \delta^{IJ}, \hspace*{1cm}
  \{\alpha_m^{AI}, \alpha_n^{BJ}\} = \frac{\i\omega_n}{2T}\delta_{m+n}\delta^{AB}
    \delta^{IJ}, \cr
  \{\theta^{A\alpha}_m, \theta^{B\beta}_n\}_D =
    \frac{\i}{4 p^+}\bar\gamma^{+\alpha\beta} \delta_{m+n}\delta^{AB},
    \hspace*{1cm}A,B=1,2. \label{Poisson-brackets}
\end{gather}
The $D$ indicates the Poisson-Dirac bracket which has to be used for the fermionic
oscillators \cite{Dirac:1950pj}. The light\-cone Hamiltonian for the closed
string, written in terms of the oscillators, is
\begin{multline}
  H_{LC} = \frac{1}{2T} p_0^2 + \half m^2T x_0^2 + 
  2\i mp^+\theta^1_0\bar\gamma^-\Pi\theta^2_0 \cr
  +\sum_{n\neq 0, A=1,2} \Big(T\alpha_{-n}^{AI}\alpha_{n}^{AI} 
    + p^+\omega_n\theta^A_{-n}\bar\gamma^-\theta^A_n\Big).
\end{multline}
Upon integration, the first of the Virasoro constraints
\eqref{Virasoro-constraints-LCG} can be written in terms of number operators
\begin{align}
  N^1 = N^2, && N^A = \sum_{n\neq 0} n\Big(
    \frac{T}{\omega_n}\alpha^{AI}_{-n}\alpha^{AI}_n
      + p^+\theta^A_{-n}\bar\gamma^-\theta^A_n\Big).
      \label{number-operator}
\end{align}
To quantize the theory, we follow the canonical quantization procedure and replace
the Poisson brackets by equal time commutation relations promoting the Fourier
modes in the expansion of the fields \eqref{eqn:sol:tension:plane-wave} to
operators. For our purpose, it is useful to introduce new, dimensionless
creation and annihilation operators for $n=1,2,\ldots$
\begin{align}
  a_0^I &= \sqrt{\frac{T}{2m}}\big(\frac{p_0^I}{T} - \i m x_0^I\big), &
  \bar a_0^I &= \sqrt{\frac{T}{2m}}\big(\frac{p_0^I}{T} + \i m x_0^I\big), \cr
  a_n^{AI} &= \sqrt{\frac{2T}{\omega_n}}\alpha_n^{AI}, &
  \bar a_n^{AI} &= \sqrt{\frac{2T}{\omega_n}}\alpha_{-n}^{AI}, \cr
  \eta_0 &= \sqrt{\frac{p^+}{2}}\big(\theta_0^1 - \i \theta_0^2\big), &
  \bar\eta_0 &= \sqrt{\frac{p^+}{2}}\big(\theta_0^1 + \i\theta_0^2\big), \cr
  \eta_n^A &= \sqrt{2p^+}\theta_n^A, & 
  \bar\eta_n^A &= \sqrt{2p^+}\theta_{-n}^A. \label{dimless-ops}
\end{align}
The Poisson brackets \eqref{Poisson-brackets} yield the commutation 
and anti-commutation relations
\begin{align}
  [a_0^I, \bar a_0^I] &= \delta^{IJ}, &
  [a_m^{AI}, \bar a_n^{BJ}] &= \delta_{mn}\delta^{IJ}\delta^{AB}, \cr
  \{\eta_0^\alpha, \bar\eta_0^\beta\} &= \frac{1}{4}\bar\gamma^{+\alpha\beta}, &
  \{\eta_m^{A\alpha},\eta_n^{B\beta}\} &= \half \bar\gamma^{+\alpha\beta}
    \delta_{mn}\delta^{AB}. \label{dimless-commutators}
\end{align}
The (normal ordered) light-cone Hamiltonian of the quantum theory in these new
oscillators reads
\begin{multline}
  H_{LC} = m(4 + e_0 +\bar a_0^I a_0^I + 2\bar\eta_0\bar\gamma^-\Pi\eta_0) \cr
    +\sum_{n=1}^{\infty} \omega_n \big(
     \bar a_n^{1I} a_n^{1I} + \bar a_n^{2I} a_n^{2I}
    +\bar\eta_n^1\bar\gamma^-\eta_n^1
    +\bar\eta_n^2\bar\gamma^-\eta_n^2 \big). \label{H-LC-ppwave}
\end{multline}
The term $4+e_0$ comes from the normal ordering and $e_0$ depends on the choice
of the fermionic vacuum. The Virasoro constraint \eqref{number-operator} becomes
a level matching for the physical states
\begin{align}
  (N^1-N^2)\ket{phys} = 0, &&
  N^A = \sum_{n=1}^\infty n \big(
    \bar a_n^{AI} a_n^{AI} + 
    \bar\eta_n^{A}\bar\gamma^-\eta_n^A \big).
\end{align}
The question is which are the physical states. The light-cone Hamiltonian can be 
rewritten making the ground state energy term explicit
\begin{align}
  & H_{LC} = E_0+
	\sum_{A=1,2}\sum_{n>0} \omega_n (a^{A I}_{n}\bar a^{A I}_n
	+\eta^A_{n} \bar\gamma^- \bar\eta^A_n ), \cr
  &E_0 = m(a^I_0\bar a^I_0+2\bar\eta_0\bar\gamma^-\Pi\eta_0+e_0).
\end{align}
 Since the vacuum state is a direct product of the bosonic and
the fermionic vacuum, it obeys
\begin{align}
  \bar a_0\ket{0} = 0,&&\bar a^{A I}_n\ket{0} = 0,&&
  \bar\eta^{A}_n\ket{0} = 0,~~n=1,2,\ldots
\end{align}
for the bosonic part and the higher order fermionic modes. The way to choose
the fermionic zero-mode vacuum can be found by introducing projected
fermionic zero modes
\begin{eqnarray}
  \eta_\pm = {\textstyle \frac{1}{\sqrt{2}}}(1\pm\Pi) \eta_0.
\end{eqnarray}
It turns out that there are exactly four different possible choices:
\begin{align}
  &\bar\eta_\pm \ket{0} = 0,&&E_0=4, \cr 
  &\eta_\pm \ket{0} = 0,&&E_0=4,\cr 
  &\bar\eta_+ \ket{0} = \eta_- \ket{0} = 0,&&E_0=8, \cr
  &\eta_+ \ket{0} = \bar\eta_- \ket{0} = 0,&&E_0=0. 
\end{align}
While the first two choices preserve the $SO(8)$ symmetry, they break the
supersymmetry of the light cone Hamiltonian. The situation is vice-versa for the
latter two: They break $SO(8)\rightarrow SO(4)\cross SO^\prime(4)$, but preserve
supersymmetry. Metsaev and Tseytlin also showed how the spectrum of states built
out of these vacua by acting with the raising and lowering operators
\eqref{dimless-ops} can be interpreted in terms of supergravity fields in the
plane wave background.

\section[The tensionless superstring in the plane wave]
{The tensionless superstring in a maximally supersymmetric plane wave background}
\begin{figure}[h]
\begin{minipage}[c]{0.45\textwidth}
\centering 
\ifpdf
  \includegraphics[width=0.8\textwidth,height=4cm,angle=270]{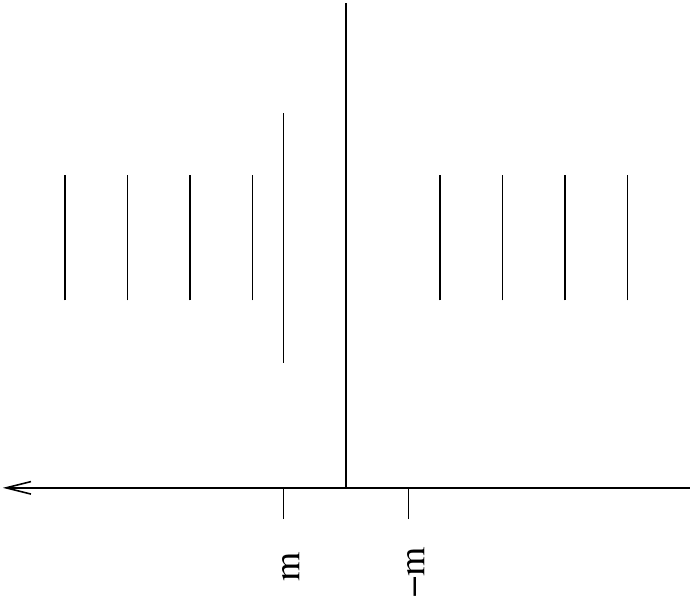}
\else
  \includegraphics[width=0.8\textwidth,height=4cm]{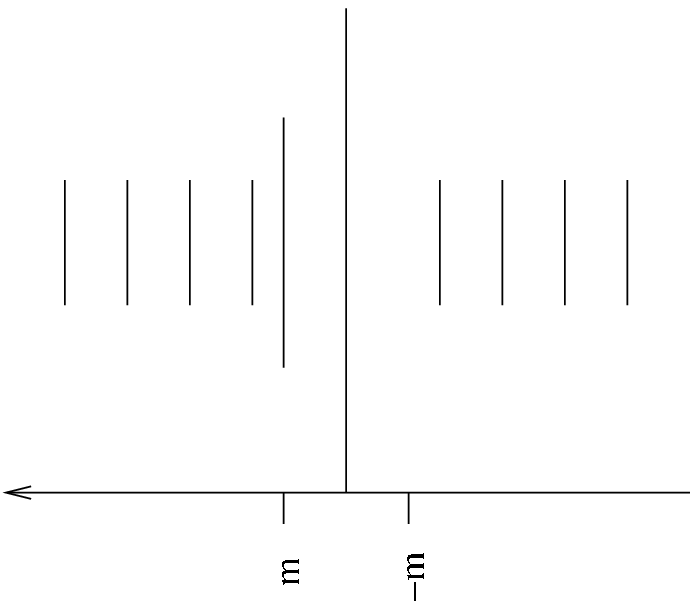}
\fi
\end{minipage} $\stackrel{}{\longrightarrow}$ 
\begin{minipage}[c]{0.45\textwidth}
\centering 
\ifpdf
  \includegraphics[width=0.8\textwidth,height=4cm,angle=270]{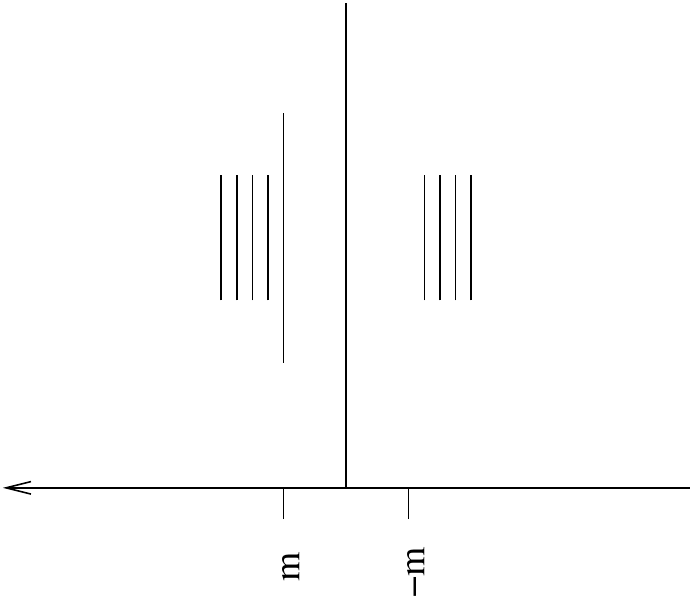}
\else
  \includegraphics[width=0.8\textwidth,height=4cm]{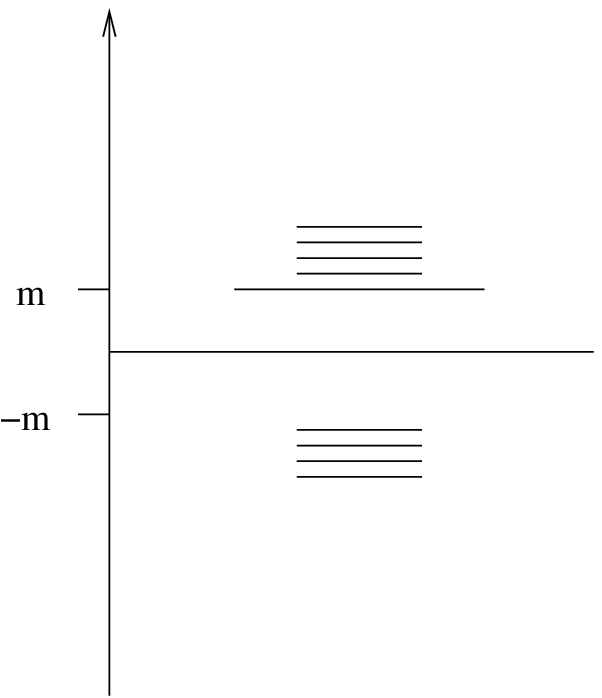}
\fi
\end{minipage}
\caption{Large $m$ or small $T$ - degeneration of the oscillator energies in the limit
$\lambda\rightarrow 0$.}
\label{figure1}
\end{figure}

The action for the tensionless string can be derived from \eqref{S-ppwave-original}
in the way discussed in section \ref{section-classical-null-action}. The rigorous derivation
is presented in [II]. Here, we start directly from the action in light\-cone 
gauge \eqref{S-ppwave-LCG} and use the shortcut. To this end, we split the tension
into $T=\lambda v^2$ with $\lambda$ being a dimensionless parameter to be
taken to zero and introduce the new worldsheet time $t=\frac{\tau}{\lambda}$.
In addition, we keep the combination $\mu=\lambda m=p^+f/v^2$ fixed. The action becomes
\begin{multline}
  S_{LCG} = \frac{v^2}{2} \int \d\sigma\d t \Big(
    \partial_t X^I \partial_t X^I 
    - \lambda^2\partial_\sigma X^I \partial_\sigma X^I - \mu^2 X^I X^I \cr
    + 2\i\frac{p^+}{v^2}\big(\theta^1\bar\gamma^- (\partial_t - \lambda^2\partial_\sigma) \theta^1
    -\theta^2\bar\gamma^-(\partial_t-\lambda^2\partial_\sigma\theta^2\big)
    -4\i \mu \frac{p^+}{v^2}\theta^1\bar\gamma^-\Pi\theta^2\Big).
    \label{S-ppwave-LCG-newdefs}
\end{multline}
The tensionless limit corresponding to $\lambda\rightarrow 0$ does not present
any difficulty and results in discarding the $\sigma$-derivatives. The result
is the light\-cone action for the tensionless string
\begin{multline}
  S_{LCG}^0 = \frac{v^2}{2} \int \d\sigma\d t \Big(
    \dot X^I \dot X^I 
    - \mu^2 X^I X^I \cr
    + 2\i\frac{p^+}{v^2}\big(\theta^1\bar\gamma^- \dot\theta^1
    -\theta^2\bar\gamma^-\dot\theta^2\big)
    -4\i \mu \frac{p^+}{v^2}\theta^1\bar\gamma^-\Pi\theta^2\Big),
    \label{S-ppwave-LCG-tensionless}
\end{multline}
the {\sl dot} indicating the derivative with respect to $t$. Since
\begin{align}
  X^+ = \frac{p^+}{T}\tau = \frac{p^+}{v^2}t,
\end{align}
$p^+$ is still the conjugate momentum for $X^+$. The action is accompanied by the
Virasoro constraints
\begin{align}
  p^+ X^{-\prime} + \i p^+(\theta^1\bar\gamma^-\partial_\sigma \theta^1
    + \theta^2\bar\gamma^-\partial_\sigma \theta^2)  = 0& \cr
  2 p^+ \dot X^- + 2i p^+  (\theta^1\bar\gamma^-\dot\theta^1
    + \theta^2\bar\gamma^-\dot\theta^2) - 4\i m p^+ \theta^1
      \bar\gamma^-\Pi\theta^2 & \cr
      -\mu^2v^2X^IX^I + v^2\dot X^I \dot X^I&=0. \label{Virasoro-constraints-LCG-0}
\end{align}
The equations of motion for $X^I$ and $\theta^A$ are
\begin{gather}
  \ddot X^I + \mu^2 X^I = 0, \cr
  \dot \theta^1 - \mu\Pi\theta^2 = 0, \hspace*{1cm}
  \dot \theta^2 + \mu\Pi\theta^1 = 0.
\end{gather}
We see that --- as expected --- $X^\mu$ behaves as a collection of 
particles with mass $\mu$ enumerated by $\sigma$.
For closed string boundary conditions, the equations of motion are solved by
\begin{multline}
   X^I_0(\sigma,t)=
    \cos( \mu t) x^I_0+\frac{1}{\mu v^2}
    \sin( \mu t) p^I_0\cr 
  +\frac{\i}{\mu}\sum_{n\neq 0}\mathop{sign}(n)\big\{
    \tilde\alpha_n^{1I}\e^{-\i(\sign(n)\mu t-2n\sigma)}
    +\tilde\alpha_n^{2I}\e^{-\i(\sign(n)\mu t+2n\sigma)}\big\}, \nonumber
\end{multline} \vspace*{-0.7cm}
\begin{multline}
  \theta^{1}_0(\sigma,t)= \cos(\mu t)\theta^{1}_0+
    \sin(\mu t)\Pi\theta^{2}_0\cr 
  +\frac{1}{\sqrt{2}}\sum_{n\neq 0}  \big\{
    \theta^{1}_n\e^{-\i(\sign(n)\mu t-2n\sigma)}
    +\i\Pi\theta^{2}_n\sign(n)
    \e^{-\i(\sign(n)\mu t+2n\sigma)}\big\}, \label{eqn:sol:tension:plane-wave-0}
\end{multline}
and similar for $\theta^2$. The Poisson brackets and the light\-cone Hamiltonian
follow in the same way as in the tensile theory. In order to quantize the
tensionless string, we can make use of the fact that the solution looks very
similar to the tensile case \eqref{eqn:sol:tension:plane-wave}. It can 
be derived as a limit of the latter as opposed to the case in flat space. We
then show that this limit survives quantization. In fact, the quantized
tensionless string is the very same limit of the tensile quantum string. To
this end, we make use of the definitions we used to obtain the tensionless
action
\begin{align}
  T=\lambda v^2, && \mu = \lambda m, && \tau = \lambda t,
\end{align}
and accompany them with $w_n = \lambda \omega_n = \sign(n)\sqrt{\mu^2+4\lambda^2n^2}$, 
where $\omega_n$ are the
frequencies entering the tensile solution \eqref{omegan-cn}. Plugging these
definitions into \eqref{eqn:sol:tension:plane-wave} yields
\begin{multline}
   X^I_\lambda(\sigma,t)=
    \cos( \mu t) x^I_0+\frac{1}{\mu v^2}
    \sin( \mu t) p^I_0\cr 
  +\i\sum_{n\neq 0}\frac{\lambda}{ w_n}\big\{
    \alpha_n^{1I}\e^{-\i(w_n t-2n\sigma)}
    +\alpha_n^{2I}\e^{-\i(w_n t+2n\sigma)}\big\}.
\end{multline}
Here, we focus on the bosonic fields $X^I$ only. The fermionic fields are treated
similarly. If in addition, the oscillators are rescaled as $\tilde\alpha_n^{AI}
= \lambda\alpha_n^{AI}$ the tensile solution looks almost like the tensionless
one except for the frequencies $w_n$. However, this is the only place where
$\lambda$ enters the solution. For $\lambda\rightarrow 0$ the spectrum becomes
degenerate $w_n\rightarrow\sign(n)\mu$ (and $c_n\rightarrow 1/\sqrt{2}$
correspondingly). In this limit, the tensile solution matches the tensionless
\begin{align}
X^I_\lambda \rightarrow X^I_0.
\end{align}
A closer look at the dimensionless bosonic creation and annihilation operators
\eqref{dimless-ops} reveals the following. 
\begin{gather}
  a_0^I = \sqrt{\frac{T}{2m}}\big( \frac{p_0^I}{T}-\i m x_0^I\big) =
          \sqrt{\frac{v^2}{2\mu}}\big( \frac{p_0^I}{v^2}-\i\mu x_0^I\big) \cr
  a_n^{AI} = \sqrt{\frac{T}{\omega_n}}\alpha_n^{AI} =
          \sqrt{\frac{2v^2}{w_n}}\tilde \alpha_n^{AI} 
          \stackrel{\lambda\rightarrow 0}{\longrightarrow}
          \sqrt{\frac{2v^2}{\mu}}\tilde\alpha_n^{AI}.
\end{gather}
The fermionic modes do not depend on $m$ and $T$. As the tension goes to zero,
the operators remain almost unchanged up to the fact that the frequencies entering
the $a_n^{AI}$'s degenerate. However, they do not enter the commutation relations
\eqref{dimless-commutators}. That is why the new, dimensionless modes were
introduced in the beginning. A direct quantization of the tensionless string
leads to the same result.  Thus, we conclude that the limit $T\rightarrow 0$
survives and commutes with quantization. In flat space, the only scale $T$ is
lost when the string becomes tensionless. Here, the background provides a second
scale with $\mu$. We should mention that the light-cone Hamiltonian for the
tensionless theory is
\begin{multline}
  H_{LC}^0 = \mu(4 + e_0 + \bar a_0^I a_0^I + 2\bar\eta_0\bar\gamma^-\Pi\eta_0) \cr
    +\mu \sum_{n=1}^{\infty}  \big(
     \bar a_n^{1I} a_n^{1I} + \bar a_n^{2I} a_n^{2I}
    +\bar\eta_n^1\bar\gamma^-\eta_n^1
    +\bar\eta_n^2\bar\gamma^-\eta_n^2 \big).
\end{multline}
The level matching condition for the physical states is
unchanged compared to the tensile case
\begin{align}
  (N^1-N^2)\ket{phys} = 0, &&
  N^A = \sum_{n=1}^\infty n \big(
    \bar a_n^{AI} a_n^{AI} + 
    \bar\eta_n^{A}\bar\gamma^-\eta_n^A \big).
\end{align}
The spectrum of the theory gets highly degenerated, since all $\omega_n$
collapse to a single value for the tensionless string. We make the following
nice observation.  The tension $T$ and the background scale $m$ enter the theory
in such a way that $mT=\mu v^2 = p^+f$ is kept constant when taking the tension
to zero. This allows for a different interpretation of the results. $m$ is the
origin of the curvature of the plane wave. Therefore, instead of considering
tensionless strings in a plane wave with finite curvature, we may change the
perspective and view the solution as a string with tension $v^2$ moving in an
infinitely curved background with $m\rightarrow \infty$ where the contribution from
the background to the energy is much higher in comparison to the splitting for
the different oscillators as shown in figure \ref{figure1} \cite{deVega:1994hu}.

\section{Tensionless strings in homogeneous plane waves}
\label{sec:hppwaves}

A question at hand is, whether the obtained results are a peculiarity of the
plane wave or if there is a generalization to more complicated
situations. To determine this we look at other types of backgrounds. 
Homogeneous plane waves that are also known as Hpp-waves are slight
generalizations of the plane wave \cite{Blau:2003rt}. These backgrounds are
parametrized by two matrices $k_{IJ}$ and $f_{IJ}$. The line element is given by
\begin{eqnarray}
  \d s^2=2\dx^+\dx^-+k_{IJ}x^Ix^J{\dx^+}\dx^++2f_{IJ}x^I\dx^J\dx^++\dx^I\dx_I.
  \label{eqn:Hppwavemetric}
\end{eqnarray}
Such a background is not maximally supersymmetric in general. It is supported by
a $B$-field given which has the $D-2$ components $B_{I+}=h_{IJ}x^J$. By a
rotation of the transverse coordinates, $k$ can be chosen to be diagonal:
$k_{IJ}=k_I\delta_{IJ}$. The type IIB string in this background is an integrable
model and was solved by Blau
et.\,al.\ \cite{Blau:2003rt} via a so-called frequency base ansatz:
\begin{eqnarray}
  X^I(\sigma,\tau)=\sum_{n=-\infty}^{\infty}X_n^I(\tau)\e^{2\i n\sigma},~~~
  X_n^I(\tau)=
    \sum_{\ell=1}^{2d}\xi_{n\ell\,}^{\vphantom{I}}
    a_{n\ell\,}^{I}\e^{\i \omega_{n\ell\,}\tau}.
\end{eqnarray}
In the quantized theory, $\xi_{n\ell\,}$ become the raising and lowering
operators. The coefficients $a_{n\ell\,}^{I}$ are eigenvectors of the matrix
\begin{eqnarray}
  M_{IJ}(\omega,n)=
    (\omega^2+k_I-4T^2n^2)\delta_{IJ}+2\i\omega f_{IJ}+4\i Tnh_{IJ}.
  \label{eqn:matrix}
\end{eqnarray}
The allowed frequencies $\omega_{n\ell}$ are determined by $\det
M(\omega_{n\ell\,},n)=0$ and the eigenvectors are given by
$M_{IJ}(\omega_{n\ell\,}^{\vphantom{I}},n)a_{n\ell\,}^{I}=0$. For the special
choice $h_{IJ}=f_{IJ}=0$ and constant $k_I=-m^2$ and for $d=2$, we get back the
plane wave solutions $\omega_{n\pm}=\pm\sqrt{m^2+4T^2n^2}$. Previously, the
frequencies in the tensionless case were obtained by letting $T\rightarrow 0$
directly in the corresponding expression. This works here as well. The
frequencies become degenerate and equal to the frequencies for $n=0$:
\begin{eqnarray}
  \omega_{n\ell\,}\rightarrow \omega_\ell,
  ~~\mbox{such that}~~ \det M(\omega_\ell)\equiv \det M(\omega_\ell,0) = 0.
\end{eqnarray}
It seems plausible that this result still holds in the corresponding quantized
theory. We leave this chapter with the open question how our results can lead to
insights in the context of tensionless strings on $AdS_5\times S^5$.

\chapter{Macroscopic Tensionless Strings}
\label{ch:T=0_Strings}\label{sec:IIB-tensionless:flat}

Tensionless strings appear at various places in string theory. In [I] we show
how they fit into the context of supergravity backgrounds. Generalizing the
results of Dabholkar and Hull which we presented in section
\ref{effective-theory}. Schwarz found a family of backgrounds to
IIB supergravity which have a macroscopic string as their source. This family is 
connected by $SL(2,\Z)$ transformations, the group under which type IIB string
theory is believed to be selfdual \cite{Schwarz:1995dk}. Today, the macroscopic
string is interpreted as a bound state of (fundamental) F-strings and D-strings,
one dimensional D-branes \cite{Witten:1995im,deAlwis:1996ze}. We derive the
background sourced by a tensionless string by accelerating Schwarz' solution to
the speed of light in a certain way. This limit resembles the gravitational
shock wave of a massless particle which was obtained in \cite{Aichelburg:1970dh}
in the same way.

We start with a review of the shock wave geometry of a massless particle. Then
we present the solution of Schwarz and show how the tensionless limit is
obtained.

\section{The gravitational shock wave}
The geometry of a pointlike particle moving at the speed of light is a
gravitational shock wave. In \cite{Aichelburg:1970dh} it was obtained by
considering a Lorentz transformation of a massive particle.  We consider the
geometry of a string traveling with the speed of light. A short introduction to
the original discussion is hence appropriate. The gravitational field of a particle
is derived from the Einstein-Hilbert action
\begin{align}
  S_{\rm EH} = \int \d^4x \sqrt{-\metricpullback}R.
\end{align}
A pointlike object of mass $m$ favors a spherical symmetric solution in its rest frame,
the Schwarzschild metric
\begin{gather}
  \ds^2 = \frac{(1-A)^2}{(1+A)^2}\dt^2 - (1+A)^4(\dx^2 + \d y^2 + \d z^2),\cr
  \hspace*{0.5cm}A=\frac{m}{2r},\hspace*{1cm}r^2=x^2+y^2+z^2.
  \label{eqn:particle-ds2}
\end{gather}
As particles moving at the speed of light tend to be massless, one might try to send
$m\rightarrow 0$, but that would recover empty Minkowski space, except for the
singularity at $r=0$. Moreover, the expectation of the gravitational field of a particle
traveling at the speed of light is rather a shock-wave front traveling alongside the
particle. The right way to approach the question is via a Lorentz transformation
and to see how the gravitational field and hence the metric behaves in the limit
of an infinite transformation. We choose to act on the $x$ and $t$ directions,
\begin{align}
  t \rightarrow t' = \gamma(t+vx),&&
  x \rightarrow x' = \gamma(x+vt),&& \gamma=\left(1-v^2\right)^{-1/2}.
  \label{eqn:lorentz}
\end{align}
After this transformation, the metric becomes
\begin{gather}
  \ds^2 = (1+A)^4(\dt^2-\dx^2-\d y^2-
  \d z^2)-\gamma^2\left[(1+A)^4-\frac{(1-A)^2}{(1+A)^2}\right](\dt-v\dx)^2,
  \nonumber \\
  A = \frac{\gamma^{-1}p}{2\sqrt{\gamma^2 x^{\prime 2}+(y^2+z^2)}}.
\end{gather}
Here, $p=\gamma m$. In order to compare a massive particle with mass $m$ at
rest with a massless particle traveling at the speed of light, we keep the
energy fixed. Especially, $p$ becomes the momentum of the massless particle.
It is tricky to take the limit $v\rightarrow 1$ because the transformation becomes
divergent, since $\gamma \rightarrow \infty$. This problem can be avoided by yet
another change of coordinates:
\begin{align}
  x''-vt'' &= x'-vt',\nonumber \\
  x''+vt'' &= x'+vt'-4p\ln\big(\sqrt{(x'-vt')^2+\gamma^{-2}}-(x'-t')\big).
\end{align}
In these new coordinates it is easy to `accelerate' the particle to the speed of light.
After going to light cone coordinates $x^-=t''-x''$, $x^+=t''+x''$, the line element
becomes
\begin{gather}
  \ds''^2=\d x^+\d x^- - \d y^2 - \d z^2 +8p\ln\sqrt{y^2+z^2} \delta(x^-)(\d
x^-)^2. \label{massless-particle-metric}
\end{gather}
In the eyes of a spectator looking in the boosted direction, this is indeed a
gravitational shock wave front. 

\section{The macroscopic IIB string}
\label{sec:I:SL2Z_multiplet}

Type IIB supergravity contains two two-form fields $\bfield^{(1)}$ and
$\bfield^{(2)}$, belonging to the NS-NS and R-R sector, respectively, two real
scalar fields, the dilaton $\dilaton$ and the axion $\chi$, the graviton and a
four-form field $C_4$ with self-dual field strength $F_5=*F_5$.  The bosonic
part of the supergravity action in the string frame reads
\begin{multline}
  S_{\rm IIB}^{string} = \frac{1}{2\kappa^2}\int \d^{10}x \sqrt{-\metric} \Big[
    \e^{-2\dilaton} \big( R+4\nabla \dilaton \cdot \nabla \dilaton 
      -\frac{1}{12}(\H^{(1)})^2 \big) \cr
      - \frac{1}{12}\big(\H^{(2)} + \chi \H^{(1)}\big)^2 
      -\frac{1}{2}(\del\chi)^2 - \frac{1}{480}\big(F_5 + \H^{(1)}\wedge\bfield^{(2)}\big)^2 \Big]
       \cr
      + \frac{1}{4\kappa^2}\int \left( C_4 +
        \frac{1}{2}\bfield^{(1)}\wedge\bfield^{(2)}\right)\wedge
        \H^{(2)}\wedge\H^{(1)}.
      \label{eqn:IIB-SUGRA-action-string} 
\end{multline}
The part coming from \eqref{eqn:space-time-action} is obvious. Type IIB theory
is self-dual under S-duality. Basically, this duality exchanges the strong and
weak coupling limit of the theory. The string coupling is given by
$g_s=\e^{\dilaton_0}$ and S-duality replaces $\dilaton$ with $-\dilaton$.
$\dilaton_0$ is the vacuum expectation value of the dilaton. In fact,
S-duality is supposed to be part of a much bigger symmetry of IIB string theory, 
namely $SL(2,\mathbbm{Z})$. However, to understand this symmetry, we would need
a non-perturbative picture of the string theory. Schwarz \cite{Schwarz:1995dk}
slightly generalized the results of \cite{Dabholkar:1990yf} and found a type IIB
background whose source is a macroscopic string that is charged under both
$\B$-fields. The four-form field charge is carried by a self-dual three-brane.
Here, we are only interested in charges carried by strings. Therefore, $C_4$ and
its field strength $F_5$ is consistently set to zero in the following. We start
with rewriting the action in the Einstein frame.
\begin{multline} 
  S_{\rm IIB} = \frac{1}{2\kappa^2}\int\d^{10}x \sqrt{-\metric} \Big[
    R - 2\big(\nabla\dilaton\cdot \nabla \dilaton -
\e^{2\dilaton}(\partial\chi)^2) \cr
  -\frac{1}{12}\big(\e^{-\dilaton}(\H^{(1)})^2 + 
  \e^{\dilaton}(\H^{(2)} + \chi \H^{(1)})^2\big) \cr
    -\frac{1}{2}(\d\chi)^2 - \frac{1}{480}\big(
    \H^{(1)}\wedge\bfield^{(2)}\big)^2 \Big]
       \cr
     + \frac{1}{8\kappa^2}\int 
\bfield^{(1)}\wedge\bfield^{(2)}\wedge 
        \H^{(2)}\wedge\H^{(1)}.\label{eqn:IIB-SUGRA-action}  
\end{multline} 
The two-forms and their three-form field strengths $\H^{(i)}$ can conveniently
be combined into two-component vectors $\calB = (\B^{(1)},\B^{(2)})$ and
$\calH$. For the sake of a better comparison to [III] we define the field
strengths with an additional factor of $2$ as compared to the introductory
chapter \ref{basics}, $H^{(i)}=2\d \B^{(i)}$. The two real scalar fields, on the
other hand, combine to one single complex scalar
$\lambda=\chi+\i\e^{-\dilaton}$ and we define the matrix
\begin{align}
  {\gen{M}}=\e^\dilaton\left(\begin{array}{cc}
    \abs{\lambda}^2&\chi\\
    \chi&1
   \end{array}\right).
\end{align}
With these ingredients, the target space action for $D=10$ IIB supergravity can
be written in the form
\begin{align}
  S^{\mathrm{IIB}}_{10}=\frac{1}{2\kappa^2}\int\d^{10}x\sqrt{-\G}\left[
  R+\frac{1}{4}\mbox{tr}(\partial {\gen{M}}\partial {\gen{M}}^{-1})-
  \frac{1}{12}{\gen{H}}^T{\gen{M}}{\gen{H}}\right]. \label{eqn:SL2Z-action}
\end{align}
The action has a global $SL(2,{\mathbbm R})$ symmetry that acts on ${\gen{M}}$,
${\gen{B}}$ and $\lambda$ as
\begin{gather}
  {\gen{M}}\rightarrow\Lambda{\gen{M}}\Lambda^T,~~ \hspace*{1cm}
  {\gen{B}}\rightarrow(\Lambda^T)^{-1}{\gen{B}},~~ \hspace*{1cm} 
  \lambda \rightarrow \frac{a\lambda+b}{c\lambda+d},\cr 
  \Lambda=\left(\begin{array}{cc}a&b\\c&d\end{array}\right),~~
  a,b,c,d \in \R.
  \label{eqn:SL2R-symmetry}
\end{gather}
Schwarz found the following $SL(2,{\mathbbm Z})$ set of backgrounds 
\cite{Schwarz:1995dk}. \begin{gather}
\ds^2 = A_q^{-3/4}\left(-\dt^{\,2} + (\dx^1)^2 \right)
        + A_q^{1/4}\,\dx^I\dx_I, \hspace*{1cm} 
     A_q = 1 + \frac{\Delta_q^{1/2} Q}{3r^6},\cr 
  \Delta_q^{1/2} = \vec{q}^T{\gen{M_0}}^{-1}\vec{q} = 
  \e^{\phi_0}(q_1-q_2\chi_0)^2 + \e^{-\phi_0}q_2^2.
\label{eqn:SL2Z-ds2:coeffs}
\end{gather}
Here, $r^2 = x^Ix_I$ is the spacial distance from the string and $x^1$
parametrizes the longitudinal direction of the string. $\phi_0$ and $\chi_0$ are
the vacuum expectation values of $\phi$ and $\chi$ and $\gen{M}$ is built out of
them in the obvious way. At first sight, this metric has an
$SL(2,{\mathbbm R})$ symmetry, but the restriction to $SL(2,{\mathbbm Z})$
follows from the Dirac quantization condition and that $q_1$ and $q_2$ are
relative prime integers, when measured in terms of the fundamental
$\calB_{\mu\nu}$ charge $Q$. If they are not relative prime, the solution can be
decomposed and interpreted as the geometry of multiple strings. Schwarz noticed
that the symmetry naturally prefers both $\calB$ charges $\vec{q}=(q_1,q_2)$ to
be present. The solution is completed by the fields
\begin{align}
\gen{B}_{01} = {\gen{M}}^{-1}\vec{q}\, \Delta_q^{-1/2} A_q^{-1},&&
\lambda = \frac{q_1\chi_0- q_2 |\lambda_0|^2 + \i q_1 \e^{-\phi_0}A_q^{1/2}}
                 {q_1 - q_2\chi_0 +\i q_2\e^{-\phi_0}A_q^{1/2}}. \label{eqn:SL2Z-B}
\end{align}
Here, $\lambda_0$ is the vacuum expectation value of $\lambda$.  The singularity
at $r=0$ is interpreted as an infinitely long source string with a slightly
modified sigma model action \cite{Schwarz:1995dk, deAlwis:1996ze,Cederwall:1997ts}
\begin{align}
S = -\frac{T_q}{2}\int \d^2\xi \Big(
     \partial^a X^\mu \partial_a X^\nu \G_{\mu\nu}
     + \epsilon^{ab}\partial_a X^\mu \partial_b X^{\nu}
     {\gen{B}}^T_{\mu\nu}\vec{q} +\ldots \Big), \label{eqn:sourceaction}
\end{align}
where the string tension is given by
\begin{align}
  T_q = \Delta^{1/2}_q Q = Q\sqrt{\e^{\phi_0}(q_1-q_2\chi_0)^2 +
        \e^{-\phi_0}q_2^2}. \label{T_q}
\end{align}

The action is a generalization of the sigma model action \eqref{S_NLSM_LCG} in
the same way \eqref{eqn:SL2Z-action} generalized \eqref{eqn:space-time-action}.
The background fields in \eqref{eqn:sourceaction} are actually string
condensates that arise as string loop effects \cite{deAlwis:1996ze}. 


This background is interpreted in terms of bound states
of open, F(undamental) and D-strings \cite{Witten:1995im} in the following way. 
The elementary, or fundamental 
string is a source for the NS-NS two-form $B^{(1)}$ but not for the R-R form $B^{(2)}$. We
can say it has charge $\vec{q}=(1,0)$. R-R charges are on the other hand carried
by D-branes. These are hyperplanes on which open strings can end, but much more
important is the fact that they have their own worldvolume dynamics
\cite{Cederwall:1996pv}. A one-brane in a conventional background has the same worldsheet
structure as the elementary string and it is therefore natural to call it a
D-string. It carries R-R charge only, and thus it is reasonable to interpret it
as the $\vec{q}=(0,1)$ partner of the fundamental string. Considering higher
charges, we can look at combined objects of F- and D-strings with charge
$\vec{q}=(q_1, q_2)$. If $T$ is the tension of the fundamental string and
$T_D$ that of a D-string, they are related by 
\begin{align}
  T_D = g_s^{-1}T,
\end{align}
where $g_s= \e^{\phi_0}$ is the string coupling constant. In the absence of the
R-R-field, the tension of a $(q_1, q_2)$-string is given by
\begin{align}
  T_{q_1,q_2} = T\sqrt{q_1^2 + g_s^{-2}q_2^2}.
\end{align}
This can be compared to \eqref{T_q}, taking into account the relation between
the string tension in the Einstein and the string frame, $T_{\rm
Einst}=g_s^{1/2}T_{\rm string}$. S-duality exchanges $g_s$ with $g_s^{-1}$ and is
part of the $SL(2,\mathbbm{Z})$ symmetry. Effectively, it interchanges the two
types of strings.  At weak coupling $g_s\rightarrow 0$, the D-strings is much
heavier than the F-strings. This can thus be interpreted as a theory of weakly
coupled F-strings. At strong coupling however, the situation is vice-versa and
the D-strings might now be seen as the weakly coupled objects. A F-string
carries the fundamental charge of the NS-NS two-form, while the D-strings carry
fundamental charge under the R-R two-form. Thus, a fundamental string has
$\calB$-charge $(1,0)$, while a D-string has $\calB$-charge $(0,1)$. Now, one
might assume a bound state of $p$ F-strings and $q$ D-strings. In the weak
coupling regime, this can be interpreted in the following way: The F-string may end
on the D-string with one of its endpoints. Such a state is allowed, but not
supersymmetric until this point drifts away to infinity. The D-string remains
but now it does not only carry its own R-R-charge, but the NS-NS-charge of the
F-string as well.

\section{The Tensionless $SL(2,\Z)$ String}
From Schwarz' solution we derive the geometry sourced by a tensionless string by
considering an infinite Lorentz transformation.  In comparison to the particle
the situation in \eqref{eqn:SL2Z-ds2:coeffs} is somewhat different. This starts
with the fact that the string is an extended object. However, we only have to
consider velocities orthogonal to the string.  Without loss of generality, we
perform a Lorentz transformation in the $z=x^9$
direction 
\begin{align}\label{lorentztransf}
  t'= \gamma (t+vz),&&
  z'=\gamma(z+vt),&& \gamma=\left(1-v^2\right)^{-1/2}.
\end{align}
There are certain subtleties in taking this limit. The exact derivation is
found in [I]. As in the particle case, we want to take $v\rightarrow 1$ while
keeping the energy constant. This is achieved by introducing a rescaled
fundamental charge $Q_0 = \gamma Q$ which is kept constant. The scalars $\phi$
and $\chi$ tend to their (constant) vacuum expectation values, while the tension
\begin{align}
  T_q = \Delta_q^{1/2}Q = \gamma^{-1}\Delta_{q}^{1/2}Q_0^\phn
\end{align}
vanishes. After going to light-cone coordinates $x^-=z'-t',~x^+=z'+t'$, the
metric becomes
\begin{align}\label{boostedmetric}
(\ds')^2 = \d x^+\d x^- + (\dx^1)^2 + \d r^2 + r^2\d\Omega
          + \frac{\pi\Delta_{q}^{1/2} Q_0^\phn}{8r^5} 
            \delta\left(x^-\right) (\d x^-)^2.
\end{align}
Here, $r^2=(x^2)^2 + \ldots (x^8)^2$. The metric is still invariant under the
$SL(2,{\mathbbm Z})$ transformations \eqref{eqn:SL2R-symmetry}. It is the
ten dimensional analogue of \eqref{massless-particle-metric} and has the 
structure of a plane wave metric
\begin{align}
  \ds^2 = \d x^+ \d x^- + K(x^2, \ldots, x^8, x^-)(\d x^{-})^2 + \sum_{i=1}^8
   (\d x^I)^2.
\end{align}
Concerning the ${\gen{B}}$ field, the Lorentz transformation generates four
non-zero components that diverge in the limit $v\rightarrow 1$. This problem is
overcome by considering the gauge transformation
\begin{align}
{\gen{B}}_{01}\rightarrow  
     {\gen{B}}_{01} - {\gen{M}}^{-1} \vec{q}\, \Delta_q^{-1/2}
   = {\gen{M}}^{-1} \vec{q}\,\Delta_q^{1/2} \left(A_q^{-1} - 1 \right). 
\end{align}
This gauge fixed ${\calB}$ vanishes as $v\rightarrow 1$, and hence does 
${\calH}=\d {\calB}$.  The energy momentum tensor becomes localized at the
position of the string
\begin{align}\label{eqn:Tmunu}
  T_{--}=\frac{1}{24}\pi^3\Delta^{1/2}_{q,0} Q_0^\phn\delta(r)\delta(x^-).
\end{align}
All other components vanish.
This is the energy momentum tensor for a tensionless string localized along the
$X^1$-direction. It can be directly derived from the action \eqref{eqn:S_T=0} of
the tensionless string. 
\begin{eqnarray}
  T_{\mu\nu}(x^I)=
    \del_\tau X_\mu\del_\tau X_\nu\delta^{(8)}(X^I-x^I),
\end{eqnarray}
where the eight-dimensional delta function covers the space transverse to the 
direction of the boost, \cf \cite{Gurses:1974cm}. Here, we already integrated
out the world-sheet directions. This implies that $X^-$ and $X^1$ are fixed to
the values of $x^-$ and $x^1$. Since the string is located at $x^I=x^-=0$, the
only non-vanishing contribution arises from 
\begin{eqnarray}
  X_-=\G_{-+}X^+=X^+\propto \tau.
\end{eqnarray}
From this, we obtain
\begin{eqnarray}
  T_{--}=\del_\tau X_-\del_\tau X_-\delta(r)\delta(x^-)\propto
  \delta(r)\delta(x^-),
\end{eqnarray}
We conclude that \eqref{boostedmetric} is the background geometry generated by a
tensionless string.

    \chapter{From Complex Geometry to Generalized Complex Geometry}\label{GCG}

Generalized complex geometry is a relatively new concept for the description of
the geometry of a manifold. It originates in the context of generalizing the
notion of Calabi-Yau manifolds to include $\B$-field fluxes. These generalized
Calabi-Yau manifolds play an important role in the context of compactification
with fluxes \cite{Grana:2005jc}. It was introduced by Hitchin
\cite{Hitchin:2004ut} and then studied in great detail by his student Gualtieri
\cite{Gualtieri:2003dx}. 

Generalized complex geometry combines the tangent bundle and the cotangent
bundle of a manifold and considers the complex geometry on the direct sum
$E=TM\oplus T^*M$. In this way it unifies complex and symplectic geometry into a
single framework. This makes it very interesting from the physics point of view.
Phase space is a prominent example of a symplectic geometry --- we saw in
chapter \ref{basics} that the symplectic structure gives rise to the Poisson
bracket in the context of sigma models. To continue in this direction,
generalized complex geometry puts the metric and the $\B$-field on an equal
footing with the (ordinary) complex structures. This makes it an elegant notion
to describe the relation between worldsheet supersymmetry of sigma models and
the geometry of their target spaces.

This chapter provides the basic notions of generalized complex geometry that are
needed to understand the relation to supersymmetric sigma models. It is not
intended and it does not claim to be a full introduction to the topic. For this
purpose, we refer to Gualtieri's thesis \cite{Gualtieri:2003dx}.

\section{Complex Geometry}
Before introducing generalized complex geometry, we start with a review of some
facts of complex geometry. For a more detailed introduction, we refer to
\cite{Nakahara:1990th}. A manifold $M$ is almost complex, if it can be equipped
with an endomorphism on its complexified tangent bundle $J \in {\rm
End}(TM\otimes \mathbbm{C})$ satisfying $J^2 = -1$. We denote the $\pm\i$
eigenbundles of $J$ by $\L$ and $\bar \L$.  $\L$ is called integrable if it is
involutive in the sense
\begin{align}
  X,Y\in \L \Rightarrow [ X, Y ] \in \L, \label{integrability-pi}
\end{align}
where $[X,Y]$ is the Lie bracket. In this case, $J$ is called a complex
structure and $M$ is a complex manifold. On a complex manifold, there exists a
chart of local holomorphic and antiholomorphic coordinate frames $\partial_\mu =
(\partial_m, \partial_{\bar m})$ with holomorphic and anti-holomorphic
transition functions such that $J$ is diagonal in these coordinates,
\begin{align}
  J^\mu_{\nu} = \left( \begin{array}{cc}~\i \delta^m_n&~0\\~0&-\i
\delta^{\bar m}_{\bar n} \end{array}\right).
\end{align}
Every complex manifold has an even number of real dimensions, say $2D$.
The integrability condition \eqref{integrability-pi} can also be expressed using
the Nijenhuis torsion for $J$:
\begin{align}
  N(J)[X,Y] = [JX,JY] - J[JX, Y] - J[X, JY] + J^2[X,Y],~X,Y \in TM.
\end{align}
In local coordinates, the Nijenhuis tensor reads $N(J)^\mu_{\nu\sigma} =
J^\rho_{\nu} J_{[\rho\sigma]}^\mu - (\nu\leftrightarrow \sigma)$. The
integrability condition \eqref{integrability-pi} is equivalent to $N(J)=0$.
Fur
further convenience, we also introduce the Nijenhuis concomitant of two
complex structures $I$ and $J$
\begin{align}
  N(I,J) = \half\big([IX,JY] - J[IX, Y] - I[X, JY] + IJ[X,Y] + (I\leftrightarrow J)\big). 
  \label{concomitant}
\end{align}
In particular, $N(J,J) = 2 N(J)$. The product $IJ$ of two complex structures $I$
and $J$ is integrable if $N(I,J)=0$ \cite{Ako:1968, Magri:1984}. 

There are various types of complex manifolds. A few of them shall be presented
here. A complex manifold $M$ is hermitian if it admits a metric $\G_{\mu\nu}$
that is hermitian with respect to the complex structure $J$:
\begin{align}
  J^\rho_\mu \G_{\rho\sigma} J^{\sigma}_\nu = \G_{\mu\nu}. \label{JgJ=g}
\end{align}
This implies that $\omega_{\mu\nu} = \G_{\mu\rho}J^{\rho}_\nu$ is a two-form
of type $(1,1)$ with respect to the complex structure. It is the Kähler form
for $J$. For closed $\omega$ the manifold is Kähler.

$M$ is called bi-hermitian if it admits two complex structures $J_\pm$ and a 
metric that is hermitian with respect to both in the sense \eqref{JgJ=g}, such
that the
complex structures are covariantly constant with respect to a connection
involving the torsion three-form $\H=\d \B$ of the manifold
\begin{align}
  \nabla^{(\pm)} J_\pm = 0, && \Gamma^{(\pm)} = \Gamma^{(0)} \pm T,
\end{align}
where $\Gamma^{(0)}$ is the Levi-Civita connection and
$T^\rho{}_{\mu\nu} = \G^{\rho\sigma}\H_{\sigma\mu\nu}$ is the
Bismut connection for $\H$. This implies that the Nijenhuis concomitant 
\eqref{concomitant} of $J_+$ and $J_-$ vanishes. The Kähler forms $\omega_\pm = \G
J_\pm$ are related to $H$ via
\begin{align}
  \H_{\mu\nu\rho} = \pm J^\kappa_{\pm\mu} J^\lambda_{\pm\nu} J^\sigma_{\pm\rho}(\d
  \omega_\pm)_{\kappa\lambda\sigma}. \label{H=JJJdw}
\end{align}
 An implication of \eqref{H=JJJdw} is
\begin{align}
  \H_{\kappa\lambda\sigma} = \pm J^\mu_{\pm[\kappa}
   J^\nu_{\pm\lambda|}     \H_{\mu\nu|\sigma]}.  
\end{align}

A hermitian manifold that admits two anticommuting complex structures $I$ and $J$ 
is called hyperhermitian.  Their product $K=IJ$ is another complex structure and $I$, $J$ and
$K$ satisfy the quaternion algebra $\mathbbm{H} = Cl_{0,2}(\mathbbm{R})$
\begin{align}
  I^2 = J^2 = K^2 = -1, && K = IJ.
\end{align}
If the two-forms $\omega_I,~\omega_J,~\omega_K$ satisfy relation \eqref{H=JJJdw}
with the same sign, then $M$ is HKT, which originally stands for `hyperKähler
with torsion'. If $H=0$, the two-forms are closed and the manifold is called
hyperKähler. Then,
\begin{align}
  \Omega = \omega_J + \i \omega_K
\end{align}
defines a (2,0)-form for $I$ and $\Omega^{D/2}$ is a top-holomorphic form for
it.

\section{Generalized Geometry}
Let $M$ be a $2D$-real dimensional manifold. An element of the bundle $E = TM\oplus T^*M$ is 
the sum of a vector field and a one-form: $\sX+\sxi \in \Gamma(E)$, where $\Gamma(E)$ is 
the space of sections of $E$. $E$ is the direct sum of the tangent and the
cotangent bundle of the manifold. There is a canonical way to define a symmetric
inner product on $\Gamma(E)$: 
\begin{align}
  \langle \sX+\sxi, \sY+\seta \rangle = \frac{1}{2}(i_\sX \seta + i_\sY \sxi), &&
  \sX+\sxi, \sY+\seta \in \Gamma(E). \label{inner-product}
\end{align}
In a local coordinate frame $(\partial_\mu, \d x^\mu)$, the inner product reads
\begin{align} 
  \langle \sX+\sxi, \sY+\seta \rangle = \sX^\mu\seta_\mu + \sY^\mu\sxi_\mu
\end{align} 
and is represented by the $4D\times 4D$ matrix
\begin{align}
  \gen{I} = \frac{1}{2}\left(\begin{array}{cc} 0&1 \\ 1&0 \end{array}\right).
\end{align}
It can be regarded as a generalized metric on $E$. We will always assume this
local coordinate frame, if we give a coordinate representation. The generalized
metric has signature $(2D,2D)$ and defines the non-compact orthogonal group
$O(2D,2D)$ by the symmetries of $\gen{I}$. The special orthogonal group
$SO(2D,2D)$ preserves the natural inner product and the orientation on $E$.

A subbundle $\L\subset E$ is isotropic with respect to the inner product if 
for all of its sections $\sX+\sxi \in \Gamma(\L)$ the following holds
\begin{align}
  \sY+\seta \in \Gamma(\L) \Rightarrow \langle \sX+\sxi, \sY+\seta \rangle = 0.
\end{align}
It is a maximally isotropic subbundle if in addition
\begin{align}
  \langle \sX+\sxi, \sY+\seta \rangle = 0 \mbox{~for all~}\sX+\sxi\in\Gamma(\L)
  \Leftrightarrow \sY+\seta \in \Gamma(\L).
\end{align}
The tangent bundle of the manifold is an example of a maximally isotropic
subspace of $E$: $TM\subset E$. For a non-vanishing section $\sX$ of $TM$, we
have
\begin{align}
  \langle \sX, \sY+\seta \rangle = i_\sX \seta.
\end{align}
This equation holds for all $\sX\in \Gamma(TM)$ if and only if $\seta=0$.
But $\sY$ is a section of $TM$. If $\L$ is maximally isotropic, then its
complement $\L^*$ in $E$ with $\L\oplus\L^*=E$ is maximally isotropic as well. It
follows that every maximally isotropic subbundle of $E$ is $2D$-dimensional.

Maximal isotropic subspaces can be identified with null spaces of pure spinors
on $M$. Based on the fact that the inner product allows us to regard $SO(2D,2D)$
as the structure group for $E$, Gualtieri proved that it always admits a
$Spin(2D,2D)$ structure. The spin bundle is isomorphic to the exterior algebra
$\wedge T^*M$. A spinor can be regarded as a formal sum of forms of different
rank. A
spinor $\varphi$ defines a subbundle $\L_\varphi \subset E$
via
\begin{align}
  \L_\varphi = \{ (\sX+\sxi)\cdot \varphi = i_X \varphi + \sxi\wedge \varphi = 0 \}.
\end{align}
This is the annihilator of $\varphi$ in $E$, the spinor's null space. By
definition, $\L_\varphi$ is isotropic. If $\L_\varphi$ is maximally isotropic, then
$\varphi$ is called a pure spinor. In general, the pure spinor can only be
defined locally. Therefore, a maximal isotropic $\L$ is identified with a pure spinor
line.

In complex geometry, integrability of the complex structures is defined with the
help of the Lie bracket. There is no Lie bracket action on $TM\oplus T^*M$. 
However, the Courant bracket \eqref{Courant-from-current} is a natural extension 
of the Lie bracket,
\begin{align}
  {[}\sX+\sxi, \sY+\seta{]}_c = {[}\sX,\sY{]} + L_\sX\seta - L_\sY \sxi -
  \frac{1}{2}\d(i_\sX\seta - i_\sY\sxi).
\end{align}
We drop the index $c$ indicating the Courant bracket from now on, when there is
no risk to confuse it with the Lie bracket. The Courant bracket does not satisfy the
Jacobi identity but it shares a lot of properties with the Lie bracket, e.g.\  
diffeomorphism invariance. It has an additional family of automorphisms,
parametrized by closed two-forms $\B\in \Omega^2(M)_{cl}$. This $\B$-field
transformation acts on the sections of $E$ as
a shearing transformation on $T^*M$:
\begin{align}
  \e^\B(\sX+\sxi) = \sX+ (\sxi + i_\sX \B). \label{b-transformation}
\end{align}
It is a symmetry of the inner product \eqref{inner-product} and it
defines an automorphism of the Courant bracket, since
\begin{align}
  {[} \e^\B(\sX+\sxi), \e^\B(\sY+\seta) {]} = \e^\B{[}\sX+\sxi, \sY+\seta{]} +
i_\sX i_\sY \d \B.
  \label{Courant-b-transform}
\end{align}
The last term vanishes since $\B$ is closed. Integrability in generalized
complex geometry is defined in the same way as in complex geometry, except that
the Lie bracket is replaced by the Courant bracket.  A maximally isotropic $\L$
that is closed under the Courant bracket 
\begin{align}
  \sX+\sxi, \sY+\seta \in \Gamma(\L) \rightarrow [\sX+\sxi, \sY+\seta] \in
  \Gamma(\L)
\end{align}
is said to be involutive or integrable. In that case, $\L$ is called a Dirac
structure. By \eqref{Courant-b-transform}, integrability of $\L$ is equivalent
to integrability of the $\B$-transformed bundle
\begin{align}
  \L_\B = \e^\B \L.
\end{align}
There exists a twisted version of the Courant bracket. Let $\H$ be a closed
three form, then the twisted Courant bracket is defined by
\begin{align}
  {[}\sX+\sxi, \sY+\seta{]}_\H = {[}\sX+\sxi, \sY+\seta{]} + i_\sX i_\sY \H.
\end{align}
Besides generalized geometry, one can also define twisted generalized
geometry, where the Courant bracket is replaced by its twisted version. 
Such a twist can be achieved by a transformation with a $\B$-field that is not
closed. This observation provides a convenient technical trick for deriving
results in twisted generalized geometry. Computations are much easier in the
untwisted case and performing such a $\B$-field transformation gives the
corresponding results in the twisted geometry.

\section{Generalized Complex Structures}
An almost generalized complex structure is a maximally isotropic complex
subbundle $\L\subset E\otimes \mathbbm{C}$ such that $\L\oplus \bar \L = E\otimes
\mathbbm{C}$. To make a connection to the notion of complex geometry, an almost
generalized complex structure can equally well be defined as an endomorphism
${\genJ} \in {\rm End}(E\otimes \mathbbm{C})$ that is both complex and
symplectic: 
\begin{align}
  {\genJ}^2 = -1, &&
  {\genJ}^t \gen{I} {\genJ} = \gen{I}. \label{genJ-defs}
\end{align}
We call ${\genJ}$ an almost generalized complex structure, and its $+\i$ 
eigenbundle is the maximally isotropic $\L$. ${\genJ}$ is integrable and called a
generalized complex structure, if $\L$ is integrable.  Twisted generalized
complex structures are defined analogously but with the Courant bracket replaced
by its twisted version. In local coordinates, such a generalized complex
structure can be written as a $4D\times 4D$ matrix
\begin{align}
  {\genJ} = \genmatrix{-J&P\\L&J^t},
\end{align}
where the components of the matrix are regarded as maps between the four
possible combinations of the tangent and the cotangent bundle $J:TM\rightarrow
TM$, $P:T^*M \rightarrow TM$, $P:T^*M\rightarrow TM$, $L:TM\rightarrow T^*M$.
Since $\genJ$ is symplectic \eqref{genJ-defs}, $P$ and $L$ are skew-symmetric.
This allows us to view $L$ as a two-form and $P$ as a bi-vector. $L$ is not to
be confused with the Lie derivative of a vector field $L_\sX$. The
$\B$-transformation acts on ${\genJ}$ as
\begin{align}
  {\genJ}_\B = {\gen{U}}_\B {\genJ} {\gen{U}}_\B^{-1}, &&
  {\gen{U}}_\B = \genmatrix{1&0\\ \B&1}    
\end{align}
and is a symmetry of the inner product \gen{I}.
The two basic examples of generalized complex structures are provided by the
embeddings of an ordinary complex structure $J$ and a symplectic structure
$\omega$ on $M$ in the notion of generalized complex geometry. They correspond
to diagonal and off-diagonal generalized complex structures, respectively:
\begin{align}
  \genJ_J = \genmatrix{-J&~0\\~0&J^t}, &&
  \genJ_\omega =\genmatrix{~0&-\omega^{-1}\\~\omega&~0}.
\end{align}
The pure spinor line bundles for these two examples are given by
\begin{align}
  \varphi_J = \e^\B\Omega,&& \varphi_\omega = \e^{\B+\i\omega},
\end{align}
where $\B\in \Omega^{2}(M)_{cl}$ and $\Omega$ is the top holomorphic form
corresponding to the complex structure $J$.

Locally, a generalized complex manifold can always be decomposed into a complex
and a symplectic part, amounting to choosing local coordinates in which the
generalized complex structure $\genJ$ splits into complex and symplectic parts.
This is a generalization of the Newlander-Nierenberg theorem for complex
manifolds and the Darboux theorem for symplectic ones.

\section{Generalized Kähler Geometry}
Generalized Kähler Geometry is defined by two commuting generalized complex
structures $\genJ_1, \genJ_2$. Vanishing of the commutator implies, that the
product of these two generalized complex structures is a generalized product
structure:
\begin{align}
  \genG = -\genJ_1\genJ_2, && \genG^2 = 1.
\end{align}
$\genG$ is the equivalent of the product structure $\hat\G$ in complex
geometry. It is called the generalized metric but is not to be mistaken for
$\gen{I}$. $\genG$ commutes with the two complex structures by construction
\begin{align}
  [\genG, \genJ_{1,2}] = 0. \label{GJ-comm}
\end{align}
It has signature $(2D,2D)$ and splits $E$ into positive and negative definite
eigenbundles. Ordinary Kähler geometry is included in generalized Kähler
geometry. Let the metric $\G_{\mu\nu}$ be Kähler with respect to the complex
structure $J^\mu_\nu$ and $\omega_{\mu\nu}$ be the corresponding Kähler form. If
$\genJ_J$, $\genJ_\omega$ are the two generalized complex structures of the
example in the previous section then ${[}\genJ_J, \genJ_\omega{]}$ commute by
construction. The generalized metric is given by
\begin{align}
  \genG = -\genJ_J \genJ_\omega = \genmatrix{~0&~\G^{-1}\\~\G&~0}. \label{genG=gg-1}
\end{align}
Equation \eqref{GJ-comm} expressed in terms of ordinary complex geometry
translates into hermiticity of $\G_{\mu\nu}$ with respect to the complex
structure $J^\mu_\nu$. 

Generalized Kähler is equal to bi-hermitian geometry. The map between
these two different formulations is given by
\begin{align}
  \genJ_{1,2} = \frac{1}{2}\genmatrix{-(J_+\pm J_-)&-(\omega^{-1}_+ \mp
  \omega^{-1}_-) \\ \omega_+ \mp \omega_- & J_+^t \pm J_-^t}. \label{genJ12}
\end{align}
The components $J_\pm$ are complex structures that can be understood as follows.
Since $\genG$ commutes with the generalized complex structures, the projection
of $\genJ_{1,2}$ onto the positive and negative eigenspaces of $\genG$ define
complex structure $J_\pm$ and there are exactly two ways for choosing the
relative sign between them, $J_+\pm J_-$. The generalized metric in the case of
a non-vanishing two-form is obtained from \eqref{genG=gg-1} via a
$\B$-transformation
\begin{align}
  \genG_\B = {\gen{U}}_\B^t \genG {\gen{U}}_\B =
    \genmatrix{\G^{-1}\B & \G^{-1} \\ \G + \B\G^{-1}\B & \B\G^{-1}}.
\label{genGB}
\end{align}
Of course, the $\B$-transformation twists the Courant bracket accordingly.  The
notion of generalized Kähler geometry has major advantages in comparison to the
much longer known bi-hermitian formulation. The geometric objects such as metric
and the two complex structures are treated in a unified way and the $\B$-field
enters through the $\B$-transformations. We call the triple $\{\GJ_1, \GG_\B,
\GJ_2\}$ a generalized Kähler structure.

Instead of considering the generalized metric \eqref{genGB} to incorporate the
contribution of the $\B$-field, we can also twist the Courant bracket by
$\H=\d\B$ only and define the generalized metric as in \eqref{genG=gg-1}. In
this way we arrive at twisted generalized Kähler geometry.

Due to the relation to bi-hermitean geometry, generalized Kähler geometry is
an important tool for the study of enhanced supersymmetry in the context of
supersymmetric non-linear sigma models. This relation is elaborated on in detail in
the next chapter.

\section{Generalized Hyperkähler Structure}
Hyperkähler geometry is included in generalized Kähler geometry. Let $I, J, K$
be the three complex structures and $\omega_I$, $\omega_J$ and $\omega_K$ their
Kähler forms. The relation can be seen by choosing $J_+=I$ and $J_-=J$ in
\eqref{genJ12}. 

We define generalized hyperkähler geometry in a different way. Provided three
anticommuting generalized Kähler structures $\GJ_i$, $i=1,2,3$ and a generalized
metric $\GG$, we define generalized hyperkähler geometry by imposing the
relations 
\begin{align}
  [\GG,\GJ_i] = 0, && \{\GJ_i,\GJ_j\} = -\delta_{ij}.
\end{align}
This implies that $\GJt_i = \GG\GJ_i$ are three additional generalized complex
structures. Each of the triples $\{\GJ_i, \GG, \GJt_i\}$ define a generalized
Kähler structure and the six generalized complex structures and the generalized
metric form a representation of the algebra of bi-quaternions $Cl_{2,1}(\mathbb{R})$:
\begin{align}
  \{ \GJ_i, \GJ_j \} = -2\delta_{ij}, &&
  \{ \GJ_i, \GJt_j \} = -2\delta_{ij}\GG, &&
  [ \GJ_i, \GG ] = 0. \label{bi-quaternions}
\end{align}
This definition coincides with the ones in \cite{Huybrechts:2003ak,Goto:2005}.
If we decompose the generalized complex structures according to \eqref{genJ12}
we find two sets of complex structure $J_{+i}$ and $J_{-i}$. They anticommute
among themselves and the metric $\G_{\mu\nu}$ is HKT with respect to both of the
triples
\begin{align}
  \{J_{+i},J_{+j}\} = \{J_{-i},J_{-j}\} = -\delta_{ij}, && J^t_{\pm i}\G J_{\pm
i} = \G, \nabla^{(\pm)}J_{\pm i} = 0.
\end{align}
This is a special case of a bi-hypercomplex geometry.

\chapter{Supersymmetric Sigma Models}\label{susysm}
The relation between supersymmetry and geometry is very intriguing. In chapter
\ref{basics} we saw how superspace is non-trivial even in the ``flat'' case. In
the context of sigma models, the geometry of the target space is determined by
the amount of supersymmetry on the sigma model worldsheet. Gates, Hull and
Ro\v{c}ek showed that a sigma model with manifest $N=(1,1)$ supersymmetry has
its supersymmetry enhanced to $N=(2,2)$ if the target space is bi-hermitean
\cite{Gates:1984nk}. The different target space geometries have been studied and 
classified for around twenty years by now \cite{Hitchin:1986ea,Howe:1988cj,
Lindstrom:2006ee}. 

Even though the possible target space geometries were known it was first the
introduction of generalized complex geometry that provided a clean 
mathematical concept to deal with these geometries. We already discussed that
bi-hermitian geometry is generalized
Kähler, but the map  \eqref{genJ12} between these two descriptions is 
involved. This triggered the question of how this map can be understood in the
context of sigma models. Much work has been done in this direction and by now the
picture is rather clear \cite{Lindstrom:2006ee,Zabzine:2006uz}: A phase space 
description favors the notion of generalized complex geometry
\cite{Zabzine:2005qf}. In [IV] we show that $N=(2,2)$ supersymmetry of the
Hamiltonian of the Gates-Hull-Ro\v{c}ek sigma model leads us directly to
generalized Kähler geometry. From the physics point of view, the map between
generalized Kähler geometry and bi-hermitian geometry can be derived from the
equivalence of the Hamiltonian and the Lagrangian treatment of the sigma model.
In [V] we elaborate this point of view and show the relation between
$N=(4,4)$ supersymmetry and generalized hyperkähler geometry.

In the Lagrangian formulation the additional supersymmetry closes only on-shell
while it is off-shell in the Hamiltonian formulation. Off-shell supersymmetry
for the action can be established by introducing auxiliary fields or by directly
considering certain manifest $N=(2,2)$ formulations. Recently, it has been shown that 
generalized Kähler geometry is in one-to-one correspondence with manifest
$N=(2,2)$ supersymmetric sigma models, where the Lagrangian serves as the
generalized Kähler potential that encodes the generalized Kähler geometry
\cite{Lindstrom:2005zr,Maes:2006bm}. For a generalized sigma model including
auxiliary fields we are not lead directly to generalized Kähler geometry
\cite{Lindstrom:2004iw}. In [III] we elaborate this and show that
supersymmetry favors geometrical objects beyond generalized complex geometry.

The chapter starts out with a review on the possible target space geometries and
their relation to the supersymmetry of the sigma model. We continue with a
description of manifest $N=(2,2)$ supersymmetry before turning to the phase
space formulation of the sigma model and the results of [IV,V].  We conclude
this chapter with a discussion of the generalized supersymmetric sigma model.

\section{Preliminaries}
Throughout this chapter, we use the notion of supersymmetry in terms of
superfields introduced in chapter \ref{basics}. The action for the
$N=(1,1)$ supersymmetric sigma model is a straightforward generalization of the
ordinary sigma model action 
\begin{align}
  S=\int \d^2 \xi \d^2 \theta D_+\Phi^\mu D_-\Phi^\nu \big(\G_{\mu\nu}(\Phi) +
\B_{\mu\nu}(\Phi)\big).
\label{S-GHR}
\end{align}
By construction, the action is invariant under the supersymmetry transformation
\begin{align}
  \delta_0(\eps)\Phi^\mu = -\i(\eps^+ Q_+ + \eps^- Q_-)\Phi^\mu.
\end{align}
Under certain circumstances, \eqref{S-GHR} has additional, non-manifest
supersymmetries \cite{Gates:1984nk}. By dimensional arguments, such
transformations have to be of the form
\begin{align}
  \delta_1(\eps) \Phi^\mu = \i\eps^+D_+\Phi^\nu J^\mu_{+\nu}(\Phi) +
  \i\eps^-D_-\Phi^\nu J^\mu_{-\nu}(\Phi). \label{deltaPhi}
\end{align}
Otherwise, the transformation would involve a dimensionful parameter. If this
is a supersymmetry for \eqref{S-GHR}, the action is invariant under the
transformation. Being a supersymmetry the transformation satisfies the
algebra 
\begin{gather}
  {[}\delta_0(\eps_1), \delta_1(\eps_2){]} \Phi^\mu = 0, \cr
  {[}\delta_1(\eps_1), \delta_1(\eps_2){]} \Phi^\mu = 2 \eps^+_1 \eps^+_2
   \partial_\+\Phi^\mu + 2 \eps^-_1 \eps^-_2 \partial_\- \Phi^\mu.
\label{2ndsusy}
\end{gather}
For physical (on-shell) solutions, \eqref{2ndsusy} may be fulfilled up to the
equations of motions for $\Phi^\mu$. This is exactly the case, when $J_\pm$ are
two complex structures and the target space geometry is bi-hermitian
\begin{align}
    J_{\pm i \mu}^\rho G_{\rho\sigma} J_{\pm i \nu}^\sigma = G_{\mu\nu}, &&
    \nabla^{(\pm)}_\rho J^\mu_{\pm\nu} = 0, \label{bi-herm2}
\end{align}
where the connections are given by $\Gamma^{(\pm)} = \Gamma^{(0)} \pm 
G^{-1}H$. Off-shell supersymmetry is achieved if the two complex structures
commute and $\hat\G=-J_+J_-$ is an integrable product structure with
$\hat\G^2=1$.  There is an alternative possibility that is similar to the
component formulation of worldsheet supersymmetry in section \ref{ws-susy}. We
can add auxiliary superfields $\S_{\pm\mu}$. These fields anticommute and
transform as a worldsheet spinor. Under supersymmetry, they mix with $\Phi^\mu$.

There are a number of other possibilities
for extended supersymmetry. The following table summarizes some of the
corresponding geometries.
\begin{center}
\begin{tabular}[t]{l@{\quad}|@{\quad}l}
  $N$ & target space geometry \\ \hline
  $(0,0)$, $(1,0)$, $(1,1)$ & Riemannian \\
  $(2,0)$, $(2,1)$ & hermitian, without $H$: Kähler \\
  $(2,2)$ & bi-hermitian, without $H$: bi-Kähler or Kähler \\
  $(4,0)$, $(4,1)$ & HKT, QKT, without $H$: hyperkähler \\
  $(4,4)$ & bi-hypercomplex
\end{tabular} \\[0.5ex]
{\small {\sl Table 6.1:} {The amount of supersymmetry restricts the target space
geometry.}
}
\end{center}

\section{Manifest $N=(2,2)$ supersymmetry}
\label{manifest-formulation} 

The focus in this chapter is on extended off-shell $N=(2,2)$ supersymmetry of the
$N=(1,1)$ supersymmetric sigma model. One way to achieve this is to start from a
manifest formulation and rewrite it in terms of $N=(1,1)$ superfields. In this
way, two of the supersymmetries become non-manifest.

Manifest $N=(2,2)$ supersymmetry is introduced by extending the worldsheet with
four Grassmann directions instead of two. We denote these directions by
$\theta^\pm$ and $\bar\theta^\pm$. The corresponding spinorial derivatives are
$\mathbbm{D}_\pm$ and $\bar{\mathbbm{D}}_\pm$. A general superfield depends on
all four Grassmann directions. However, $N=(2,2)$ supersymmetry is implemented
by constraints on the superfields. There are three types of $N=(2,2)$
superfields \cite{Lindstrom:2005zr,Maes:2006bm}. A chiral superfield $\chir$
is constrained by $\bar{\mathbbm{D}}_\pm \chir = 0$ and a twisted chiral field
$\twisted$ by $\mathbbm{D}_+\twisted = \bar{\mathbbm{D}}_-\twisted = 0$. There
is a doublet of semichiral superfields $\mathbbm{X}$ and
$\bar{\mathbbm{Y}}$\footnote{The semichiral superfields are sometimes also
called left-/right-chiral.} \cite{Buscher:1987uw,Sevrin:1996jr} and there are the
corresponding antichiral fields.  We collect the different fields and their
constraints in table 6.2.
\begin{center}
\begin{tabular}[t]{l@{\quad}|@{\quad}l}
  Type & Constraint \\ \hline
  chiral & $\bar{\mathbbm{D}}_\pm \chir = 0$ \\ antichiral & $\mathbbm{D}_\pm
\bar\chir = 0$ \\
  twisted chiral & $\mathbbm{D}_+ \twisted = \bar{\mathbbm{D}}_- \twisted = 0$
\\
  twisted antichiral & $\mathbbm{D}_- \bar\twisted = \bar{\mathbbm{D}}_+
\bar\twisted = 0$ \cr
  semichiral & $\mathbbm{D}_+\bar{\mathbbm{X}} = \bar{\mathbbm{D}}_-\mathbbm{Y} =
0$ \\
  semi-antichiral & $\bar{\mathbbm{D}}_+\mathbbm{X} =
{\mathbbm{D}}_-\bar{\mathbbm{Y}} =
0$
\end{tabular} \\[0.5ex]
{\small {\sl Table 6.2:} {The different types of $N=(2,2)$ superfields.}}
\end{center}

To connect to the previous section, it is useful to write the $N=(2,2)$
superfields in their $N=(1,1)$ supersymmetric components. We define the
$N=(1,1)$ covariant derivatives $D_\pm$ and the corresponding supercharges
$\hat Q_\pm$ by
\begin{align}
  D_\pm = \mathbbm{D}_\pm + \bar{\mathbbm{D}}_\pm,&&
  \hat Q_\pm = \i(\mathbbm{D}_\pm - \bar{\mathbbm{D}}_\pm).
\end{align}
The component fields are then
\begin{align}
  \lambda &\eq \lambda|, & \chi &\eq \chi|, \cr
  X &\eq \mathbbm{X}|, &  Y &\eq \mathbbm{Y}|, \cr
  \Psi_- &\eq \hat Q_-\mathbbm{X}|, & \Upsilon_+ &\eq \hat Q_+\mathbbm{Y}|.
\end{align}
When reducing to $N=(1,1)$ superfields, two of the supersymmetries become
non-manifest.  The non-manifest supersymmetry transformations for these fields
are found by writing the constraints for the $N=(2,2)$ superfields in terms of
the component fields. For the chiral superfields, this reads
\begin{align}
  \hat Q_\pm \chir = \i D_\pm \chir, && \hat Q_\pm \bar\chir = -\i D_\pm
   \bar\chir.
\end{align}
For $p$ chiral fields $\chir^a$ and $\bar p$ antichiral fields $\chir^{\bar a}$,
it is convenient to introduce notation $A=(a,\bar a)$ and the complex structure
\begin{align}
  J_B^A = \genmatrix{\i\delta^a_b&0\\ 0&-\i\delta^{\bar a}_{\bar b}}.
\end{align}
The transformation for $\chir^A$ is simply $\hat Q_\pm \chir^A =
J^A_B D_\pm \chir^B$. Similarly, we introduce coordinates $A^\prime =
(a^\prime,\bar a^\prime)$ for twisted chiral fields $\twisted^{a'}$ and
twisted antichiral fields $\twisted^{{\bar a}'}$ as well as 
$M=(m,\bar m)$ and $M'=(m',\bar m')$ for the semichiral fields $\mathbbm{X}^M$ and
$\mathbbm{Y}^{M'}$. With the complex structures defined in the obvious way, the
non-manifest supersymmetry transformations are
\begin{align}
  \hat Q_\pm\lambda^A &= J^A_B D_\pm \lambda^B, & \hat Q_\pm\chi^{A'} &= \mp J^{A'}_{B'}
      D_\pm\chi^{B'}, \cr
  \hat Q_+ X^M &= J^M_N D_+ X^N, & \hat Q_+ \Psi_-^M &= J^M_N D_+\Psi^N_-, \cr
  \hat Q_+ Y^{M'} &= \Upsilon_+^{M'}, & \hat Q_+ \Upsilon_+^{M'} &= -\i
  \partial_\+ Y^{M'},
\cr
  \hat Q_- X^M &= \Psi_-^M, & \hat Q_- \Psi_-^M &= -\i \partial_\- X^M, \cr
  \hat Q_- Y^{M'} &= - J^{M'}_{N'} Y^{N'}, & \hat Q_- \Upsilon_+^{M'} &= - J^{M'}_{N'} D_-
\Upsilon^{N'}.
\end{align}
A general $N=(2,2)$ action is a functional of the constrained superfields. It
has the form
\begin{align}
  S=\int \d^2\xi \d^2\theta \d^2\bar\theta
  K(\chir^A,\twisted^{A'},\mathbbm{X}^M, \mathbbm{Y}^{M'}). \label{S-N22}
\end{align}
The corresponding component action is obtained by the relation
\begin{align}
  S& = \int \d^2\sigma \d^2\theta \d^2\bar\theta K = 
  \int \d^2\sigma \mathbbm{D}^2_+ \mathbbm{D}^2_- K \\ &=
  \int \d^2\sigma D_+ D_- Q_+ Q_- K| =
  \int \d^2\sigma \d^2\theta \tilde K,
\end{align}
with $\tilde K(\chir, \twisted, X, Y, \Psi_-, \Upsilon_+) = Q_+Q_-
K(\chir,\twisted, \mathbbm{X},\mathbbm{Y})|$. Comparing to the previous section,
we can conclude that the chiral and twisted chiral superfields describe
situations where the two complex structures commute. Semichiral superfields, on
the other hand, describe situations, where the complex structures do not commute
\cite{Lindstrom:2004hi}. All of these situations lead to generalized Kähler
geometry where $K$ is the generalized Kähler potential \cite{Lindstrom:2005zr}.
Recently, it has been shown that a manifest $N=(2,2)$ supersymmetric sigma model
cannot contain any other type of manifest $N=(2,2)$ superfields
\cite{Maes:2006bm}.  Generalized Kähler geometry can be fully described in terms
of chiral, twisted chiral and semichiral superfields. To get a feeling for the
reduction to $N=(1,1)$, consider the topological model
\begin{align}
  S=\int \d^2\xi \d^2\theta \d^2\bar\theta K(\mathbbm{X}, \mathbbm{\bar X}).
\end{align}
With $S_{-M}=\omega_{MN}\Psi_-^N$ and
\begin{align}
  \omega_{MN}=\genmatrix{&\i \del_m\del_{\bar n}K \\ -\i \del_{\bar m}\del_n K&},
\end{align}
the action reduces to
\begin{align}
  S=-\frac{\i}{4}\int\d^2\xi \d^2\theta D_+ X^M S_{-M}.
\end{align}
If the fields are collected into $\Phi^\mu = (\lambda^A,\chi^{A'},X^M,Y^{M'})$ and
$S_\pm$ are defined similar as above, the action \eqref{S-N22} can be brought into the form
\begin{align}
  S=\int \d^2\xi \d^2\theta \Big( D_+\Phi^\mu D_-\Phi^\nu \E_{\mu\nu} + S_{-M}
S_{+N'}\eps^{MN'}\Big). \label{S-manifest-N22}
\end{align}
The tensors $\E_{\mu\nu}$ and $\eps^{MN'}$ are determined by the generalized
Kähler potential $K$ \cite{Lindstrom:2004hi}.

\section{Enhanced supersymmetry in $N=1$ phase space}
\label{enhanced-susy-in-phase-space}

We will now turn to the Hamiltonian treatment of the $N=(2,2)$ supersymmetric
sigma model. The supersymmetric version of phase space corresponds to the
cotangent bundle of the superloop space ${\cal L}M =\{\phi^\mu:
S^{1,1}\rightarrow M\}$. Here, $S^{1,1}$ is a supercircle with coordinates
$\sigma,~\theta$ where $\theta$ is the Grassmann-valued direction. We have to
reverse the parity on the fibers in order to get the right statistics and denote
the cotangent bundle by $\Pi T^*{\cal L}M$. The conjugate momenta are worldsheet
fermions. In a local coordinate frame, we have a superfield
$\phi^\mu(\sigma,\theta)$
and its conjugate momentum $S_\mu(\sigma,\theta)$. Their expansion in $\theta$ is 
\begin{align}
  \phi^\mu(\sigma) = X^\mu(\sigma) + \theta \lambda^\mu(\sigma), &&
  S_\mu(\sigma) = \psi_\mu(\sigma) + \i\theta P_\mu(\sigma),
\end{align}
such that $P_\mu$ is the momentum conjugate to $X^\mu$. Here, we follow the
notation of \cite{Zabzine:2005qf}. The symplectic structure on $\Pi T^* {\cal
L}M$ is defined  such that the restriction to the bosonic part coincides with
\eqref{symplectic-form}
\begin{align}
  \omega = \i \int_{S^{1,1}}\d\sigma \d\theta (\delta S_\mu\wedge \delta
\phi^\mu ). 
\label{omega-susy}
\end{align}
If we perform the Berezin integral, then
\begin{align}
  \omega = \int \d\sigma \big(\delta X^\mu \wedge \delta P_\mu -\i
\delta\psi_\mu \wedge \delta \lambda^\mu\big).
\end{align}
The part that involves the bosonic fields is indeed equal to
\eqref{symplectic-form}. The symplectic structure yields the (super-)Poisson
bracket
\begin{align}
  \{F, G\} = \i \int \d\sigma\d\theta  F\Big( 
   \frac{\overleftarrow{\delta}}{\delta S_\mu} 
   \frac{\overrightarrow{\delta}}{\delta \phi^\mu}
  -\frac{\overleftarrow{\delta}}{\delta \phi^\mu}
   \frac{\overrightarrow{\delta}}{\delta S_\mu}
   \Big) G.
\end{align}
This Poisson bracket satisfies the appropriate graded versions of antisymmetry,
the Leibniz rule and the Jacobi identity. On this phase space, there are two 
operators, a spinorial derivative and a corresponding supercharge
\begin{align}
  D = \partial_\theta + \i \theta \partial, &&
  Q = \partial_\theta - \i \theta \partial. \label{D-and-Q}
\end{align}
They satisfy the algebra
\begin{align}
  D^2 = \i\del, && Q^2 = -\i\del, && \{D, Q\} = 0.
\end{align}
The generator for the manifest supersymmetry is defined by
\begin{align}
  {\Qop}(\eps) = -\int \d\sigma \d\theta \eps S_\mu Q\phi^\mu. \label{Q_1}
\end{align}
It acts on the fields through the Poisson bracket
\begin{align}
  \delta_1(\eps)\phi^\mu = \{\phi^\mu, {\Qop}_1(\eps)\} = -\i\eps Q\phi^\mu &&
  \delta_1(\eps)S_\mu = \{S_\mu, {\Qop}_1(\eps)\} = -\i \eps Q S_\mu.
\label{delta_1}
\end{align}
Taking the Poisson bracket of ${\Qop}_1$ with itself yields the generator of
worldsheet translations
\begin{align}
  \{{\Qop}(\eps),{\Qop}(\tilde\eps)\} = {\Pop}(2\eps\tilde\eps), &&
  {\Pop}(a) = \int \d\sigma \d\theta a S_\mu\partial \phi^\mu.
\end{align}
Any additional generator of supersymmetry transformations ${\Qop}_1$ has to be
of the form \cite{Zabzine:2005qf}
\begin{align}
  {\Qop}_1(\eps) = -\frac{1}{2}\int\d\sigma \d\theta \eps \big( 2D\phi^\mu
S_\nu J_\mu^\nu(\phi) + D\phi^\mu D\phi^\nu L_{\mu\nu}(\phi) + S_\mu S_\nu
P^{\mu\nu}(\phi) \big). \label{Q_2}
\end{align}
It has to satisfy the Poisson brackets
\begin{align}
  \{{\Qop}_1(\eps), {\Qop}_1(\tilde \eps)\} = {\Pop}(2\eps\tilde\eps),&&
  \{{\Qop}(\eps), {\Qop}_1(\tilde \eps)\} = 0. \label{QQ-Poissonbracket}
\end{align}
Here, we assume that the supersymmetry does not have central charges. It is
shown in \cite{Zabzine:2005qf} that these conditions are satisfied if and only if the
tensors in \eqref{Q_2} group together into a generalized complex structure
$\genJ$ and the target space manifold is generalized complex. The
transformations on the fields are given by
\begin{align}
  \delta(\eps)\phi^\mu = \{\phi^\mu, \Qop_1(\eps)\} =& \i\eps\big(D\phi^\nu J^\mu_\nu
- S_\nu P^{\mu\nu}\big), \cr
  \delta(\eps)S_\mu = \{ S_\mu, \Qop_1(\eps)\} =& \i\eps\big(
     D(S_\nu J^\nu_\mu) - \half S_\nu S_\rho P^{\nu\rho}{}_{,\mu} + D(D\phi
   L_{\mu\nu}) \cr & + S_\nu D\phi^\rho J^\nu_{\rho,\mu} - \half D\phi^\nu D\phi^\rho
  L_{\nu\rho,\mu}\big). \label{Q1-components}
\end{align}
For later use we
observe that $\Qop_1$ can be written in a very compact way using the symmetric
inner product \eqref{inner-product}. It makes explicit use of $\genJ$
\begin{align}
  \Qop_1(\eps) = -\frac{1}{2}\int\d\sigma\d\theta \eps \langle \Evec,\genJ
\Evec \rangle. \label{Q-2-alt}
\end{align}
$\Evec$ is given by
\begin{align}
  \Evec=\matrix{c}{D\phi \\ S}. \label{Lambda}
\end{align}

If the geometry is twisted by a non-vanishing three-form $\H$, the symplectic
form gets twisted as in the bosonic case \eqref{omega-H} and the generators are
modified.  This modification can be generated by a $\B$-transformation with
$\H=\d \B$ replacing
\begin{align}
  S_\mu \rightarrow S_\mu - \B_{\mu\nu}D\phi^\mu.
\end{align}
The generator of manifest supersymmetry becomes
\begin{align}
  \Qop(\eps) = -\int\d\sigma\d\theta \eps
(S_\mu-B_{\mu\nu}D\phi^\nu)Q\phi^\mu.
\end{align}
Since also the Poisson bracket gets twisted, the form of the transformations on
the fields \eqref{delta_1} remains unchanged. We denote the twisted Poisson
bracket by $\{\cdot,\cdot\}_H$. It is given by
\begin{align}
  \{F, G\}_H = \i \int \d\sigma\d\theta  F\Big( 
   \frac{\overleftarrow{\delta}}{\delta S_\mu} 
   \frac{\overrightarrow{\delta}}{\delta \phi^\mu}
  -\frac{\overleftarrow{\delta}}{\delta \phi^\mu}
   \frac{\overrightarrow{\delta}}{\delta S_\mu}
  +2\frac{\overleftarrow{\delta}}{\delta S_\nu}H_{\mu\nu\rho}
   \frac{\overrightarrow{\delta}}{\delta S_\rho} 
   \Big) G.
\end{align}
A more detailed description for the case $H\neq 0$ is part of [Zab06a,IV].

\section{The Poisson sigma model - A first application}
In the phase space formulation, time evolution is generated by the Hamiltonian.
It is thus natural to study the condition under which a Hamiltonian is
invariant under the additional supersymmetry. A relatively simple example is the
Wess Zumino (WZ)-Poisson sigma model that plays an important role in the context of
deformation quantization \cite{Schaller:1994es,Bonechi:2005mw}. Its $N=1$ supersymmetric
version is given by the action
\begin{align}
  S_{PSM} = \int\d^2\xi \d\theta \Big( S_\mu D\phi^\mu + \frac{1}{2} 
  S_\mu S_\nu \Pi^{\mu\nu} \Big). \label{S_PSM}
\end{align}
If $\Pi$ is a Poisson structure, it satisfies the Jacobi identity 
\begin{align}
  \Pi^{[\mu\nu}{}_\sigma \Pi^{\rho]\sigma} = 0.
\end{align}
This relation allows for a special choice of local coordinates, in which 
$\Pi$ becomes block diagonal and constant
\begin{align}
  \Pi = \matrix{ccc}{0&1& \\ -1&0& \\ & & 0}.
\end{align}
These coordinates are called Casimir-Darboux coordinates. This simplifies the
local analysis around regular points. $\Pi$ can be twisted by a three form $H$
making it a WZ-Poisson structure that spoils the Jacobi identity
\begin{align}
  \Pi^{[\mu\nu}{}_\sigma \Pi^{\rho]\sigma} =
\Pi^{\mu\kappa}\Pi^{\nu\lambda}\Pi^{\rho\sigma}H_{\kappa\lambda\sigma}.
\end{align}
For such a model, the phase space is constrained by the equations of motion
\begin{align}
  {\cal C}: D\phi^\mu + \Pi^{\mu\nu}S_\nu = 0. \label{PSM-eom}
\end{align}
This is a first class constraint for $S_\mu$. In fact, the left hand side is the
supersymmetric version of a current of the form \eqref{Current}. The canonical
Hamiltonian for the WZ-Poisson sigma model vanishes and thus, the Hamiltonian is
given by the constraint  \cite{Alekseev:2004np}
\begin{align}
  {\Hop}(S,\phi) = \int\d\sigma \d\theta \Lambda_\mu \big(D\phi^\mu +
\Pi^{\mu\nu} S_\nu\big).
\end{align}
The superfields $\Lambda_\mu(\sigma,\theta)$ act as Lagrange multipliers for
the constraint. 
The condition that $\Pi$ is a WZ-Poisson structure is equal to the physical
constraint that ${\cal C}$ is preserved by Hamilton dynamics
\begin{align}
  \{ D\phi^\mu+\Pi^{\mu\nu}S_\nu, \Hop \}|_{\cal C} = 0.
\end{align}
By construction, the Hamiltonian is invariant under the manifest supersymmetry
\eqref{Q_1}. We saw in the previous section that the phase space admits $N=2$ 
supersymmetry if the target space geometry is generalized complex. All that
remains is to find the conditions under which ${\Hop}$ is invariant under the
transformation ${\Qop}_2$ in the constrained phase space, i.e.
\begin{align}
  \{{\Hop},{\Qop}_1(\eps)\}|_{\cal C} = 0. 
\end{align}
These conditions were derived and studied by Calvo \cite{Calvo:2005ww}.
The solution to this equation involves the Dirac structure associated 
to $\Pi$
\begin{align}
  L_\Pi = \{ \sX + \sxi \in TM\oplus T^*M, \xi|_{\cal C} = \Pi(X) \}.
\end{align}
The WZ-Poisson sigma model admits $N=2$ supersymmetry in phase space if and only
if $L_\Pi$ is involutive with respect to the generalized complex structure
${\genJ}$ associated to the second supersymmetry transformation ${\Qop}_1$:
\begin{align}
  {\genJ} (L_\Pi) \subset L_\Pi.
\end{align}

\section{The sigma model Hamiltonian} \label{sigma-model-hamiltonian}
In [IV] we study supersymmetry of the Hamiltonian that corresponds to the sigma
model \eqref{S-GHR} 
\begin{align}
  S = \int \d^2\xi \d^2\theta D_+\Phi^\mu D_-\Phi^\nu (\G_{\mu\nu}(\Phi) +
\B_{\mu\nu}(\Phi)).
\end{align}
To derive the Hamiltonian we reformulate the sigma model in terms of $N=1$
components of the $N=(1,1)$ superfields $\Phi^\mu$ by integrating out one of the
fermionic directions after a proper coordinate transformation. To this extent,
we define
\begin{align}
  \theta^{0,1} = \frac{1}{\sqrt2}(\theta^+\mp\i \theta^-), && D_{0,1} =
\frac{1}{\sqrt2}(D_+\pm \i D_-), && 
  Q_{0,1} = \frac{1}{\sqrt2}(Q_+\pm \i Q_-)
\,. \label{theta-and-D}
\end{align}
With these definitions the action reads
\begin{multline}
  S = -\frac{1}{4}\int \d^2\xi \d\theta^1 \d\theta^0 \Big( 2 D_0\Phi^\mu 
  D_1\Phi^\nu \G_{\mu\nu}  \cr 
  + (D_1\Phi^\mu D_1\Phi^\nu - D_0\Phi^\mu D_0\Phi^\nu)\B_{\mu\nu}  \Big).
\end{multline}
We define the component $N=1$ superfields
\begin{align}
  \phi^\mu = \Phi^\mu|_{\theta^0=0}, &&
  S_\mu = \G_{\mu\nu} D_0\Phi^\nu |_{\theta^0=0}.
\end{align}
and abuse notation to write $\G_{\mu\nu}(\phi) = \G_{\mu\nu}(\Phi)|_{\theta^0 =
0}$ and $\B_{\mu\nu}(\phi) = \B_{\mu\nu}(\Phi)|$. We denote $D=D_1|$ and
$\partial = \partial_\sigma$. With this prescription, $D$ is equal to the
definition in \eqref{D-and-Q}. The phase space action is obtained by performing
the $\d\theta^0$ integral
\begin{align}
  S = \int \d^2\sigma \d\theta (S_\mu -\B_{\mu\nu}D\phi^\nu\big)
  \partial_0\phi^\mu
  - \int \d t {\Hop}(S,\phi), \label{S=theta-H}
\end{align}
where $\Hop$ is the Hamiltonian defined as
\begin{align}
 {\Hop}(S,\phi) = \frac{1}{2}\int \d\sigma \d\theta \Big(
	\i\partial\phi^\mu D\phi^\nu \G_{\mu\nu} + S_\mu D S_\nu \G^{\mu\nu} +
        S_\mu D\phi^\nu S_\rho \G^{\rho\sigma}\Gamma^\mu_{\nu\sigma} 
\cr
        -\frac{1}{3}S_\mu S_\nu S_\rho \H^{\mu\nu\rho} + D\phi^\mu D\phi^\nu
        S_\rho \H_{\mu\nu}{}^{\rho} \Big). 
\label{GHRhamiltonian}
\end{align}
$\Gamma^\mu{}_{\nu\rho}$ is the Levi-Civita connection for $\G_{\mu\nu}$.
The last two terms that depend on $\H_{\mu\nu\rho}$ do not appear in the bosonic
sigma model Hamiltonian. In fact, those two terms do not have purely bosonic
components.  To find the form of the supersymmetry transformation, we
introduce $\eps^{0,1}=\frac{1}{\sqrt2}(\eps^+\mp\i\eps^-)$ and write
\eqref{deltaPhi} in the form
\begin{align}
  \delta(\eps)\Phi^\mu = -\i(\eps^0 Q_0 +\eps^1 Q_1)\Phi^\mu
\end{align}
The term with $\eps = \eps^1$ gives rise to the manifest supersymmetry of the
fields $\phi^\mu$ and $S_\mu$ with $Q=Q_1|_{\theta^0=0}$. The part
involving $\eps^0$ on the other hand is not a source for a manifest
supersymmetry. It gives rise to the non-manifest supersymmetry transformations
\begin{align}
  \tilde\delta_0(\eps) \phi^\mu &= \eps \G^{\mu\nu} S_\nu, \\
  \tilde\delta_0(\eps) S_\mu &= \i\eps \G_{\mu\nu} \partial \phi^\nu +\eps S_\nu S_\rho
   \G^{\nu\sigma}\Gamma^\rho_{\mu\sigma}. \label{delta_0}
\end{align}
In the derivation, terms corresponding to time evolution were dropped. There is
no obvious way to write down a generator for this transformation, since it
cannot be of the form \eqref{Q_2}.
 
The additional supersymmetry transformation $\Qop_1(\eps)$ yields a twisted
generalized complex structure $\GJ$ due to the presence of $\H$.
The Hamiltonian admits enhanced supersymmetry if the target space geometry is
twisted generalized Kähler, since
\begin{align}
  \{ \Hop, \Qop_1(\eps) \}_H = 0 \label{HQ-PB}
\end{align}
implies that the twisted generalized complex structure $\genJ$ commutes with the
generalized metric given in \eqref{genG}
\begin{align}
  \genG = \genmatrix{0&\G^{-1} \\ \G& 0}.
\end{align}
Consequently, the Hamiltonian is invariant under two extra supersymmetries with
generators $\Qop_1(\eps)$ and $\tilde{\Qop}_1(\eps)$ of the form \eqref{Q_2}.
These generators satisfy \eqref{QQ-Poissonbracket} and in addition
\begin{align}
  \{{\Qop}_1(\eps), \tilde{\Qop}_1(\tilde\eps)\}_H = 2\eps\tilde\eps {\Hop}.
  \label{QQ=H}
\end{align}
Since the two generalized complex structures commute with the generalized metric
$\genG$, the two extra supersymmetries commute with the non-manifest
supersymmetry \eqref{delta_0}. Recently, \cite{Malikov:2006rm} provided a rigid
mathematical proof of the derivation of the Hamiltonian and its supersymmetries.

In conclusion, the sigma model Hamiltionian \eqref{GHRhamiltonian} is $N=(2,2)$
supersymmetric if the target space geometry is twisted generalized Kähler. If
$\H=0$, the target space geometry is generalized Kähler. From the physics point
of view the relation between generalized Kähler and bi-hermitian geometry is
thus given by the equivalence of the Hamiltonian and Lagrange formulation of the sigma
model. This can be seen by rewriting the supersymmetry transformation
\eqref{deltaPhi} in the $N=1$ component fields
\begin{align}
  \delta(\eps)\phi^\mu &= \big(\eps^+D_+ \Phi^\nu J^\mu_{+\nu} + \eps^-D_- \Phi^\nu
    J^\mu_{-\nu}\big)|\theta^0=0 \cr
  &=\frac{i}{2}\eps^1\big(D\phi^\nu(J^\mu_{+\nu} + J^\mu_{-\nu}) + S_\nu
    ((\omega_+^{-1})^{\mu\nu} - (\omega_-^{-1})^{\mu\nu})\big) \cr
  &\phantom{=}+\frac{i}{2}\eps^0\big(D\phi^\nu(J^\mu_{+\nu} - J^\mu_{-\nu}) - S_\nu
    ((\omega_+^{-1})^{\mu\nu} + (\omega_-^{-1})^{\mu\nu})\big).     
\end{align}
From this, we identify the tensors in \eqref{Q1-components} and find exactly the
relation \eqref{genJ12} between $J_\pm$ and the generalized complex structures
$\GJ_{1,2}$.


\section{Topological twists}
Our picture of the $N=(2,2)$ supersymmetric sigma model can be used to discuss
topological twists and the corresponding topological field theories in a very
natural way.  The generators of supersymmetry in phase space can be associated
to BRST transformations by converting them to odd generators
\cite{Zabzine:2005qf}.  This is formally done by setting the odd parameter
$\eps$ to one in \eqref{Q_1} and \eqref{Q_2}. This does not change the algebra
that $\Qop$ and $\Qop_1$ satisfy. The linear combination
\begin{align}
  \gen{q} = {\Qop}(1) + \i {\Qop}_1(1)
\end{align}
is nilpotent. A generalized complex structure can therefore be associated to
an odd differential $\gen{s}$ on $C^{\infty}(\Pi T^*{\cal L}M)$.
\begin{align}
  \gen{s}\phi^\mu = \{\gen{q},\phi^\mu\} &&\gen{s}S_\mu =\{\gen{q},S_\mu\}.
\end{align}
In the case of generalized Kähler geometry, we can define two such operators
$\gen{s}_1$, $\gen{s}_2$ by considering the two generators $\Qop_2(1)$ and 
$\tilde{\Qop}_1(1)$.  Due to the relation \eqref{QQ=H}, the Hamiltonian
\eqref{GHRhamiltonian} is BRST exact and can be written in two ways
\begin{align}
  {\Hop} = -\frac{\i}{2}\int\d\sigma\d\theta \gen{s}_1 \tilde \Q_1(1) =
   -\frac{\i}{2}\int\d\sigma\d\theta \gen{s}_2 \Q_1(1). \label{H_gf}
\end{align}
Let us focus on the first version. The topological field theory is localized at
the fixed points of the BRST transformations. Purely bosonic fixed points are given 
by the first class constraint
\begin{align}
   v^\mu p_\mu + \xi_\mu \partial X^\mu = 0, \label{TFT-constraint}
\end{align}
where $v+\xi\in\Gamma(\L)$ is a section of the $+\i$-eigenbundle of $\genJ$. This
theory was originally discussed in \cite{Alekseev:2004np} and later reexamined by
\cite{Bonelli:2005ti}. It covers the topological A- and B-model. The phase space action 
that corresponds to \eqref{H_gf} is
\begin{align}
  S = \int\d^2\xi \d\theta\Big(\big(S_\mu -
B_{\mu\nu}D\phi^\nu\big)\partial_0\phi^\mu + \i\half \gen{s}_1 \tilde{\Qop}_1(1)
\Big).
\end{align}
This is the gauge fixed action for the theory defined by \eqref{TFT-constraint}.
One of the complex structures defines the topological field theory and the
operator ${\gen{s}}_1$ while the other is used for the gauge fixing. The first
term in the action can be interpreted as a topological term.  The two
possibilities of distributing the two generalized complex structures correspond
to the two non-equivalent ways of twisting the $N=(2,2)$ sigma model. An
extensive discussion of topological strings and generalized complex geometry can
be found in \cite{Pestun:2006rj}.

\section{$N=(4,4)$ supersymmetric Hamiltonian}
In this section, we focus on the results of [V] and show how to generalized
hyperkähler geometry from $N=(4,4)$ supersymmetry similar to the discussion in
section \ref{sigma-model-hamiltonian}. We saw that for a generalized complex
target manifold, the phase space admits $N=2$ supersymmetry and that the
Hamiltonian \eqref{GHRhamiltonian} is $N=(2,2)$ supersymmetric on a
(twisted) generalized Kähler manifold. We start with discussing $N=4$ 
supersymmetry in phase space and show that the necessary condition for this is a
generalized hypercomplex manifold that admits three generalized complex
structures satisfying the algebra of quaternions
\begin{align}
  \{\GJ_1,\GJ_2\} = 0, && \GJ_3 = \GJ_1\GJ_2.
\end{align}
According to the discussion of $N=2$ supersymmetry, the additional generators of 
supersymmetry besides the manifest are of the form
\eqref{Q_2}
\begin{align}
  {\Qop}_i(\eps) = -\frac{1}{2}\int\d\sigma \d\theta \eps \big( 2D\phi^\mu
S_\nu J_{i\mu}^\nu(\phi) + D\phi^\mu D\phi^\nu L_{i\mu\nu}(\phi) + S_\mu S_\nu
P^{\mu\nu}_i(\phi) \big).
\end{align}
These are generators of supersymmetry transformations if we can relate them to
generalized complex structures $\GJ_i$. If we denote the generator of manifest
supersymmetry by $\Qop_0(\eps)\equiv \Qop(\eps)$ then we require that the
$\Qop_i(\eps)$ satisfy the supersymmetry
algebra
\begin{align}
  \{ \Qop_i(\eps), \Qop_j(\tilde \eps) \} = \delta_{ij}\Pop(2\eps\tilde\eps), &&
   i=0,1,2,3.
\end{align}
We do not consider a Hamiltonian at this stage. The Poisson brackets involving
$\Qop_0(\eps)$ and those with $i=j$ imply that $\GJ_i$ are (integrable)
generalized complex structures as in the previous discussion. The remaining
brackets translate into conditions for the generalized complex structures
\begin{align}
  \{\GJ_1, \GJ_2 \} = 0,&& \GJ_3 = \GJ_1\GJ_2, &&
  \Xop{N}(\GJ_1,\GJ_2) = 0.
\end{align}
This coincides with the definition of generalized hypercomplex geometry 
The two generalized complex structures anticommute and their (generalized) Nijenhuis
concomitant vanishes. It is defined as in \eqref{concomitant} but with the Lie
bracket replaced by the Courant bracket. Actually, if $\GJ_1$ and $\GJ_2$ are
integrable, then vanishing of the Nijenhuis concomitant implies integrability
for $\GJ_3$. We conclude that the phase space admits $N=4$ supersymmetry if and
only if the manifold is generalized hypercomplex.

Next, we show that invariance of the Hamiltonian \eqref{GHRhamiltonian} under the
three additional supersymmetries $\Qop_i$ requires a generalized hyperkähler
manifold. For this, we combine the discussion of $N=4$ supersymmetry in phase
space with the discussion that lead to $N=(2,2)$ supersymmetry. The Hamiltonian
is $N=4$ supersymmetric if
\begin{align}
  \{ \Qop_i(\eps), \Xop{H}\} = 0,&& i=0,1,2,3.
\end{align}
We compare this to \eqref{HQ-PB} and find that each of the three additional
supersymmetry generators gives rise to a generalized Kähler structure since the
corresponding generalized complex structures commute with the generalized metric
$\GG$
\begin{align}
  [ \GJ_i, \GG ] = 0,&& i=1,2,3.
\end{align}
As a consequence, $\GJt_i = \GG\GJ_i$ are three additional generalized complex
structures. They correspond to supersymmetry transformations $\tilde\Qop_i$
according to the discussion of in section \ref{sigma-model-hamiltonian} such
that
\begin{align}
  \{ \Qop_i(\eps), \tilde\Qop_j(\eps) \} = 2\i\eps\tilde\eps\delta_{ij}\Xop{H}.
\end{align}
Using anticommutativity of the $\GJ_i$, it is not difficult to show that 
$\GJ_i$, $\GJt_i$ and $\GG$ satisfy the relations of a generalized hyperkähler
structure \eqref{bi-quaternions}.

In conclusion, the Hamiltonian \eqref{GHRhamiltonian} is $N=(4,4)$
supersymmetric if and only if the target manifold is generalized hyperkähler, or twisted
generalized hyperkähler for the case $H\neq 0$. 

\section{Twistor space for generalized complex structures}

In this section, we define the twistor space of generalized complex structures
that is associated to the $N=(4,4)$ supersymmetry of the sigma model
Hamiltonian. The idea of a twistor space is to encode the geometric properties
of the target manifold $M$ in the holomorphic structure of a larger manifold,
the twistor space. The original idea goes back to Penrose \cite{Penrose:1976}
and Salamon \cite{Salamon:1982,Salamon:1986}. We here follow the same approach
as in the definition of the twistor space for hyperkähler geometry \cite{Hitchin:1986ea}. 
Twistor spaces of generalized complex structures and generalized Kähler
structure are also discussed in \cite{Davidov:2005, Davidov:2006} in order to
find examples of generalized complex and generalized Kähler structures that are
not induced by complex, symplectic and Kähler structures.  Before discussing the
twistor space for the generalized hyperkähler geometry, we first review the
results for hyperkähler geometry.  Given a hypercomplex structure $J_1$, $J_2$,
$J_3$ the linear combination
\begin{align}
  K = c^1 J_1 + c^2 J_2 + c^3 J_3
\end{align}
is a complex structure if $\vec{c}$ lies on the unit sphere: $c^2 = 1$. This
sphere can be identified with $\mathbb{C}P^1$. $\mathbb{C}P^1$ is usually
represented as $\mathbb{C}^2$ with coordinates $(\zeta,\tilde\zeta)$ and the
identification $(\zeta,\tilde\zeta)\simeq(\lambda\zeta,\lambda\tilde\zeta)$ for
$\lambda\neq 0$. Therefore, we can cover it with two sheets of coordinates
$(\zeta, 1)$ and $(1,\tilde \zeta)$ such that $\tilde \zeta = \zeta^{-1}$ in the
overlapping region. In these coordinates,
\begin{align}
 K = \frac{1-\zeta\bar\zeta}{1+\zeta\bar\zeta}J_1 +
    \frac{\zeta+\bar\zeta}{1+\zeta\bar\zeta}J_2 +
     \i\frac{\zeta-\bar\zeta}{1+\zeta\bar\zeta}J_3.
\end{align}
The twistor space of complex structures is the product space $M\times S^2$, such
that at any point $p\in M$, $S^2$ parametrized the space of complex structures
on $T_p M$. A complex structure for the whole manifold is then given by the pair
\begin{align}
  {\mathbf{K}}_\zeta = \left(\frac{1-\zeta\bar\zeta}{1+\zeta\bar\zeta}J_1 +
    \frac{\zeta+\bar\zeta}{1+\zeta\bar\zeta}J_2 +
    \i\frac{\zeta-\bar\zeta}{1+\zeta\bar\zeta}J_3, I\right),
\end{align}
where $I$ is the ordinary complex conjugation on the sphere. This construction
allows to define hyperkähler geometry in terms of an abstract parameter space.

We now define the twistor space of generalized complex structures in a
completely analogous way. Given the six generalized complex structures $\GJ_i$ and
$\GJt_i$ of the previous section, we find that the linear combinations that 
define generalized complex structures are given by the relation
\begin{align}
  \GK = \tsfrac{1}{2}(c^i+d^i)\GJ_i + \tsfrac{1}{2}(c^i-d^i)\GJt_i, &&
  \vec{c}^2 = \vec{d}^2 = 1. \label{GK=}
\end{align}
The space of generalized complex structures for a generalized hyperkähler
structure is parametrized by $Z=S^2\times S^2$. In $\mathbb{C}P^1\times
\mathbb{C}P^1$ coordinates $z,w$, the vectors $\vec{c}$ and $\vec{d}$ are given
by
\begin{align}
  \vec{c}=\left(\frac{1-z\bar z}{1+z\bar z}, \frac{z+\bar z}{1+z\bar z},
   \frac{\i(z-\bar z)}{1+z\bar z}\right),&&
  \vec{d}= \left(\frac{1-w\bar w}{1+w\bar w}, \frac{w+\bar w}{1+w\bar w},
   \frac{\i(w-\bar w)}{1+w\bar w}\right).
\end{align}
Since the generalized complex structures $\GJ_i$, $\GJt_i$ are a realization of
the bi-quaternionic algebra, it follows that $\GK^2=-1$ and $\tilde\GK = \GG\GK$
where $\GG$ is the generalized metric. The generalized metric $\GG$ acts on the 
parameter space by letting $\vec{d}\rightarrow -\vec{d}$. In the $\mathbb{C}P^1$
coordinate $w$, this corresponds to the anti-podal map
\begin{align}
\tau_w : w \rightarrow -\tilde w^{-1}
\end{align}
that changes the orientation of the $w$-sphere. The ordinary complex structures
for the two spheres $I_z$ and $I_w$ define a complex structure $I$ for $Z$. This
complex structure induces a generalized complex structure on $TZ\oplus T^*Z$ by
\begin{align}
  {\hat\GJ} = \left( \begin{array}{cc}-I&0 \\ 0& I^t \end{array} \right).
\end{align}
A generalized complex structure for the combined space $M\times S^2\times S^2$
is then given by 
\begin{align}
  {\bm J} = (\GK(z,w), \hat\GJ). \label{bfJ}
\end{align}
It is an interesting question, if $\gen{I}$ can be chosen in a more
general way in this context. Generalized complex structures for $S^2\times S^2$
were explicitly defined in \cite{Hitchin:2005cv}.
 
The triples $\{\GK, \GG, \GKt = \GG\GK\}$ form generalized Kähler structures.
The two spheres parametrize the space of ordinary left- and right-complex
structures on $TM$. We can clarify this by introducing 
\begin{align}
  \GJ_i^{(\pm)} = \frac{1}{2}\big(\GJ_i\pm\GJt_i\big) = \frac{1}{2}\big(1 \pm
  \GG\big) \GJ_i.
\end{align}
These are the projections of the generalized complex structures on the $\pm$
eigenspaces of $\GG$. Explicitly and with relation \eqref{genJ12}, they are
given by 
\begin{align}
  \GJ_i^{(\pm)} = \frac{1}{2}\left(\begin{array}{cc}
    -J_{\pm i}& -\omega^{-1}_{\mp i} \\
    \omega_{\mp i}& J_{\pm i}^t
  \end{array}\right).
\end{align}
With this, \eqref{GK=} becomes
\begin{align}
  \GK = c^i\GJ_i^{(+)} + d^i\GJ_i^{(-)}.
\end{align}
We indeed find that $\vec{c}$ and $\vec{d}$ parametrize the two sets of complex
structures $J_{+i}$ and $J_{-i}$.

It remains to show that ${\bm J}$ as defined in \eqref{bfJ} is indeed a
generalized complex structure. In order to see this, we reformulate the previous
discussion in the pure spinor language. Let $\pi$ be the pure spinor line
associated to the generalized complex structure $\GJ_1$ and $\varphi$ be a local
pure spinor representative such that for the sections
$\sX+\sxi$ of the $+\i$ eigenbundle,
\begin{align}
  (\sX+\sxi)\cdot \varphi = i_X\varphi + \xi\wedge\varphi = 0.
\end{align}
Since $\GJ_1$ is integrable, the spinor is pure and satisfies
\begin{align}
  \d \varphi = (\sX+\sxi)\cdot \varphi \label{dvarphi}
\end{align}
for some section $\sX+\sxi$ of $TM\oplus T^*M$. Given that $\varphi$ is a pure spinor 
for the $+\i$ eigenspace of $\GJ_1$, then
\begin{align}
\sigma = (1+\tsfrac{1}{2}z\GJ_{3}^{(+)}+\tsfrac{1}{2}w\GJ_{3}^{(-)})\cdot\varphi
\end{align}
is a pure spinor for $\GK$. Since $\GJ_i$ and $\GJt_i$ are integrable by
assumption, $\GK$ is integrable as well. This follows from the fact that the
Nijenhuis concomitants vanish. Especially, for fix $z,w$, 
\begin{align}
  \d\sigma|_{z,w} = (\sX+\sxi)\cdot\sigma \label{dphi}
\end{align}
for some $\sX+\sxi\in \Gamma(TM\oplus T^*M)$. The bar indicates that the derivative
is taken for fixed values of $z$ and $w$. The generalized complex structure
$\hat\GJ$ is integrable by construction. We can associate to it a pure spinor
$\eta$ such that $(A+b)\cdot \eta = 0$ for sections $A+b$ of $TZ\oplus T^*Z$.
Explicitly, $\eta$ is the top-holomorphic form $\eta=\d z \wedge \d w$. Since
$\sigma$ is holomorphic in $z,w$, the spinor $\rho =
\sigma\wedge \eta$ satisfies 
\begin{align}
  \d(\sigma\wedge\eta) &= \d\sigma|_{z,w}\wedge\eta + (-1)^{|\sigma|}\sigma\wedge\d\eta
  + \d z\wedge \nabla_{\partial_z}\sigma \wedge \eta
  + \d w\wedge \nabla_{\partial_w}\sigma \wedge \eta. \label{dphiwedgeeta}
\end{align}
$\rho$ is a spinor. It is an element of the exterior algebra
$\wedge T^*(M\times S^2\times S^2) = (\wedge T^*M)\wedge (\wedge T^*S^2) \wedge
(\wedge T^*S^2)$. By construction the last two terms in \eqref{dphiwedgeeta}
vanish such that
\begin{align}
  \d \rho &= (X+\xi)\cdot\sigma \wedge\eta + 
    (-1)^{|\sigma|}\sigma\wedge(A+b)\cdot\eta \cr
  &= (X+\xi+A+b)\cdot\rho.
\end{align}
$\rho$ is a pure spinor for the almost generalized complex structure ${\bm
J}=(\GK({z,w}), \hat\GJ)$ and we conclude that ${\bf J}$ is integrable. 

The case for $H\neq 0$ follows in the same way, except that \eqref{dvarphi} and
\eqref{dphi} are replaced by their twisted versions
\begin{align}
  (\d+\H\wedge)\varphi = (\sX+\sxi)\cdot\varphi.
\end{align}

\section{Generalized Supersymmetric Sigma Models}
We already mentioned that the $N=(2,2)$ supersymmetric non-linear sigma model
with action \eqref{S-GHR} yields generalized Kähler geometry and can be
parametrized completely in terms of chiral, twisted chiral and semichiral
$N=(2,2)$ superfields. For the latter case the additional supersymmetry closes
on-shell unless the action is complemented by an auxiliary term taking care of
the additional components of the semichiral fields. For a sigma model with
additional auxiliary fields in general, one is not guided uniquely to
generalized Kähler geometry. This may have various reasons. On one possibility
we elaborate in Paper III, namely a possible geometry beyond the generalized
complex case. Up to field redefinitions, the most general action involving
auxiliary spinorial fields $S_{\pm\mu}$ was introduced in
\cite{Lindstrom:2004eh} 
\begin{align}
  S = -\frac{1}{4}\int \d^2\sigma \d^2\theta  \Big(S_{[+\mu} D_{-]}\Phi^\mu +
      S_{+\mu}\E^{\mu\nu}S_{-\nu} + 2D_+\Phi^\mu
D_-\Phi^\nu(\B_{\mu\nu}-b_{\mu\nu})\Big). 
  \label{SDPhi-action}
\end{align}
We refer to this model as the generalized non-linear sigma model. $b_{\mu\nu}$
is a globally defined two form on $M$, while $\B_{\mu\nu}$ is only locally
defined in general and is the origin for the WZ-term with $H=\d B$. We are
not interested in the difference between $\B$ and $b$ and set them equal to each
other throughout this section. Extended supersymmetry for this action was first
considered in \cite{Lindstrom:2004eh}. The solution makes heavy use of the field
equations for $S_{\pm\mu}$. Also, $\E^{\mu\nu}$ was supposed to be invertible,
in which case one may perform a coordinate transformation similar to a
$\B$-transformation
\begin{align}
  S_{\pm\mu}\rightarrow S_{\pm\mu} + \E_{\mu\nu}D_{\pm}\Phi^\nu
\end{align}
and bring the action into the form
\begin{align}
  S = -\frac{1}{4}\int \d^2\sigma \d^2\theta \Big(S_{+\mu} \E^{\mu\nu} S_{-\nu} + 
  2 D_+\Phi^\mu D_-\Phi^\nu \E_{\mu\nu}\Big). \label{SES-action}
\end{align}
The field equations for $S_{\pm\mu}$ are $S_{\pm\mu}=0$ and like in the
introductory discussion of supersymmetry, it is consistent to substitute them
into the action and recover the original sigma model \eqref{S-GHR} with
$\E=\G+\B$. The price to pay is that the extended supersymmetry only closes
on-shell.  If $\E^{\mu\nu}$ is a Poisson tensor, \eqref{SDPhi-action} becomes
the $N=(1,1)$ supersymmetric version of the Poisson sigma model. It has been
shown \cite{Bergamin:2004sk} that the solution of \cite{Lindstrom:2004eh} works
even in this case, despite the fact the existence of a metric $\G_{\mu\nu}$ was
a crucial point in its derivation. The most general form of an additional
supersymmetry is given by \cite{Lindstrom:2004eh}
\begin{align}
  \delta^{\ppm}(\eps^\pm)\F^\mu=&
     \eps^{\pm}\left(D_{\pm}\F^\nu
     J^{\ppm\mu}_{\nu}-\S_{\pm\nu}P^{\ppm\mu\nu}\right)
     \cr
  \delta^{\ppm}(\eps^\pm) \S_{\pm\mu}=&
    \eps^{\pm}\left(D_\pm^2\F^\nu
    L^{\ppm}_{\mu\nu}-D_{\pm}\S_{\pm\nu}K^{\ppm \nu}_{\mu}
    +\S_{\pm\nu}\S_{\pm\sigma}N^{\ppm\nu\sigma}_{\mu}\right.\cr
  &\left. \qquad \qquad +D_{\pm}\F^\nu D_{\pm}\F^\rho
    M^{\ppm}_{\mu\nu\rho}+D_{\pm}\F^\nu \S_{\pm\sigma}
    Q^{\ppm\sigma}_{\mu\nu}\right) \cr
  \delta^{\ppm}(\eps^\pm) \S_{\mp\mu}=&
    \eps^{\pm}\left(D_{\pm}\S_{\mp\nu}R^{\ppm\nu}_{\mu}+
    D_{\mp}\S_{\pm\nu}Z^{\ppm\nu}_{\mu}+D_{\pm}D_{\mp}\F^\nu
      T^{\ppm}_{\mu\nu}\right.
    \cr
  &\left. \qquad \qquad +\S_{\pm\rho}D_{\mp}\F^\nu U^{\ppm\rho}_{\mu\nu}
    +D_{\pm}\F^\nu \S_{\mp\rho} V^{\ppm\rho}_{\mu\nu}\right.\cr
  &\left. \qquad \qquad +D_{\pm}\F^\nu D_{\mp}\F^\rho X^{\ppm}_{\mu\nu\rho}
    +\S_{\pm\nu}\S_{\mp\rho}Y^{\ppm\nu\rho}_{\mu}\right).
  \label{general-susy}
\end{align}
With these, the transformations are given by, e.g.
\begin{gather}
  \delta(\eps)\F^\mu = (\delta^{(+)}(\eps^+) + \delta^{(-)}(\eps^-))\F^\mu.
\end{gather}
As in \eqref{Q_2} the form of the transformation is constrained by dimensional
arguments, i.e., it must not contain any dimensionful parameter. The involved
tensors are subject to a number of conditions if these transformations are to
satisfy the supersymmetry algebra and in order to yield a symmetry of the
action. If one
disregards the third transformation in \eqref{general-susy}, the remaining two
lines recover the form of the transformations for the cases that yield
generalized complex geometry. While in that case, the conditions for closure of the
algebra have a geometric meaning, here, no such
interpretation is known yet. If only one half of the extended supersymmetry is
considered, say only the $\delta^{(+)}$-transformations, it has been shown
\cite{Lindstrom:2004iw} that generalized complex geometry is a solution. The authors
found it intriguing and rather curious that the tensors involved in the
transformation \eqref{general-susy} seem to rearrange themselves into a
$6D\times 6D$ matrix rather than the $4D\times 4D$ generalized complex matrices
\begin{align}
{\GenJ}^{\p} = \genMatrix{%
             \ph{-}J^\p&-P^\p&\ph{-}0 \\
             -L^\p&\ph{-}K^\p&\ph{-}0 \\
             \ph{-}T^\p&-Z^\p&\ph{-}R^\p
             } \label{genJ+}
\end{align}
and that some of the non-differential conditions could be rewritten in a form
resembling an almost complex structure
\begin{align}
  \GenJ^{\p 2}=-1.
\end{align}
We proceed from here in a bottom-up approach and try to mimic the
concept of generalized complex geometry as best as we can to find a solution for
a very simple case of \eqref{SDPhi-action}. Ignoring all differential
conditions for the moment and guided by the outcome of the previous discussion, 
we arrange the tensors in $6D\times 6D$ matrices. It is worth noticing that
while $D_\pm\F$ live on $TM$, both auxiliary fields $\S_{\pm}$ live
on the cotangent bundle $T^*M$. It is useful to define two copies of the
cotangent bundle $T^*M_\pm$, where the index indicate the copy that $S_\pm$ is
living on, respectively. We define the bundle $E=TM\oplus(T^*M_+\oplus T^*M_-)$.
A look at \eqref{genJ+} reveals that it is written in local coordinates for 
$E$. In addition to $\genJ^{(\pm)}$, we introduce
\begin{align}
  &{\GenJ}^\m = \genMatrix{%
             \ph{-}J^\m&\ph{-}0&-P^\m \\
             \ph{-}T^\m&\ph{-}R^\m&-Z^\m \\
             -L^\m&\ph{-}0&\ph{-}K^\m
             } \label{genJ-}.
\end{align}
The non-differential conditions from the supersymmetry algebra imply that 
$\GenJ^\ppm$ are almost complex structures on $E$ that commute.
\begin{align}
  \GenJ^{\ppm 2} = -1, && [\GenJ^\p, \GenJ^\m] = 0.
\end{align}
The non-differential conditions that come from invariance of the action, say
\eqref{SDPhi-action}, can be understood with the help of the matrix
\begin{align}
  {\Gen{G}} = \genMatrix{0&1&-1\\ 1&0&\E\\ -1&\E^{\,t}&0}. \label{genG}
\end{align}
This matrix has to be bi-hermitian
\begin{align}
  \GenJ^{\ppm t}{\Gen{G}}\GenJ^{\ppm} = {\Gen{G}}.
\end{align}
We are tempted to refer to $\Gen{G}$ as a generalized metric on $E$. This
interpretation is supported by the fact that it encodes the action.
However, in general, it does not have maximum rank and thus fails to be a
suitable candidate. The following observation is worth mentioning. The upper
left $4D\times 4D$ submatrix of $\Gen{G}$ is equal to the generalized metric
$\gen{I}$ that appears in generalized complex geometry. This corresponds to
projecting $E$ onto $TM\oplus T^*M_+$. Under this projection, $\GenJ^\ppm$
reduce to generalized complex structures on the reduced bundle. A similar
argument holds for the projection onto $TM\oplus T^*M_-$.  The differential
conditions are a lot more involved. However, if $\E^{\mu\nu}\equiv \Pi^{\mu\nu}$
is a symplectic structure then integrating out the auxiliary fields yields the
action
\begin{align}
  S = -\frac{1}{2}\int \d^2\sigma \d^2 \theta D_+\Phi^\mu D_-\Phi^\nu \B_{\mu\nu},
  \label{S-B}
\end{align}
where $\B_{\mu\nu}$ is the closed two-form given by the inverse of $\Pi$.
Clearly, this model isi trivial. The main issue that prevents
finding a
geometric condition for the admission of enhanced supersymmetry is a proper
language of integrability similar to the notion of the Courant bracket on $E$.
Therefore, the main purpose of [III] is to look for hints that point in the
right direction.  For this, we make one additional assumption. We assume that
$P^\ppm$ are invertible Poisson tensors. This implies that $J^\ppm$, $K^\ppm$
and $R^\ppm$ are covariantly constant complex structures. This and the remaining
differential conditions resulting from the algebra and invariance of the action
are satisfied provided that the almost complex structures $\genJ^\ppm$ are
`covariantly constant' with respect to a certain connection matrix
\begin{gather}
  \nabla\GenJ^\ppm = \partial \GenJ^\ppm + \GenJ\cdot\genGamma -
\genGamma\cdot\GenJ, \cr \label{nablagenJ}
  \genGamma = {\rm diag}(\Gamma^{(J)}, \Gamma^{(K)}, \Gamma^{(R)}).
\end{gather}
The components of $\genGamma$ are connections that are related to each other
through $\Pi$, $P^{\ppm}$ and their inverses. This resembles the
situation in the Gates-Hull-Ro\v{c}ek case for ordinary supersymmetric sigma models.
But it is still not clear how to correctly interpret this relation as an
integrability condition for $\GenJ^\ppm$. For this model, the
connection matrices $\genGamma$ are flat and have a vanishing generalized
Riemann tensor in the sense
\begin{align}
  \Gen{R} = \d\genGamma - \genGamma\circ \genGamma = 0.
\end{align}
$\B$-transformations are an important ingredient in generalized
complex geometry and could be interpreted as gauge transformations for the
geometry. These transformations have an equivalent in the geometry of the
generalized sigma model. For $\B\in \Omega^2_{cl}(M)$, we define an automorphism
of the bundle $E$ by
\begin{align}
  \Gen{U}_\B=\genMatrix{1&0&0\\-\B&1&0\\-\B&0&1}.
\end{align}
It transforms the complex structure matrices according to
\begin{align} 
  \GenJ{\ppm} \rightarrow \Gen{U}_\B \GenJ^\ppm \Gen{U}_\B^{-1}. \label{e:J=UJU}
\end{align}
If $\genGamma$ transforms as a connection under this `gauge' transformation, then
\eqref{nablagenJ} is invariant. This strengthens its interpretation as an
integrability condition for $\GenJ^{\ppm}$. The full geometric picture remains
unclear, since we lack a proper concept of integrability for
these objects.

The manifest formulation of $N=(2,2)$ sigma models involving semichiral fields
provides a way to gain a better understanding of the presented geometric framework
and its relation to generalized complex geometry. The action
\eqref{S-manifest-N22} is a special case of \eqref{SES-action}. Expectedly,
generalized complex and especially generalized Kähler geometry fits into the
above picture as a certain subclass. In [III], we elaborate this 
and consider the special case of a toy model that involves semichiral
superfields only and has the action
\begin{gather}
  S = -\int \d^2\xi \d^2\theta \d^2\bar\theta \Big( 
    \chiral{X}{\achiral{Y}} - {\achiral{X}}\chiral{Y} \Big)
    = \int \d^2\xi \d^2\theta \d^2\bar\theta \Big(
      \chiral{X}^M B_{MN^\prime} \chiral{Y}^{N^\prime}\Big).
  \label{e:N=2.2action}
\end{gather}
If we reduce this action to $N=(1,1)$ superfields according to section
\ref{manifest-formulation} and make a proper field redefinition, this action
can be rewritten as \cite{Lindstrom:2004hi}
\begin{gather}
  S = - \frac{1}{4}\int \d^2\xi \d^2\theta
    \Big( \S_{+\mu}\B^{\mu\nu}\S_{-\nu} + 
          D_+\phi^\nu \B_{\mu\nu}D_-\phi^\nu \Big),
\end{gather}
where $\B_{\mu\nu}$ is constant and antisymmetric. This implies that the second
term vanishes, however, it is kept here for clarity. In comparison to
\eqref{e:N=2.2action}, this action contains twice as many spinorial fields. As a
consequence of the constraints for the semichiral superfields, half of them are
constrained by
\begin{align}
  \S_{-M} &= \S_{+M^\prime} = 0, \label{e:SFEQ}
\end{align}
There is a second interpretation of this constraint. It can be interpreted as
the field equations for $S_\pm$: Effectively, half of the spinorial fields are
integrated out by means of their equations of motion. In the complex structure
matrices, it is consistent to neglect the entries corresponding to these
components.
Effectively, $\GenJ^\ppm$ collapse to generalized complex structures: 
\begin{gather}
 \begin{split}
   \Gen{J}^\p_{6D} = \genMatrix{
     \ph{-}J^\p&-P^\p&\ph{-}0\\
     \ph{-}0&\ph{-}K^\p&\ph{-}0\\
     \ph{-}0&-Z^\p&\ph{-}R^\p
   } \longrightarrow \gen{J}^\p_{4D} =
   \genmatrix{
     \ph{-}\hat J^\p&-\hat P^\p \\
     \ph{-}0&\ph{-}\hat K^\p
   }.
 \end{split}
\end{gather}
In terms of $M,M^\prime$ coordinates, this reads
\begin{gather}
  \gen{J}^\p_{4D} = \matrix{cc|cc}{
    \ph{-}J^\p&\ph{-}0&\ph{-}0&\ph{-}0 \\
    \ph{-}0&\ph{-}J^{\p\prime}&-P^\p&\ph{-}0 \\
    \hline
    \ph{-}0&\ph{-}0&\ph{-}K^\p&\ph{-}0 \\
    \ph{-}0&\ph{-}0&-Z^\p&\ph{-}R^{\p\prime}
  } \label{e:JGCG}
\end{gather}
where we identified the tensors with their components that survive the collapse,
e.g.\ $K^\alpha_\beta \rightarrow K^A_B$.  There is a similar reduction for
$\Gen{J}^\m_{6D} \rightarrow \gen{J}^\m_{4D}$. This result coincides with the
derivation in \cite{Lindstrom:2004hi}. 

In section \ref{enhanced-susy-in-phase-space} we gave a shorthand notation for
the additional generators of supersymmetry \eqref{Q-2-alt}.
Especially, the $\B$-transformation reduces to $\Evec \rightarrow
\genU_{\B}\Evec$. Here, the situation is different, there are two
derivatives $D_\pm$ and two corresponding auxiliary fields $S_\pm$. 
A way around this problem is to promote the matrices to operators 
\begin{align}
  \genop{J}^\p &= 
	\genMatrix{%
		\ph{-}J D_+   &-P         &\ph{-}0 \\
		-L D_+^2      &\ph{-}K D_+&\ph{-}0 \\
		\ph{-}T D_+D_-&-Z D_-     &\ph{-}R D_+
		}
\end{align}
and define $\Lambda$ by $\Lambda=(\phi,S_+,S_-)^t$. Similarly, we can define
$\genop{J}^\m$ and promote $\Gen{G}$ to an operator. Then, for example, the
transformations \eqref{general-susy} can be written in a very compact way
\begin{align}
  \delta^{\ppm} \Lambda = \eps^\pm \genop{J}^\ppm \Lambda,
\end{align}
up to terms involving third rank tensors.

We conclude that our results strongly point towards a geometrical interpretation
beyond generalized complex geometry though the lack of a proper notion for this
case presents a major obstacle for elaborating further in this direction.

    \chapter*{Deutsche Zusammenfassung}
\addcontentsline{toc}{chapter}{Deutsche Zusammenfassung}

\selectlanguage{\german}
\let\pskip\parskip
\parskip 1ex
Stringtheorie ist eines der faszinierendsten Teilgebiete der modernen
theoretischen Physik. Sie vereint zwei Konzepte, die auf den ersten Blick
unvereinbar erscheinen: Gravitation und Quantenmechanik. Das macht sie zu \glqq
dem\grqq\ Kandidaten für eine Theorie, die alle Naturgesetze beschreibt. Während
Quantenfeldtheorien auf der einen Seite in der Lage sind, die
elektromagnetische, die schwache und die starke Wechselwirkung in hinreichender
Genauigkeit zu beschreiben, haben wir durch Einsteins allgemeine
Relativitätstheorie ein Verständnis der Gravitation für verhältnismäßig große Abstände.
Die Stringtheorie vereint diese zwei Konzepte mittels einer auf den ersten Blick
einfachen und naiven Idee: Warum sollten die fundamentalen Bausteine der Natur
nicht eindimensionale Objekte, Strings, statt punktförmige Teilchen sein?

In der vorliegenden Dissertation werden zwei Aspekte der Stringtheorie näher
betrachtet: Spannungslose Strings und supersymmetrische Sigmamodelle.

In der Teilchenphysik spielen masselose Teilchen eine wichtige Rolle. Das Photon
beispielsweise ist der Träger der elektromagnetischen Kraft. Zudem können
Teilchen bei sehr hohen kinetischen Energien als nahezu masselos angesehen
werden. In der Stringtheorie ist das Äquivalent zur Teilchenmasse die Spannung
$T$ des Strings, dessen Masse pro Einheitslänge. Dem spannungslosen String wird
eine ähnliche Rolle zugesprochen wie den masselosen Teilchen. Er taucht erstmals
in der Literatur auf im Zusammenhang mit der Diskussion von Strings, die sich
wie masselose Teilchen mit Lichtgeschwindigkeit bewegen, jedoch haben wir bis
heute nur ein sehr grobes Verständnis seiner eigentlichen Natur. Auch ihm wird
eine entscheidende Rolle bei der Beschreibung hochenergetischer Strings
zugesprochen. Beispielsweise können wir uns einen String vorstellen, der mit
wachsender Winkelgeschwindigkeit rotiert. Nach und nach wird sich der Großteil
der Energie des Strings um dessen Endpunkte konzentrieren, während der
überwiegende Teil spannungslos wird. Der String zerfällt bildlich gesprochen in
eine Ansammlung freier Teilchen, die sich jedoch nur orthogonal zum String
bewegen können.

Der spannungslose String unterscheidet sich in vielfältiger Weise von einem
\glqq gewöhnlichen\grqq\ String mit nicht-verschwindender Spannung. Die
verschwindende Spannung führt zu einer Erweiterung der Symmetrie der Raumzeit,
des sogenannten Zielraumes, in den wir uns die Weltfläche, die der String
überstreicht, eingebettet denken. In der quantentheoretischen Betrachtung wird
der Unterschied noch drastischer. So kollabiert das Spektrum des spannungslosen
String zu einem einheitlichen Masseniveau: Alle Anregungen des Strings sind
masselos. Insbesondere gilt dies auch für tachyonische Anregungen, die für
gewöhnlich instabil sind und aus dem physikalischen Spektrum entfernt werden
müssen, da sie eine imaginäre Masse besitzen. Der spannungslose String besitzt
keine kritische Dimension. Eine Quantisierung ist für jede beliebige
Raumzeit-Dimension möglich und nicht auf zehn bzw. 26 Dimensionen begrenzt wie
im Falle nicht-verschwindender Spannung.  Jedoch wird die erweiterte
Raumzeitsymmetrie nur in $D=2$ Dimensionen bewahrt, während eine Quantisierung
in allen anderen Fällen in einem topologischen Spektrum resultiert. Man
vermutet, dass der spannungslose String die noch ungebrochene Phase der
Stringtheorie beschreibt, in der alle Zustände gleichberechtigt sind, und dass
zu geringeren Energien hin ein Phasenübergang stattfindet, in dem sich die
verschiedenen Energieniveaus ausbilden.

Das Angregungsspektrum des spannungslosen Strings enthält Zustände mit hohem
Spin. Das legt die Vermutung eines Zusammenhangs zu der sogenannten Higher Spin
Gauge-Theorie (\glqq Höhere-Spin-Eichtheorie\grqq ) nahe. Diese Relation
lässt sich am einfachsten im Zusammenhang mit der AdS/CFT-Korrespondenz
verstehen. Bei der Betrachtung eines Hologramms sieht man ein dreidimensionales
Bild, dessen Information auf einer zweidimensionalen Fläche enthalten ist.
Übertragen auf die Stringtheorie besagt dieses sogenannte holographische Prinzip
in seiner bekanntesten Version, dass Stringtheorie in einem Anti-de Sitter Raum
äquivalent ist zu einer konformen Feldtheorie auf dem Rand dieses Raumes. Diese
Korrespondenz wurde seitdem die Vermutung ihrer Existenz im Jahre 1997 erstmals
aufgestellt wurde, immer wieder getestet. Das hat zu so erstaunlichen
Ergebnissen geführt, wie dass gewisse Sektoren der Stringtheorie mit Hilfe der
dualen Beschreibung exakt lösbare Modelle sind, die sich mit Methoden der
Festkörperphysik lösen lassen. Jedoch gibt es bis heute --- meines Wissen nach
--- keinen direkten Beweis für die Korrespondenz. Sie relatiert die
Kopplungskonstante auf der Feldtheorieseite zur Spannung $T$ in der
Stringtheorie. Daher entspricht der spannungslose String einer freien
Feldtheorie, die die Existenz von Higher Spin-Feldern zulässt.

Fünfdimensionaler Anti-de Sitter ist Teil eines zehndimensionalen Raumes, in
dem der sogenannte Typ-IIB String konsistent quantisiert werden kann,
$AdS_5\times S^5$. Leider ist die Betrachtung von Strings in diesem Hintergrund ein
schwieriges Unterfangen, und besonders zu deren Quantisierung ist nicht viel
bekannt. $AdS_5\times S^5$ ist einer von drei bekannten Hintergründen für
Typ-IIB Strings, die maximal supersymmetrisch sind. Das bedeutet, dass sie 32
Supersymmetrien besitzen. Neben $AdS_5\times S^5$ sind dies der flache, leere
Raum sowie ein erst kürzlich entdeckter, sogenannter Plane-Wave Hintergrund, der
eine Reihe von Eigenschaften mit $AdS_5\times S^5$ gemeinsam hat, aber deutlich
einfacher ist. Wie sich herausstellt, lässt sich dieser Hintergrund in einer
bestimmten Weise als Grenzfall von $AdS_5\times S^5$ ableiten. Seine Einfachheit
ermöglicht es, den geschlossenen Typ-IIB String zu betrachten und für den Fall
der Lichtkegeleichung, in dem nur die transversalen Freiheitsgrade in Betracht
gezogen werden, zu lösen und zu quantisieren.

Die Art, wie die Stringtheorie die Geometrie des Raumes beeinflusst ist sehr
verblüffend. Wir haben schon im Rahmen von Kompaktifizierungen angesprochen,
dass aus Konsistenzgründen der interne, sechsdimensionale Raum von einer
bestimmten Art sein muss. Die Geometrie ist dadurch bestimmt, dass wir unser
vierdimensionalen Teilraum $N=1$ supersymmetrisch sein soll. Wenn der interne
Raum zudem Kähler sein soll, bleibt nur eine Möglichkeit, nämlich, dass es sich
um eine Calabi-Yau-Mannigfaltigkeit handelt. Auch wenn man wusste, dass es neben
Kähler auch andere Möglichkeiten gab, so wurden diese für lange Zeit nicht in
Betrachtung gezogen. Für ein Sigmamodell mit Supersymmetrie auf der Weltfläche,
der Fläche, die ein String in der Zielmannigfaltigkeit, sprich
Raumzeit, überstreicht, wird die Geometrie des Zielraums durch die Dimension der
Weltfläche und der Anzahl Supersymmetrien bestimmt. Beispielsweise besitzt das
$N=(1,1)$ supersymmetrische Sigmamodell die doppelte Anzahl Supersymmetrien wenn
die Zielmannigfaltigkeit bi-hermitesch ist. Auch wenn die Geometrien
klassifiziert sind, so wurden die Fälle, die nicht Kähler waren doch für lange
Zeit als für die Stringtheorie weniger bedeutend eingestuft. Erst in neuerer Zeit wurde mit
generalisierter komplexer Geometrie ein neues mathematisches Konzept entwickelt,
das komplexe und symplektische Geometrie vereint und gleichmäßig zwischen ihnen
interpoliert. Es bietet genau den richtigen Rahmen, um
die Verbindung zwischen Weltflächensupersymmetrie und der Geometrie der
Zielmannigfaltigkeit näher zu untersuchen. So stellte sich heraus, dass eine
Untermenge dieser neuen Geometrien, die sogenannten generalisierten Kähler
Geometrien, identisch ist mit der bi-hermiteschen Geometrie und zudem eine
vollständige Beschreibung im Rahmen von manifester $N=(2,2)$ Supersymmetrie
besitzt. Generalisierte Calabi-Yau Geometrie ist eine weitere Unterkategorie,
die heutzutage bei Flusskompaktifizierungen eine wichtige Rolle spielt.
Letztlich kann generalisierte komplexe Geometrie in der Lage sein, eine
mathematische Erklärung für Spiegelsymmetrie zu liefern. Generalisierte komplexe
Geometrie vereint die topologischen A- und B-Modelle in einem einzigen Modell.

Es folgt eine Zusammenfassung der Originalarbeiten, die dieser Dissertation
zugrunde liegen.

\parskip \pskip

\subsubsection*{Artikel I}
Im ersten Artikel beschreiben wir, wie spannungslose Strings zu
Supergravitationslösungen, Hintergründe, auf denen Stringtheorie konsistent ist,
führen. Zu diesem Zweck betrachten wir die Geometrie eines makroskopischen Typ-IIB Strings
im Grenzfall, in dem der String sich mit Lichtgeschwindigkeit bewegt. Dies führt
dazu, dass die Spannung des Strings verschwindet und die Geometrie ähnlich wie
im Teilchenfall als eine gravitationelle Schockwelle beschrieben werden kann. 

\subsubsection*{Artikel II}
Wir studieren den spannungslosen, geschlossenen Typ-IIB String in der maximal
supersymmetrischen Plane-Wave-Geometrie. Die Lösung ist ähnlich wie im Falle
nicht-verschwindender Spannung. Auch die Quantisierung des spannungslosen
Strings ist im Gegensatz zum flachen Raum unproblematisch. Dies ist auf die
Existenz eines Parameters zurückzuführen, der mit der Krümmung des Hintergrundes
zusammenhängt. Wir können zeigen, dass sich der spannungslose String direkt aus
dem spannungsbehafteten String im Grenzfall verschwindender Spannung ableiten
lässt und konstatieren, dass dieser Grenzwert mit der Quantisierung kommutiert.

\subsubsection*{Artikel III}
Im dritten Artikel diskutieren wir die Bedingung, unter der ein generalisiertes
Sigmamodell mit zwei Supersymmetrien zusätzliche Supersymmetrien besitzt. Wir
finden, dass sich die dabei involvierten Tensoren in natürlicher Weise zu
Objekten gruppieren, die eine geometrische Interpretation jenseits der
generalisierten komplexen Geometrie nahelegen. Aufgrund unseres unzulänglichen
Verständnisses dieser Art von Geometrie sind wir bei unseren Betrachtungen an
ein sehr einfaches Sigmamodell gebunden und können nur die wesentlichen
geometrischen Objekte identifizieren, sowie zeigen, wie generalisierte komplexe
Geometrie in diese Beschreibung eingebettet ist.

\subsubsection*{Artikel IV}
Wir erklären die Beziehung zwischen generalisierter Kählergeometrie und
bi-hermitescher Geometrie von einem physikalischen Standpunkt aus. Wir zeigen,
dass generalisierte Kählergeometrie die Bedingung für $N=(2,2)$ erweiterte
Supersymmetrie im Phasenraum ist. Damit lässt sich die Relation zwischen
generalisierter Kählergeometrie und bi-hermitescher Geometrie mit der
Äquivalenz zwischen Hamilton- und Lagrangeformalismus beschreiben. Als Anwendung
unserer Resultate beschreiben wir topologische Twists.

\subsubsection*{Artikel V}
In diesem Artikel, der auf den Ergebnissen des vorherigen basiert, studieren wir 
die Bedingung für $N=(4,4)$ Supersymmetrie in der Hamiltonformulierung des
Sigmamodells. Wir finden eine Definition für generalisierte Hyperkählergeometrie
und definieren den Twistorraum der generalisierten komplexen Strukturen.

\selectlanguage{\english}

    \chapter*{Svensk Sammanfattning}
\addcontentsline{toc}{chapter}{Svensk Sammanfattning}
\selectlanguage{swedish}

\let\pskip\parskip
\parskip 1ex

Strängteori är en av dem mest fascinerande ämnen som utvecklats inom den moderna
teoretiska fysiken. Den förenar två koncept som inte verkar passa ihop:
gravitation och kvantmekanik. Detta gör strängteori till en potentiell kandidat
för en teori som beskriver naturens alla lager. Medans kvantfältteori å ena
sidan kan beskriva den elektromagnetiska, den svaga samt den starka växelverkan
tillräckligt exakt, så förstår vi gravitationen genom Einsteins allmänna
relativitetsteori som gäller på förhållandevis stora avstånd. Strängteori förenar
dessa två koncept genom en idé som verkar lika enkel som naiv: Varför skall
inte naturens fundamentala byggstenar vara endimensionella objekt, strängar
istället för punktformiga partiklar?

I denna avhandling betraktas två olika aspekter av strängteorin: spänningslösa
strängar och supersymmetriska sigmamodeller.

Masslösa partiklar har en viktig roll inom partikelfysiken. Fotonen till exempel
transporterar den elektromagnetiska kraften. Partiklar som rör sig med väldigt
höga kinetiska energier kan anses som nästan masslösa. Massens ekvivalent inom
strängteori är spänningen $T$, strängens massa per enhetslängd. Den
spänningslösa strängen tilldelas en roll motsvarande masslösa partiklar. För
första gången diskuteras den i samband med strängar som rör sig likt masslösa partiklar
med ljushastigheten, men än idag är vår förståelse av den spänningslösa
strängens natur rätt så grov.  Som masslösa partiklar anses spänningslösa
strängar vara viktiga för förståelsen av strängteorin vid höga energier. Till
exempel kan vi tänka oss en sträng som roterar med en växande vinkelhastighet.
Ju högre vinkelhastigheten blir, desto mer koncentreras strängens energi  kring
dess ändpunkter medans den stora delen av strängen blir spänningslös. Strängen
sönderfaller och blir en samling partiklar som är bundna att röra sig ortogonalt
mod strängen.

Den spänningslösa strängen skiljer sig från den ``vanliga'', spänningsfulla
strängen på flera olika sätt. Den försvinnande spänningen ger upphov till en
utvidgad symmetri av rumtiden, själva målrummet som vi tänker oss strängens
världsytan vara inbäddad i. I den kvantteoretiska diskussionen av strängen blir
skillnaden ännu mera drastiskt. Den spänningslösa strängens spektrum kollapsar
till en enhetlig massnivå: alla excitationer blir masslösa. Detta gäller även
för tachyoniska excitationer som brukar vara instabila och måste elimineras ur
det fysikaliska spektrumet på grund av att dem har en imaginär massa. Den
spänningslösa strängen har ingen kritisk dimension. Kvantiseringen är möjligt i
alla rumstidsdimensioner och inte bara i tio eller 26 dimensioner som för den
spänningsfulla strängen. Den utvidgade rumtidssymmetrien bevaras dock bara för
$D=2$ dimensioner, medans kvantiseringen annars leder till ett topologiskt
spektrum. Den spänningslösa strängen tros vara en obruten fas inom strängteorin
då alla tillstånd är fortfarande jämställda och att det finns en fasövergång mot
lägre energier som ger upphov till dem olika energinivåer.

Den spänningslösa strängens excitationssprektrum innehåller tillstånd med högre
spinn. Det tyder mot ett samband med högre spinn gaugeteori. Detta går enklast
att förstå i sambandet med AdS/CFT-korrespondensen. 
Antagligen har alla någon gång sett ett hologram: Informationen av ett
tredimensionellt objekt sparas på en tvådimensionell yta. I strängteorin säger
detta holografiska princip i dess mest kända version att strängteori i ett
Anti-de-Sitter rum är ekvivalent med en konform fältteori som lever på detta
rummets rand.  Sedan dess upptäckt 1997 testades korrespondensen på flera
olika sett och levererade några intressanta resultat som att vissa sektorer av
strängteorin är integrabla modeller som via den duala beskrivningen kan lösas med metoder av
kondenserade materiens fysiken. En fullständig bevis saknas dock än idag.
Korrespondensen relaterar gaugeteorins kopplingskonstant och strängens
spännings. Den spänningslösa strängen svarar mot en fri fältteori som tillåter
just högre spinn gaugefält.

Femdimensionellt anti-de Sitter rum är del av en tiodimensionell bakgrund för
typ-IIB strängteori, $AdS_5\times S^5$. Tyvärr ligger speciellt kvantiseringen
av strängteorin i denna bakgrund utanför vår nuvarande förmåga. $AdS_5\times S^5$
är en av tre kända bakgrunder för typ-IIB strängteorin som är maximalt
supersymmetrisk: den bevarar 32 supersymmetrier. De andra två bakgrunden brevid 
$AdS_5\times S^5$ är det tomma, plana rummet samt en såkallad planvågsnbakgrund
som upptäcktes för för ett par år sedan. Den delar en del egenskapar med
$AdS_5\times S^5$ men är mycket enklare än den sistnämda. Det visar sig även att
planvågsbakgrunden är en viss gräns av $AdS_5\times S^5$. Dess enkelhet gör det
möjligt att diskutera och kvantisera den slutna typ-IIB strängen i denna
bakgrund.

Det sättet på det strängteorin påverkar rumtidsgeometrin är väldigt
fascinerande. När vi pratade om kompaktifiering så diskuterade vi redan att det
interna sexdimensionella rummet måste vara av en viss typ. Geometrin bestäms
genom att vi kräver $N=1$ supersymmetri i vårt fyradimensionella rum. När det
interna rummet är dessutom Kähler så finns det bara en enda möjlighet: Det måste
vara en Calabi-Yau mångfalt. Även om det var känd att det fanns flera andra
alternativ så diskuterades dem inte på allvar under en lång tid. För en
sigmamodell med supersymmetrin på världsytan bestäms geometrin genom världsytans
dimension och antalet supersymmetrier. Den $N=(1,1)$ supersymmetriska
sigmamodellen till exempel har två ytterligare supersymmetrier om målmångfalden
är bihermitsk. Även om dessa geometrier var kända så klassades de för det mesta
som mindre viktigt i samband med strängteorin. Först för några år sedan
utvecklades ett nytt matematiskt koncept som förenar komplexa och symplektiska
geometrin och dessutom interpolerar mellan dem två. Det är det rätta verktyget
för studier av världsytesupersymmetriens relation till målmångfaldens geometri.
Det visade sig att en viss delmängd av dessa nya geometrier, de så kallade
generaliserade Kähler geometrier, är identiska med den bi-hermitska geometrin
och att det finns en fullständigt beskrivning av generaliserad Kähler geometri
med hjälp av manifest $N=(2,2)$ supersymmetri. Generaliserad Calabi-Yau geometri
är en annan viktig delmängd som idag är viktig i samband med
flödeskompaktifieringar. Slutligen finns det potential för att generaliserad komplex
geometri kan ge en matematisk förklaring av spegelsymmetrin, eftersom den
förenar den topologiska A- och B-modellen i en enda modell.

Vi slutar med en sammanfattning av originalarbeten som denna avhandling grundar
på.

\subsubsection*{Artikel I}
I den första artikeln beskriver vi hur spänningslösa strängar ger upphov till
bakgrundslösningar till typ-IIB supergravitation. Det gör vi genom att betraktar geometrin som
härstammar från en makroskopisk sträng i den gränsen då strängen rör sig med
ljushastighet. I denna gräns försvinner strängens spänning och geometrin liknar
en gravitaionell chockvåg.

\subsubsection*{Artikel II}
Vi studerar den spänningslösa, slutna typ-IIB strängen i den maximalt
supersymmetriska planvågsgeometrin. Lösningen liknar det spänningsfulla fallet.
Även kvantiseringen är inte problematiskt till skillnad från det plana rummet.
Detta hänger ihop med existensen av en parameter som är relaterad till
bakgrundens krökning. Vi visar även att den spänningslösa strängen fås i en viss
gräns av det spänningsfulla fallet och konstaterar att gränsen kommuterar med
kvantiseringen.

\subsubsection*{Artikel III}
I tredje artikeln diskuterar vi villkoret för en generaliserad sigmamodell med
två supersymmetrier att ha ytterligare supersymmetrier. Vi ser dem
involverade tensorerna grupperar sig på ett naturligt sätt till geometriska
objekt som tyder mod en tolkning bortom generaliserad komplex geometri. På grund
av att vi inte har tillräckligt förståelse av denna typ av geometri är vi bundna
till en väldigt enkel sigmamodell där vi kan bara identifiera dem väsentliga
geometriska objekt samt förklara hur generaliserad komplex geometri är inbäddad
i denna beskrivning.

\subsubsection*{Artikel IV}
Vi förklarar relationen mellan generaliserad Kähler geometri och bi-hermitsk
geometri ur en fysikalisk synvinkel. Vi visar att generaliserad Kähler geometri
uppstår som villkor för $N=(2,2)$ supersymmetri i fasrummet. Relationen mellan
generaliserad Kähler geometri och den bi-hermitska geometrin kan därför
tolkas genom ekvivalensen av Hamilton- och Lagrangebeskrivningen av
den supersymmetriska sigmamodellen. I diskussionen av topologiska twists hittar
vi en första tillämpning av våra resultat.

\subsubsection*{Artikel V}
I denna artikel diskuterar vi villkoret för $N=(4,4)$ supersymmetri i
Hamiltonformuleringen av sigmamodellen. Byggande på den förra artikeln hittar vi
en definition av generaliserad hyperkähler geometri och definierar twistorrummet
för dem generaliserade komplexa strukturer.

\parskip \pskip
\selectlanguage{\english}

    \chapter*{Acknowledgments} 
At this place, I would like to take the opportunity to thank a number of people
for their support during the last four years. First of all, I would like to thank 
my supervisor Ulf Lindström. This thesis would not have been possible without all 
his support and encouragement. 

I would like to thank the former and present graduate students and postdocs
the whole department for making the past four years such a fabulous
and enjoyable time with all the discussions and entertaining moments.
Especially, I would like to thank my roommate Jonas for the nice athmosphere in
our office and all our collaborations. I also thank Rolf for teaching me how to
have practical use of gravity when standing on top of a mountain in the middle
of the winter.

I also thank my other collaborators Linus and Maxim. Especially Maxim was a
great source of inspiration during the last year of my studies.  Special thanks
go to Niklas, Laura and Maxim for reviewing and commenting on this thesis. 

Last but not least, I want to thank my parents Gisela and Wilhelm for all their
support and patience and especially for bringing me in close contact to physics
already in very early years. I would have not been where I am now without them.


    \backmatter
    \nocite{*} 
    \bibliographystyle{alpha}
    \bibliography{Thesis}

\newcommand{\etalchar}[1]{$^{#1}$}
\begin{thebibliography}{BFOHP02b}

\bibitem[AS71]{Aichelburg:1970dh}
P.~C. Aichelburg and R.~U. Sexl.
\newblock On the gravitational field of a massless particle.
\newblock {\em Gen. Rel. Grav.}, 2:303--312, 1971.

\bibitem[AS05]{Alekseev:2004np}
Anton Alekseev and Thomas Strobl.
\newblock Current algebra and differential geometry.
\newblock {\em JHEP}, 03:035, 2005.

\bibitem[AY68]{Ako:1968}
M.~Ako and K.~Yano.
\newblock On certain operators associate with tensor fields.
\newblock {\em Kodai. Math. Sem. Rep.}, 20:414, 1968.

\bibitem[BCZ05]{Bonechi:2005mw}
Francesco Bonechi, Alberto~S. Cattaneo, and Maxim Zabzine.
\newblock Geometric quantization and non-perturbative {P}oisson sigma model.
\newblock 2005.

\bibitem[BDVH76]{Brink:1976sc}
L.~Brink, P.~Di~Vecchia, and Paul~S. Howe.
\newblock A locally supersymmetric and reparametrization invariant action for
  the spinning string.
\newblock {\em Phys. Lett.}, B65:471--474, 1976.

\bibitem[Ber05]{Bergamin:2004sk}
L.~Bergamin.
\newblock Generalized complex geometry and the poisson sigma model.
\newblock {\em Mod. Phys. Lett.}, A20:985--996, 2005.

\bibitem[BFOHP02a]{Blau:2001ne}
Matthias Blau, Jose Figueroa-O'Farrill, Christopher Hull, and George
  Papadopoulos.
\newblock A new maximally supersymmetric background of {IIB} superstring
  theory.
\newblock {\em JHEP}, 01:047, 2002.

\bibitem[BFOHP02b]{Blau:2002dy}
Matthias Blau, Jose Figueroa-O'Farrill, Christopher Hull, and George
  Papadopoulos.
\newblock Penrose limits and maximal supersymmetry.
\newblock {\em Class. Quant. Grav.}, 19:L87--L95, 2002.

\bibitem[BFOP02]{Blau:2002mw}
Matthias Blau, Jose Figueroa-O'Farrill, and George Papadopoulos.
\newblock Penrose limits, supergravity and brane dynamics.
\newblock {\em Class. Quant. Grav.}, 19:4753, 2002.

\bibitem[BLR88]{Buscher:1987uw}
T.~Buscher, U.~Lindström, and M.~Ro\v{c}ek.
\newblock New supersymmetric sigma models with {W}ess-{Z}umino terms.
\newblock {\em Phys. Lett.}, B202:94, 1988.

\bibitem[BMN02]{Berenstein:2002jq}
David Berenstein, Juan~M. Maldacena, and Horatiu Nastase.
\newblock Strings in flat space and pp waves from ${N} = 4$ super {Y}ang
  {M}ills.
\newblock {\em JHEP}, 04:013, 2002.

\bibitem[BNRA89]{Barcelos-Neto:1989ms}
J.~Barcelos-Neto and M.~Ruiz-Altaba.
\newblock Superstrings with zero tension.
\newblock {\em Phys. Lett.}, B228:193, 1989.

\bibitem[Bon03]{Bonelli:2003zu}
Giulio Bonelli.
\newblock On the covariant quantization of tensionless bosonic strings in {AdS}
  spacetime.
\newblock {\em JHEP}, 11:028, 2003.

\bibitem[BOPT03]{Blau:2003rt}
Matthias Blau, Martin O'Loughlin, George Papadopoulos, and Arkady~A. Tseytlin.
\newblock Solvable models of strings in homogeneous plane wave backgrounds.
\newblock {\em Nucl. Phys.}, B673:57--97, 2003.

\bibitem[BZ05]{Bonelli:2005ti}
Giulio Bonelli and Maxim Zabzine.
\newblock From current algebras for p-branes to topological {M}- theory.
\newblock {\em JHEP}, 09:015, 2005.

\bibitem[C{\etalchar{+}}03]{Callan:2003xr}
Jr. Callan, Curtis~G. et~al.
\newblock Quantizing string theory in ${AdS}_5 \times {S}^5$: {B}eyond the
  pp-wave.
\newblock {\em Nucl. Phys.}, B673:3--40, 2003.

\bibitem[Cal05]{Calvo:2005ww}
Ivan Calvo.
\newblock Supersymmetric {WZ}-{P}oisson sigma model and twisted generalized
  complex geometry.
\newblock 2005.

\bibitem[CT97]{Cederwall:1997ts}
Martin Cederwall and P.~K. Townsend.
\newblock The manifestly ${SL(2,Z)}$-covariant superstring.
\newblock {\em JHEP}, 09:003, 1997.

\bibitem[CvGNW97]{Cederwall:1996pv}
Martin Cederwall, Alexander von Gussich, Bengt E.~W. Nilsson, and Anders
  Westerberg.
\newblock The {D}irichlet super-three-brane in ten-dimensional type {IIB}
  supergravity.
\newblock {\em Nucl. Phys.}, B490:163--178, 1997.

\bibitem[Dab95]{Dabholkar:1995ep}
Atish Dabholkar.
\newblock Ten-dimensional heterotic string as a soliton.
\newblock {\em Phys. Lett.}, B357:307--312, 1995.

\bibitem[dAS96]{deAlwis:1996ze}
S.~P. de~Alwis and K.~Sato.
\newblock {D}-strings and {F}-strings from string loops.
\newblock {\em Phys. Rev.}, D53:7187--7196, 1996.

\bibitem[DGHRR90]{Dabholkar:1990yf}
Atish Dabholkar, Gary~W. Gibbons, Jeffrey~A. Harvey, and Fernando Ruiz~Ruiz.
\newblock Superstrings and solitons.
\newblock {\em Nucl. Phys.}, B340:33--55, 1990.

\bibitem[Dir50]{Dirac:1950pj}
Paul A.~M. Dirac.
\newblock Generalized hamiltonian dynamics.
\newblock {\em Can. J. Math.}, 2:129--148, 1950.

\bibitem[DM06a]{Davidov:2005}
Johann Davidov and Oleg Mushkarov.
\newblock Twistor spaces of generalized complex structures.
\newblock {\em Journal of Geometry and Physics}, 56:1623, 2006.

\bibitem[DM06b]{Davidov:2006}
Johann Davidov and Oleg Mushkarov.
\newblock Twistorial construction of generalized {K}ähler manifolds.
\newblock 2006.

\bibitem[dVGN95]{deVega:1994hu}
H.~J. de~Vega, I.~Giannakis, and A.~Nicolaidis.
\newblock String quantization in curved space-times: {N}ull string approach.
\newblock {\em Mod. Phys. Lett.}, A10:2479--2484, 1995.

\bibitem[DZ76]{Deser:1976rb}
S.~Deser and B.~Zumino.
\newblock A complete action for the spinning string.
\newblock {\em Phys. Lett.}, B65:369--373, 1976.

\bibitem[ES05]{Engquist:2005yt}
J.~Engquist and P.~Sundell.
\newblock Brane partons and singleton strings.
\newblock 2005.

\bibitem[GG75]{Gurses:1974cm}
Metin Gurses and Feza Gursey.
\newblock Derivation of the string equation of motion in general relativity.
\newblock {\em Phys. Rev.}, D11:967, 1975.

\bibitem[GGRS83]{Gates:1983nr}
S.~J. Gates, Marcus~T. Grisaru, M.~Ro\v{c}ek, and W.~Siegel.
\newblock Superspace, or one thousand and one lessons in supersymmetry.
\newblock {\em Front. Phys.}, 58:1--548, 1983.

\bibitem[GHR84]{Gates:1984nk}
Jr. Gates, S.~J., C.~M. Hull, and M.~Ro\v{c}ek.
\newblock Twisted multiplets and new supersymmetric nonlinear sigma models.
\newblock {\em Nucl. Phys.}, B248:157, 1984.

\bibitem[GKP98]{Gubser:1998bc}
S.~S. Gubser, Igor~R. Klebanov, and Alexander~M. Polyakov.
\newblock Gauge theory correlators from non-critical string theory.
\newblock {\em Phys. Lett.}, B428:105--114, 1998.

\bibitem[GLS{\etalchar{+}}95]{Gustafsson:1994kr}
H.~Gustafsson, U.~Lindström, P.~Saltsidis, B.~Sundborg, and R.~van Unge.
\newblock Hamiltonian {BRST} quantization of the conformal string.
\newblock {\em Nucl. Phys.}, B440:495--520, 1995.

\bibitem[GM87]{Gross:1987kz}
David~J. Gross and Paul~F. Mende.
\newblock The high-energy behavior of string scattering amplitudes.
\newblock {\em Phys. Lett.}, B197:129, 1987.

\bibitem[Got05]{Goto:2005}
Ryushi Goto.
\newblock On deformations of generalized {C}alabi-{Y}au, hyper{K}ähler, ${G}_2$
  and ${S}pin(7)$ structures {I}.
\newblock 2005.

\bibitem[Gra06]{Grana:2005jc}
Mariana Grana.
\newblock Flux compactifications in string theory: A comprehensive review.
\newblock {\em Phys. Rept.}, 423:91--158, 2006.

\bibitem[GSW87]{Green:1987sp}
Michael~B. Green, J.~H. Schwarz, and Edward Witten.
\newblock Superstring theory.
\newblock 1987.
\newblock Cambridge, Uk: Univ. Pr. ( 1987) 469 P. ( Cambridge Monographs On
  Mathematical Physics).

\bibitem[Gua03]{Gualtieri:2003dx}
Marco Gualtieri.
\newblock Generalized complex geometry.
\newblock 2003.

\bibitem[Gue00]{Gueven:2000ru}
R.~Gueven.
\newblock Plane wave limits and {T}-duality.
\newblock {\em Phys. Lett.}, B482:255--263, 2000.

\bibitem[Hit03]{Hitchin:2004ut}
Nigel Hitchin.
\newblock Generalized {C}alabi-{Y}au manifolds.
\newblock {\em Quart. J. Math. Oxford Ser.}, 54:281--308, 2003.

\bibitem[Hit06]{Hitchin:2005cv}
Nigel Hitchin.
\newblock Instantons, poisson structures and generalized kaehler geometry.
\newblock {\em Commun. Math. Phys.}, 265:131--164, 2006.

\bibitem[HKLR87]{Hitchin:1986ea}
Nigel~J. Hitchin, A.~Karlhede, U.~Lindstrom, and M.~Rocek.
\newblock Hyperkahler metrics and supersymmetry.
\newblock {\em Commun. Math. Phys.}, 108:535, 1987.

\bibitem[HMS00]{Haggi-Mani:2000ru}
Parviz Haggi-Mani and Bo~Sundborg.
\newblock Free large ${N}$ supersymmetric {Y}ang-{M}ills theory as a string
  theory.
\newblock {\em JHEP}, 04:031, 2000.

\bibitem[HP88]{Howe:1988cj}
Paul~S. Howe and G.~Papadopoulos.
\newblock Further remarks on the geometry of two-dimensional nonlinear sigma
  models.
\newblock {\em Class. Quant. Grav.}, 5:1647--1661, 1988.

\bibitem[HS05]{Howe:2004ib}
P.~S. Howe and E.~Sezgin.
\newblock The supermembrane revisited.
\newblock {\em Class. Quant. Grav.}, 22:2167--2200, 2005.

\bibitem[Hul95]{Hull:1995nu}
C.~M. Hull.
\newblock String-string duality in ten-dimensions.
\newblock {\em Phys. Lett.}, B357:545--551, 1995.

\bibitem[Huy05]{Huybrechts:2003ak}
Daniel Huybrechts.
\newblock Generalized {C}alabi-{Y}au structures, ${K3}$ surfaces, and
  ${B}$-fields.
\newblock {\em Int. J. Math.}, 16:13, 2005.

\bibitem[ILS92]{Isberg:1992ia}
J.~Isberg, U.~Lindström, and B.~Sundborg.
\newblock Space-time symmetries of quantized tensionless strings.
\newblock {\em Phys. Lett.}, B293:321--326, 1992.

\bibitem[ILST94]{Isberg:1993av}
J.~Isberg, U.~Lindström, B.~Sundborg, and G.~Theodoridis.
\newblock Classical and quantized tensionless strings.
\newblock {\em Nucl. Phys.}, B411:122--156, 1994.

\bibitem[Joh03]{Johnson:2003gi}
C.~V. Johnson.
\newblock D-branes.
\newblock 2003.
\newblock Cambridge, USA: Univ. Pr. (2003) 548 p.

\bibitem[KG84]{Kowalski-Glikman:1984wv}
J.~Kowalski-Glikman.
\newblock Vacuum states in supersymmetric {K}aluza-{K}lein theory.
\newblock {\em Phys. Lett.}, B134:194--196, 1984.

\bibitem[KL86]{Karlhede:1986wb}
A.~Karlhede and U.~Lindström.
\newblock The classical bosonic string in the zero tension limit.
\newblock {\em Class. Quant. Grav.}, 3:L73--L75, 1986.

\bibitem[Lin04]{Lindstrom:2004eh}
Ulf Lindström.
\newblock Generalized ${N} = (2,2)$ supersymmetric non-linear sigma models.
\newblock {\em Phys. Lett.}, B587:216--224, 2004.

\bibitem[Lin06]{Lindstrom:2006ee}
Ulf Lindström.
\newblock A brief review of supersymmetric non-linear sigma models and
  generalized complex geometry.
\newblock 2006.

\bibitem[LMTZ05]{Lindstrom:2004iw}
Ulf Lindström, Ruben Minasian, Alessandro Tomasiello, and Maxim Zabzine.
\newblock Generalized complex manifolds and supersymmetry.
\newblock {\em Commun. Math. Phys.}, 257:235--256, 2005.

\bibitem[LRSS86]{Lizzi:1986nv}
F.~Lizzi, B.~Rai, G.~Sparano, and A.~Srivastava.
\newblock Quantization of the null string and absence of critical dimensions.
\newblock {\em Phys. Lett.}, B182:326, 1986.

\bibitem[LRvUZ05a]{Lindstrom:2004hi}
Ulf Lindström, Martin Ro\v{c}ek, Rikard von Unge, and Maxim Zabzine.
\newblock Generalized {K}ähler geometry and manifest ${N} = (2,2)$
  supersymmetric nonlinear sigma-models.
\newblock {\em JHEP}, 07:067, 2005.

\bibitem[LRvUZ05b]{Lindstrom:2005zr}
Ulf Lindström, Martin Ro\v{c}ek, Rikard von Unge, and Maxim Zabzine.
\newblock Generalized {K}ähler manifolds and off-shell supersymmetry.
\newblock 2005.

\bibitem[LST91]{Lindstrom:1990qb}
U.~Lindström, B.~Sundborg, and G.~Theodoridis.
\newblock The zero tension limit of the superstring.
\newblock {\em Phys. Lett.}, B253:319--323, 1991.

\bibitem[LZ04]{Lindstrom:2003mg}
Ulf Lindström and Maxim Zabzine.
\newblock Tensionless strings, {WZW} models at critical level and massless
  higher spin fields.
\newblock {\em Phys. Lett.}, B584:178--185, 2004.

\bibitem[Mal98]{Maldacena:1997re}
Juan~M. Maldacena.
\newblock The large {N} limit of superconformal field theories and
  supergravity.
\newblock {\em Adv. Theor. Math. Phys.}, 2:231--252, 1998.

\bibitem[Mal06]{Malikov:2006rm}
Fyodor Malikov.
\newblock Lagrangian approach to sheaves of vertex algebras.
\newblock 2006.

\bibitem[Met02]{Metsaev:2001bj}
R.~R. Metsaev.
\newblock Type {IIB} {G}reen-{S}chwarz superstring in plane wave {R}amond-
  {R}amond background.
\newblock {\em Nucl. Phys.}, B625:70--96, 2002.

\bibitem[MM84]{Magri:1984}
F.~Magri and C.~Morosi.
\newblock A geometric characterization of integrable {H}amiltonian systems
  through the theory of {P}oisson-{N}ijenhuis manifolds.
\newblock {\em Università di Milano, Quaderno}, S19, 1984.

\bibitem[Moh03]{Mohaupt:2002py}
Thomas Mohaupt.
\newblock Introduction to string theory.
\newblock {\em Lect. Notes Phys.}, 631:173--251, 2003.

\bibitem[MS06]{Maes:2006bm}
Joris Maes and Alexander Sevrin.
\newblock A note on ${N} = (2,2)$ superfields in two dimensions.
\newblock 2006.

\bibitem[MT02]{Metsaev:2002re}
R.~R. Metsaev and A.~A. Tseytlin.
\newblock Exactly solvable model of superstring in plane wave {R}amond-
  {R}amond background.
\newblock {\em Phys. Rev.}, D65:126004, 2002.

\bibitem[Nak90]{Nakahara:1990th}
M.~Nakahara.
\newblock Geometry, topology and physics.
\newblock 1990.
\newblock Bristol, UK: Hilger (1990) 505 p. (Graduate student series in
  physics).

\bibitem[Pen72]{Penrose:1972}
Roger Penrose.
\newblock Techniques of differential topology in relativity.
\newblock 1972.
\newblock SIAM, Philadelphia.

\bibitem[Pen76]{Penrose:1976}
Roger Penrose.
\newblock Nonlinear gravitons and curved twistor theory.
\newblock {\em Gen. Rel. Grav.}, 7:31, 1976.

\bibitem[Pes06]{Pestun:2006rj}
Vasily Pestun.
\newblock Topological strings in generalized complex space.
\newblock 2006.

\bibitem[Ple04]{Plefka:2003nb}
Jan~Christoph Plefka.
\newblock Lectures on the plane-wave string / gauge theory duality.
\newblock {\em Fortsch. Phys.}, 52:264--301, 2004.

\bibitem[Pol81a]{Polyakov:1981rd}
Alexander~M. Polyakov.
\newblock Quantum geometry of bosonic strings.
\newblock {\em Phys. Lett.}, B103:207--210, 1981.

\bibitem[Pol81b]{Polyakov:1981re}
Alexander~M. Polyakov.
\newblock Quantum geometry of fermionic strings.
\newblock {\em Phys. Lett.}, B103:211--213, 1981.

\bibitem[Pol95]{Polchinski:1995mt}
Joseph Polchinski.
\newblock Dirichlet-branes and ramond-ramond charges.
\newblock {\em Phys. Rev. Lett.}, 75:4724--4727, 1995.

\bibitem[Pol98]{Polchinski:1998rq}
J.~Polchinski.
\newblock String theory.
\newblock 1998.
\newblock Cambridge, UK: Univ. Pr. (1998) 402 p.

\bibitem[PR02]{Parnachev:2002kk}
Andrei Parnachev and Anton~V. Ryzhov.
\newblock Strings in the near plane wave background and {AdS/CFT}.
\newblock {\em JHEP}, 10:066, 2002.

\bibitem[Sal82]{Salamon:1982}
S.~Salamon.
\newblock Quaternionic {K}ähler manifolds.
\newblock {\em Invent. Math.}, 67:143, 1982.

\bibitem[Sal86]{Salamon:1986}
S.~Salamon.
\newblock Differential geometry of quaternionic manifolds.
\newblock {\em Ann. Sci. Ec. Norm. Sup. Paris}, 19:31, 1986.

\bibitem[Sal95]{Saltsidis:1995qr}
P.~Saltsidis.
\newblock Hamiltonian {BRST} quantization of the conformal spinning string.
\newblock {\em Nucl. Phys.}, B446:286--298, 1995.

\bibitem[Sav04]{Savvidy:2003fx}
G.~K. Savvidy.
\newblock Tensionless strings: Physical {F}ock space and higher spin fields.
\newblock {\em Int. J. Mod. Phys.}, A19:3171--3194, 2004.

\bibitem[Sch77]{Schild:1976vq}
Alfred Schild.
\newblock Classical null strings.
\newblock {\em Phys. Rev.}, D16:1722, 1977.

\bibitem[Sch95]{Schwarz:1995dk}
John~H. Schwarz.
\newblock An ${SL(2,Z)}$ multiplet of type {IIB} superstrings.
\newblock {\em Phys. Lett.}, B360:13--18, 1995.

\bibitem[SS94]{Schaller:1994es}
Peter Schaller and Thomas Strobl.
\newblock Poisson structure induced (topological) field theories.
\newblock {\em Mod. Phys. Lett.}, A9:3129--3136, 1994.

\bibitem[SS02]{Sezgin:2002rt}
E.~Sezgin and P.~Sundell.
\newblock Massless higher spins and holography.
\newblock {\em Nucl. Phys.}, B644:303--370, 2002.

\bibitem[ST97]{Sevrin:1996jr}
Alexander Sevrin and Jan Troost.
\newblock Off-shell formulation of ${N = 2}$ non-linear sigma-models.
\newblock {\em Nucl. Phys.}, B492:623--646, 1997.

\bibitem[Sun01]{Sundborg:2000wp}
Bo~Sundborg.
\newblock Stringy gravity, interacting tensionless strings and massless higher
  spins.
\newblock {\em Nucl. Phys. Proc. Suppl.}, 102:113--119, 2001.

\bibitem[Sza02]{Szabo:2002ca}
Richard~J. Szabo.
\newblock Busstepp lectures on string theory: An introduction to string theory
  and d-brane dynamics.
\newblock 2002.

\bibitem[Vas99]{Vasiliev:1999ba}
Mikhail~A. Vasiliev.
\newblock Higher spin gauge theories: {S}tar-product and {AdS} space.
\newblock 1999.

\bibitem[Wit96]{Witten:1995im}
Edward Witten.
\newblock Bound states of strings and p-branes.
\newblock {\em Nucl. Phys.}, B460:335--350, 1996.

\bibitem[Wit98]{Witten:1998qj}
Edward Witten.
\newblock Anti-de sitter space and holography.
\newblock {\em Adv. Theor. Math. Phys.}, 2:253--291, 1998.

\bibitem[Zab06a]{Zabzine:2005qf}
Maxim Zabzine.
\newblock Hamiltonian perspective on generalized complex structure.
\newblock {\em Commun. Math. Phys.}, 263:711--722, 2006.

\bibitem[Zab06b]{Zabzine:2006uz}
Maxim Zabzine.
\newblock Lectures on generalized complex geometry and supersymmetry.
\newblock 2006.

\bibitem[Zwi04]{Zwiebach:2004tj}
B.~Zwiebach.
\newblock A first course in string theory.
\newblock 2004.
\newblock Cambridge, UK: Univ. Pr. (2004) 558 p.

\end{thebibliography}
\end{document}